\documentclass[twocolumn,superscriptaddress,nofootinbib]{revtex4-1}
\usepackage[utf8]{inputenc} 
\usepackage[T1]{fontenc}    
\usepackage{hyperref}       
\usepackage{url}            
\usepackage{booktabs}       
\usepackage{amsfonts}       
\usepackage{nicefrac}       
\usepackage{microtype}      
\usepackage{lipsum}
\usepackage{amsmath}
\usepackage{graphicx}
\usepackage{comment}
\usepackage{xcolor}         

\definecolor{midgreen}{rgb}{0.52, 0.73, 0.4}

\usepackage{float}
\DeclareMathOperator{\sech}{sech}
\newcommand{\Lagr}{\mathcal{L}}

\hypersetup{pdftex,colorlinks=true,linkcolor=blue,citecolor=magenta,menucolor=black,urlcolor=blue,filecolor=blue}

\begin{document}

\title{Transport coefficients of the quark-gluon plasma at the critical point and across the first-order line}

\author{Joaquin Grefa}
\affiliation{Physics Department, University of Houston, Houston, Texas 77204, USA}
\author{Mauricio Hippert}
\affiliation{Illinois Center for Advanced Studies of the Universe, Department of Physics, University of Illinois at Urbana-Champaign, Urbana, Illinois 61801, USA}

\author{Jorge Noronha} 
\affiliation{Illinois Center for Advanced Studies of the Universe, Department of Physics, University of Illinois at Urbana-Champaign, Urbana, Illinois 61801, USA}

\author{Jacquelyn Noronha-Hostler}
\affiliation{Illinois Center for Advanced Studies of the Universe, Department of Physics, University of Illinois at Urbana-Champaign, Urbana, Illinois 61801, USA}

\author{Israel Portillo}
\affiliation{Physics Department, University of Houston, Houston, Texas 77204, USA}

\author{Claudia Ratti}
\affiliation{Physics Department, University of Houston, Houston, Texas 77204, USA}

\author{Romulo Rougemont}
\affiliation{Instituto de F\'{i}sica, Universidade Federal de Goi\'{a}s, Av. Esperan\c{c}a - Campus Samambaia, CEP 74690-900, Goi\^{a}nia, Goi\'{a}s, Brazil}

\date{\today}

\begin{abstract}

A bottom-up Einstein-Maxwell-dilaton holographic model is used to compute, for the first time, the behavior of several transport coefficients of the hot and baryon-rich strongly coupled quark-gluon plasma at the critical point and also across the first-order phase transition line in the phase diagram. The observables under study are the shear and bulk viscosities, the baryon and thermal conductivities, the baryon diffusion, the jet quenching parameter $\hat{q}$, as well as the heavy-quark drag force and the Langevin diffusion coefficients. These calculations provide a phenomenologically promising estimate for these coefficients, given that our model quantitatively reproduces lattice QCD thermodynamics results, both at zero and finite baryon density, besides naturally incorporating the nearly perfect fluidity of the quark-gluon plasma. We find that the diffusion of baryon charge, and also the shear and bulk viscosities, are suppressed with increasing  baryon density, indicating that the medium becomes even closer to perfect fluidity at large densities. On the other hand, the jet quenching parameter and the heavy-quark momentum diffusion are enhanced with increasing density. The observables display a discontinuity gap when crossing the first-order phase transition line, while developing an infinite slope at the critical point. The transition temperatures associated with different transport coefficients differ in the crossover region but are found to converge at the critical point. 
\end{abstract}
\maketitle
\tableofcontents

\section{Introduction}
\label{sec:introduction}

More than two decades have been devoted to the study of the quark-gluon plasma (QGP), the deconfined phase of strongly interacting QCD matter that existed in the early universe  microseconds after the big bang and that can be created in relativistic heavy-ion collision experiments at Brookhaven National Laboratory (RHIC) \cite{PHENIX:2004vcz,BRAHMS:2004adc,PHOBOS:2004zne,STAR:2005gfr} and at CERN (LHC) \cite{ATLAS:2013xzf}. Different experimental conditions allow us to explore different areas of the phase diagram of strongly interacting matter, from the very high temperatures and vanishing net-baryon density realized at the highest collision energies at the LHC, to the intermediate densities explored at RHIC in its collider and fixed target modes by decreasing the collision energy \cite{Cebra:2014sxa}, all the way up to the high density regime, which is/will be the focus of low-energy experiments such as HADES \cite{HADES:2019auv}, FAIR \cite{Friese:2006dj,Tahir:2005zz,Lutz:2009ff,Durante:2019hzd} and NICA \cite{Kekelidze:2017tgp,Kekelidze:2016wkp}, and which overlaps with the conditions achieved in neutron star mergers \cite{Dexheimer:2020zzs}.

One of the surprising features of this deconfined phase of matter is its almost-perfect fluidity, a property which emerged from the theoretical analysis of the collective behavior observed at RHIC \cite{Heinz:2013th,Luzum:2013yya}, and caused a paradigm shift in our understanding of the theory. What was anticipated as a weakly interacting gas of quarks and gluons turned out to be a strongly coupled fluid, whose transport properties cannot be easily extracted from first principles. In fact, the strongly coupled nature of the system calls for a treatment in terms of lattice QCD simulations, which are however limited in their effective calculation of dynamical quantities \cite{Meyer:2011gj,Ding:2015ona,Ratti:2018ksb,Rothkopf:2019ipj}. Besides the bulk ($\zeta$) and shear ($\eta$) viscosities \cite{Auvinen:2017fjw,Dore:2020jye,Schafer:2021csj}, which are used as input in hydrodynamical simulations of the dynamical evolution of the collision, baryon number diffusion coefficient and conductivity play an important role at finite density \cite{Denicol:2012cn,Denicol:2018wdp,An:2021wof,Du:2021zqz}. Other quantities of relevance for the dynamical description of the QGP are those related to the energy loss of light and heavy flavors in the medium, such as the heavy quark drag force and the Langevin diffusion coefficients \cite{Moore:2004tg}, and also the jet quenching parameter $\hat{q}$ associated with the medium-induced radiated energy loss of light partons \cite{Baier:1996sk}.

In this work, we use the holographic correspondence \cite{Maldacena:1997re,Gubser:1998bc,Witten:1998qj,Witten:1998zw} to study the transport properties of strongly interacting matter in a 5-dimensional Einstein-Maxwell-dilaton (EMD) model across its phase diagram. We make use of the EMD model proposed by some of us in Ref.\ \cite{Critelli:2017oub}, following a similar approach developed in the seminal works of Refs.\ \cite{Gubser:2008yx,DeWolfe:2010he,DeWolfe:2011ts}. The model not only reproduces $2+1$ flavors lattice QCD results with physical values of the quark masses at zero \cite{Borsanyi:2010cj,Borsanyi:2013bia,Bellwied:2015lba} and finite \cite{Borsanyi:2021sxv} baryon density but it also naturally incorporates \cite{Kovtun:2004de} the nearly perfect fluid behavior of the QGP. Recently, thanks to important improvements in the numerical treatment of our EMD model, we extended the coverage of its phase diagram up to values of baryon chemical potential $\mu_B\sim1100$ MeV \cite{Grefa:2021qvt}, unveiling the location of a line of first-order phase transition ending at a critical end point (CEP). While in Ref. \cite{Grefa:2021qvt} the focus was on the analysis of thermodynamic quantities, in the present work we focus on the study of transport properties of holographic QCD matter.

Here we make use of well-known holographic formulas for the transport coefficients considered, and also benefit from the numerical advances introduced in Ref. \cite{Grefa:2021qvt}, in order to determine new physical results and holographic predictions for several transport coefficients relevant for the QGP at finite temperature and baryon density, including not only the crossover and critical regions of the phase diagram, but also, for the first time, their behavior across the first order phase transition line.

For several other applications involving nonconformal dilatonic holographic models of strongly interacting matter, see e.g. Refs.\ \cite{Gursoy:2007cb,Gursoy:2007er,Gursoy:2009kk,Gursoy:2010aa,Finazzo:2014cna,Rougemont:2015oea,Rougemont:2015wca,Rougemont:2015ona,Finazzo:2015xwa,Rougemont:2016nyr,Finazzo:2016mhm,Attems:2016tby,Attems:2016ugt,Critelli:2016cvq,Demircik:2016nhr,Rougemont:2017tlu,Knaute:2017opk,Li:2017ple,Rougemont:2018ivt,Arefeva:2018hyo,Attems:2018gou,Attems:2019yqn,Arefeva:2020vae,Rougemont:2020had,Zollner:2020nnt,Attems:2020qkg,Ballon-Bayona:2020xls,Ballon-Bayona:2021tzw,Cai:2022omk,Li:2022erd,Dudal:2021jav,Jena:2022nzw,Cai:2022omk}. For holographic applications involving QCD in the Veneziano regime (V-QCD), where both the number of colors $N_c$ and the number of flavors $N_f$ go to infinity, with the ratio $N_f/N_c$ kept fixed, see e.g. Refs. \cite{Jarvinen:2011qe,Hoyos:2020hmq,Hoyos:2021njg,Jarvinen:2021jbd,Demircik:2021zll}. For microscopic calculations of transport properties in lattice QCD and other nonholographic approaches also aimed to effectively describe QCD matter, see e.g. Refs.\ \cite{Danielewicz:1984ww,Gavin:1985ph,Aarts:2002cc,Noronha-Hostler:2008kkf,Noronha-Hostler:2012ycm,Denicol:2013nua,Haas:2013hpa,Christiansen:2014ypa,Aarts:2014nba,Kadam:2014cua,Rose:2017bjz,Fotakis:2019nbq,Fotakis:2021diq,McLaughlin:2021dph,Soloveva:2021quj,Kadam:2022xcc,Kadam:2020utt}.

This work is organized as follows. In Sec. \ref{sec:EMD} we briefly review the main features of the bottom-up EMD holographic model constructed in Ref.\ \cite{Critelli:2017oub} and further analyzed with respect to its thermodynamic properties in Ref.\ \cite{Grefa:2021qvt}. The transport of baryon charge in this EMD model is presented in Sec. \ref{sec:baryon}, with the calculation of the baryon and thermal conductivities, and the baryon diffusion coefficient. In Sec. \ref{sec:energyloss} we present our results for energy loss by calculating the heavy-quark drag force, the Langevin diffusion coefficients, and the jet quenching parameter. The  shear and bulk viscosities are discussed in Sec. \ref{sec:viscosities}, while Sec. \ref{out} describes the phase diagram obtained from out-of-equilibrium observables. We summarize our main conclusions in Sec. \ref{sec:conclusion}. In Appendix \ref{sec:numerical} we present some details regarding the numerical calculation of the transport coefficients in the present EMD model.

\emph{Notation:} In this work we use a mostly plus metric signature and natural units $c=\hbar=k_B=1$.

\section{The Holographic EMD Model}
\label{sec:EMD}

The physical quantities computed in the next sections of this work have been previously evaluated by some of us in an older version of the EMD model \cite{Rougemont:2015wca,Rougemont:2015ona,Rougemont:2017tlu} at moderate baryon densities, and far from the CEP and the line of first-order phase transition. The updated EMD model we put forward in Refs.\ \cite{Critelli:2017oub,Grefa:2021qvt}, which we briefly review in this section, comprises a much more precise fitting to lattice data on QCD thermodynamics at zero baryon density \cite{Borsanyi:2013bia,Bellwied:2015lba}, which is employed to fix the free parameters of the EMD model. This improvement turned out in a much better quantitative agreement between the predictions of the EMD model at finite baryon chemical potential and the latest lattice results on QCD thermodynamics at finite baryon density \cite{Borsanyi:2021sxv}. Going beyond what has been done in previous papers, in the present work we are going to cover also the phase transition region when evaluating all the transport coefficients.

The bulk EMD action is given by \cite{DeWolfe:2010he,Critelli:2017oub}
\begin{eqnarray} \label{eq:action}
    S&=&\int_{\mathcal{M}_5} d^5 x \Lagr = \frac{1}{2\kappa_{5}^{2}}\int_{\mathcal{M}_5} d^{5}x\sqrt{-g}\times
    \nonumber\\
    &\times&\left[R-\frac{(\partial_\mu \phi)^2}{2}-V(\phi)-\frac{f(\phi)F_{\mu\nu}^{2}}{4}\right],
\end{eqnarray}
where $\kappa_{5}^{2}\equiv 8\pi G_{5}$ is the 5-dimensional gravitational constant, $g_{\mu\nu}$ is the bulk metric field with the associated Ricci scalar $R$, $A_{\mu}$ is a Maxwell field with the associated strength tensor $F_{\mu\nu}=\partial_{\mu}A_{\nu}-\partial_{\nu}A_{\mu}$, and $\phi$ is a scalar called the dilaton field, which has an associated potential $V(\phi)$ and couples to the Maxwell field through the coupling function $f(\phi)$. The dilaton field is responsible for breaking the conformal invariance of the 4-dimensional dual gauge theory living at the boundary of the higher dimensional bulk. Having QCD as the target dual gauge theory at the boundary, the conformal symmetry breaking is implemented by fixing $V(\phi)$ in the bulk such as to have the holographic equation of state quantitatively matching the corresponding lattice QCD results with $2+1$ flavors and physical values of the quark masses evaluated at vanishing chemical potential \cite{Borsanyi:2013bia,Bellwied:2015lba}. On the other hand, the boundary value of the Maxwell field is meant to introduce the baryon chemical potential $\mu_B$ in the dual gauge theory, which may be accomplished by fixing the Maxwell-Dilaton coupling function $f(\phi)$ such that the holographic second-order baryon number susceptibility at $\mu_B=0$ quantitatively matches the corresponding lattice QCD results \cite{Borsanyi:2011sw,Bellwied:2015lba}. The charged, isotropic, rotationally invariant, and asymptotically anti-de Sitter (AdS) black hole solutions can be described by the following Ansatz for the EMD fields \cite{DeWolfe:2010he,Critelli:2017oub}:

\begin{equation}
\begin{array}{rcl} \label{eq:ansatz}
     ds^2 &=& e^{2A(r)}[-h(r)dt^2+d\vec{x}^2]+\frac{dr^{2}}{h(r)}, \\
     \phi &=& \phi(r), \\
     A &=& A_{\mu}dx^{\mu}=\Phi(r)dt.
\end{array}
\end{equation}
The black hole event horizon is given by the largest root of the equation $h(r_{H})=0$, and the asymptotically AdS boundary is located at $r\rightarrow\infty$. The AdS radius is set to unity for simplicity and, in turn, an energy scaling factor $\Lambda$ is introduced to convert the quantities computed from the gravity side of the holographic duality to field theory units in MeV \cite{Critelli:2017oub}:
\begin{equation} \label{eq:thermodynamics}
\begin{array}{cc}
     &T=\frac{1}{4\pi\phi_{A}^{1/\nu}\sqrt{h_{0}^{\textrm{far}}}}\Lambda, \quad \mu_{B}=\frac{\Phi_{0}^{\textrm{far}}}{\phi_{A}^{1/\nu}\sqrt{h_{0}^{\textrm{far}}}}\Lambda, \\
      &s=\frac{2\pi}{\kappa_{5}^{2}\phi_{A}^{3/\nu}}\Lambda^{3}, \qquad \rho_{B}=-\frac{\Phi_{2}^{\textrm{far}}}{\kappa_{5}^{2}\phi_{A}^{3/\nu}\sqrt{h_{0}^{\textrm{far}}}}\Lambda^{3},
\end{array}
\end{equation}
where $T$, $\mu_B$, $s$ and $\rho_B$ are the temperature, the baryon chemical potential, and the entropy and baryon charge densities of the medium, respectively. Moreover, $h_0^{\textrm{far}}$, $\Phi_0^\textrm{far}$, $\Phi_2^\textrm{far}$, and $\phi_A$ are asymptotic coefficients extracted from the ultraviolet, near-boundary expansions of the EMD fields, and $\nu \equiv d-\Delta$, where $d=4$ is the number of spacetime dimensions of the boundary gauge theory and $\Delta\approx 2.73294$ is the effective scaling dimension of the gauge field theory operator dual to the bulk dilaton field (see Refs.\ \cite{Critelli:2017oub,Grefa:2021qvt} for details).

As mentioned above, the free parameters $\kappa_5^2$ and $\Lambda$, and the free functions $V(\phi)$ and $f(\phi)$ of the EMD model were fixed in Ref.\ \cite{Critelli:2017oub} by matching the holographic equation of state and the second order baryon susceptibility evaluated at $\mu_B=0$ with the corresponding lattice QCD results from Refs.\ \cite{Borsanyi:2013bia,Bellwied:2015lba}, which yields

\begin{equation}\label{eq:free_functions}
    \begin{array}{rcl}
         V(\phi)&=& -12\cosh(0.63\,\phi)+0.65\,\phi^{2}-0.05\,\phi^{4}+0.003\,\phi^{6},  \\
         \\
         \kappa_{5}^{2} &=& 8\pi G_{5}=8\pi(0.46), \qquad \Lambda=1058.83\, \textrm{MeV},  \\
         \\
         f(\phi) &=& \frac{\sech(c_{1}\phi+c_{2}\phi^{2})}{1+c_{3}}+\frac{c_{3}}{1+c_{3}}\sech(c_{4}\phi),
    \end{array}
\end{equation}
where $c_1=-0.27$, $c_2=0.4$, $c_3=1.7$, and $c_4=100$.

\begin{figure}
    \centering
    \includegraphics[width=0.49\textwidth]{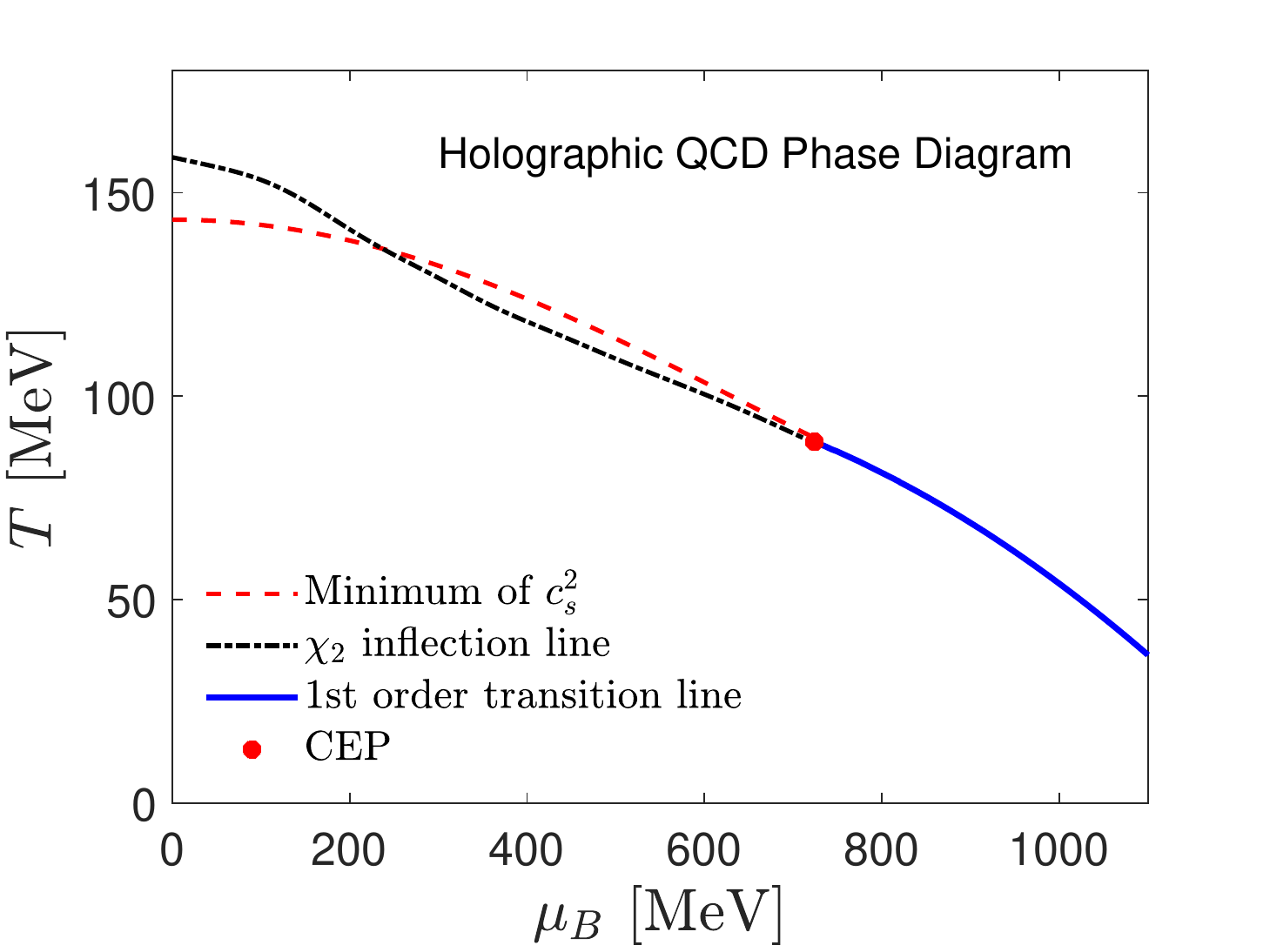}
    \includegraphics[width=0.49\textwidth]{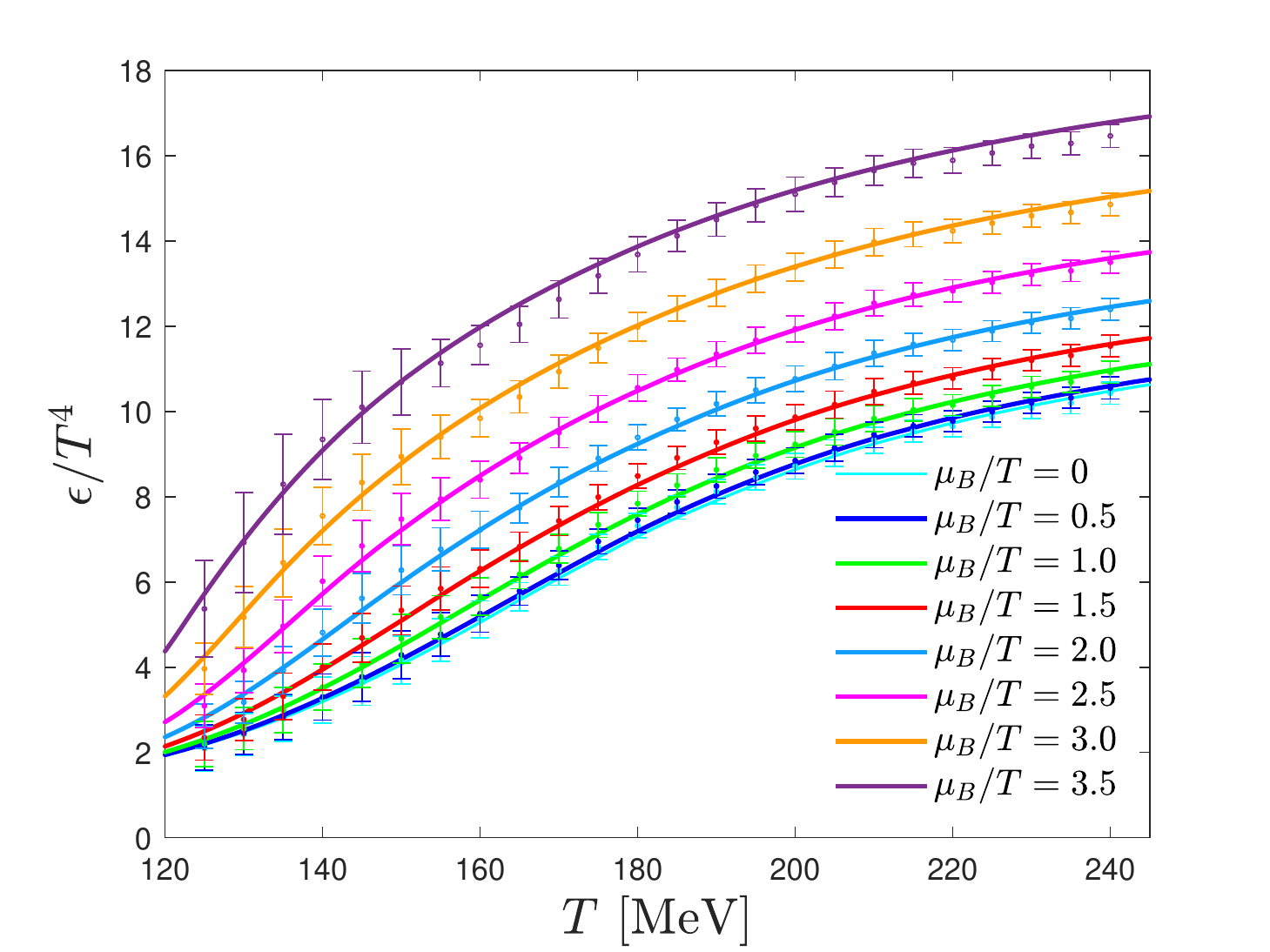}
    \caption{Top panel: holographic phase diagram showing the location of the minimum of the square of the speed of sound at constant entropy per baryon number $c_{s}^{2}$ and of the inflection point of the second order baryon susceptibility $\chi_{2}^{B}$ that were chosen to characterize the crossover region, and the line of first-order phase transitions ending at the CEP. Bottom panel: normalized energy density as a function of the temperature for different values of $\mu_{B}/T$ and its comparison with state-of-the-art lattice QCD results from \cite{Borsanyi:2021sxv}.}
    \label{fig:phase_diagram}
\end{figure}

This EMD model has been shown in Ref.\ \cite{Grefa:2021qvt} to quantitatively agree with state-of-the-art lattice QCD thermodynamics at finite baryon density \cite{Borsanyi:2021sxv}. In Ref.\ \cite{Critelli:2017oub} this EMD model provided the following prediction for the location of the QCD CEP, $(T^{c},\mu_B^{c})\sim (89,724)$ MeV, and in Ref.\ \cite{Grefa:2021qvt} this CEP was shown to be the end point of a line of first-order phase transitions lying at larger values of $\mu_B$. The phase diagram of this EMD model is depicted in Fig.\ \ref{fig:phase_diagram}, where we also display the normalized energy density at finite baryon density predicted by the EMD model compared to the corresponding lattice result from Ref.\ \cite{Borsanyi:2021sxv}.

We note that, even though our EMD model constructed in Ref.\ \cite{Critelli:2017oub} had been the first effective model in the literature to be shown to correctly \emph{predict} the behavior of QCD thermodynamics at finite baryon density (indeed, 4 years before the publication of the latest lattice QCD results from Ref.\ \cite{Borsanyi:2021sxv}), very recently another successful and similar EMD model \cite{Cai:2022omk} has been proposed with a different set of free parameters. The EMD model of Ref.\ \cite{Cai:2022omk} was matched at zero baryon density to the lattice equation of state from Ref.\ \cite{HotQCD:2014kol}, instead of the lattice results from Ref.\ \cite{Borsanyi:2013bia} used to fix the free parameters of our EMD model. Even though both sets of lattice results at zero baryon density quantitatively agree within error bars for most values of temperature, at high $T$ some quantities start to disagree even when considering the error bars (see Fig.\ 6 of Ref.\ \cite{HotQCD:2014kol}). The recent EMD model of Ref.\ \cite{Cai:2022omk} also produces a good quantitative agreement with the latest lattice QCD results at finite $\mu_B$ from Ref.\ \cite{Borsanyi:2021sxv}, which looks competitive with ours \cite{Critelli:2017oub,Grefa:2021qvt}, even though it predicts a CEP at a significantly different location in the phase diagram: $(T^c,\mu_B^c)\sim(105,558)$ MeV. This indicates that the latest lattice results available at finite $\mu_B$ \cite{Borsanyi:2021sxv} cannot distinguish between the two predictions, since both regard a region of the QCD phase diagram with values of temperature lower than the ones achieved in these lattice simulations. 

The holographic evaluation of the transport coefficients which we are going to consider for the EMD model of Refs.\ \cite{Critelli:2017oub,Grefa:2021qvt} in the course of the next sections makes use of well-known formulas already derived in the literature. For the sake of brevity, in what follows we refer the interested reader to the appropriate references where those derivations are presented in detail. However, before proceeding to the actual calculations with such holographic formulas, we briefly comment on the general reasoning involved.

In the context of the hot and dense medium described by the present isotropic EMD model, there is an $SO(3)$ rotation symmetry which organizes into different irreducible representations the diffeomorphism and gauge invariant combinations of the perturbations of the bulk EMD fields with zero spatial momentum at the linearized level \cite{DeWolfe:2011ts}. The gauge and diffeomorphism invariant perturbation in the $SO(3)$ singlet channel is holographically related to the bulk viscosity of the dual gauge theory at the boundary, which we shall discuss in Sec. \ref{sec:bulk}. The perturbation in the $SO(3)$ triplet channel is related to the baryon conductivity, which we analyze in Sec. \ref{sec:baryoncond}. Finally, the perturbation in the $SO(3)$ quintuplet channel is related to the shear viscosity, which we investigate in Sec. \ref{sec:shear}. Since these diffeomorphism and gauge invariant perturbations transform under different irreducible representations of the $SO(3)$ rotation symmetry group of the isotropic fluid, they cannot mix at the linearized level and one needs to solve a decoupled equation of motion for each of these fluctuations \cite{DeWolfe:2011ts}. 

In the case of the perturbation associated with the shear viscosity, it can be shown \cite{DeWolfe:2011ts} that it gives $\eta/s = 1/4\pi$ for any values of $T>0$ and $\mu_B\ge 0$, as it is well known for any holographic model which is isotropic, rotationally invariant, and has at most two derivatives of the metric field in the bulk gravity action \cite{Kovtun:2004de,Buchel:2003tz}. However, the natural dimensionless combination appearing in the hydrodynamic expression for the viscous part of the energy-momentum tensor of the boundary gauge theory at finite baryon density is not $\eta/s$, but rather $\eta T/(\epsilon+P)$ \cite{Liao:2009gb,Denicol:2013nua}, where $\epsilon$ and $P$ are the energy density and the pressure of the fluid, respectively. The combination $\eta T/(\epsilon+P)$ reduces to $\eta/s$ at $\mu_B=0$ but acquires a nontrivial behavior at finite baryon density, as we shall see in Sec. \ref{sec:shear}.

Concerning the parton energy loss, the associated observables to be considered in Sec. \ref{sec:energyloss} are calculated in the holographic framework by considering a probe Nambu-Goto (NG) action for a classical string on top of the background solutions for the bulk fields \cite{Gubser:2006bz,Herzog:2006gh,Casalderrey-Solana:2006fio,Gubser:2006qh,Gubser:2006nz,Liu:2006ug,Liu:2006he,Casalderrey-Solana:2007ahi,Gursoy:2009kk,Gursoy:2010aa,DEramo:2010wup}. The NG action is proportional to the square root of the `t Hooft coupling, $\sqrt{\lambda_t}$, which in a bottom-up setup as ours is taken as an extra free parameter which should be fixed by some phenomenological input. In Sec. \ref{sec:energyloss} we leave the value of $\sqrt{\lambda_t}$ unspecified, such that when comparing the holographic energy loss with the results from other approaches, one may consider different values for the `t Hooft coupling following different prescriptions (see e.g. Ref.\ \cite{Liu:2006ug}).

We close this section by remarking that, since the EMD background \eqref{eq:ansatz} supports nontrivial profiles for the Maxwell and dilaton fields, one could consider, in principle, the coupling of these background fields to the string described by the probe NG action. However, following Refs.\ \cite{Rougemont:2015wca,Critelli:2016cvq,Rougemont:2020had}, in this work these couplings are assumed to be small corrections to the NG action in the gauge/gravity duality where the `t Hooft coupling is assumed to be large and are, thus, neglected. Concerning the calculations of the heavy-quark drag force (to be discussed in Sec. \ref{sec:dragforce}) and the Langevin diffusion coefficients for heavy flavors (see Sec. \ref{sec:langevin}), one could minimally couple the string endpoint at a flavor brane close to the boundary to the Maxwell field on top of it, as done e.g. in Ref.\ \cite{Kiritsis:2011ha} in the case of finite mass quarks. However, we shall consider in our analysis infinitely heavy probe quarks, and in such a case the minimal coupling term is of order $0$ in the `t Hooft coupling, while the NG action is of order $1/2$. Consequently, the minimal coupling term between the Maxwell field and the string is suppressed for infinitely heavy quarks. 

Finally, concerning the calculation of the jet quenching parameter $\hat{q}$ associated with the energy loss of light partons (to be considered in Sec. \ref{sec:jet}), the minimal coupling term plays no role at all, since the contributions coming from each string endpoint (both located at the boundary in this calculation) cancel each other out. Moreover, as a working hypothesis, we assume that the coupling term between the effective 5-dimensional dilaton field and the Ricci scalar induced on the probe string worldsheet is also of order 0 in the `t Hooft coupling (as in the 10-dimensional case) and, therefore, its contribution is also taken to be formally suppressed relative to the NG action. 

\section{Transport of Baryon Charge}
\label{sec:baryon}

In this section we present our results regarding the transport of baryon charge in a hot and dense QGP, by analyzing the baryon and thermal conductivities, and also the baryon diffusion coefficient. Studies of baryon transport coefficients at large densities are especially urgent to understand the dynamical behavior of baryons at large densities. A large baryon diffusion, for instance, can transport more baryons to midrapidity \cite{Denicol:2018wdp}. Initial studies have also considered the influence of criticality on baryon diffusion \cite{Du:2021zqz}. However, a word of caution is that baryon diffusion is not the only form of diffusion that affects the QGP.  In fact, there is an entire diffusion matrix involving baryon, strange, and electric (BSQ) conserved charges that can influence the dynamical evolution \cite{Greif:2017byw,Fotakis:2019nbq,Fotakis:2021diq}. However, in our current framework we only study the baryon conserved charge.

\subsection{Baryon conductivity}
\label{sec:baryoncond}

As discussed in Refs.\ \cite{DeWolfe:2011ts,Rougemont:2015ona,Rougemont:2017tlu}, the equation of motion (EOM) for the relevant linearized homogeneous perturbation in the $SO(3)$ triplet channel of the EMD model, $a(r,\omega)$, is given by
\begin{eqnarray}\label{eq:condEOM}
   \!\!\!\!\!\! a''(r,\omega)&+&\left(2A'(r)+\frac{h'(r)}{h(r)}+\frac{f'(\phi)}{f(\phi)}\phi'(r)\right)a'(r,\omega)
    \nonumber \\
  &+&\frac{e^{-2A(r)}}{h(r)}\left(\frac{\omega^2}{h(r)}-f(\phi)\Phi'(r)^{2}\right)a(r,\omega)=0,
\end{eqnarray}
where $\omega$ is the frequency of the plane wave Ansatz for the perturbation\footnote{We take the wave number equal to zero in the homogeneous regime.} and the prime denotes a derivative with respect to the holographic coordinate $r$, except for $f'(\phi)\equiv \partial_\phi f(\phi)$.

Equation \eqref{eq:condEOM} needs to be solved numerically over the EMD background fields $A(r)$, $h(r)$, $\phi(r)$, and $\Phi(r)$ with in-falling wave condition at the black hole horizon (which in our numerical calculations is located at $r=r_{\textrm{start}}=10^{-8}$ \cite{Critelli:2017oub,Grefa:2021qvt}), and normalized to unity at the boundary (which in our numerical calculations is located at $r=r_{\textrm{max}}\sim 2$ --- $10$, depending on how a given background solution asymptotes to AdS in the ultraviolet \cite{Critelli:2017oub,Grefa:2021qvt}). These conditions may be implemented by setting
 \begin{equation}\label{eq:condWaveCond}
     a(r,\omega)\equiv\frac{r^{-i\omega}P(r,\omega)}{r_{\textrm{max}}^{-i\omega}P(r_{\textrm{max}},\omega)},
 \end{equation}
where $P(r,\omega)$ must be a regular function at the horizon. The EOM for $P(r,\omega)$ can be obtained by substituting (\ref{eq:condWaveCond}) into  (\ref{eq:condEOM}). The procedure for the numerical integration of $P(r,\omega)$ is similar to what is done in the case of the black hole background fields as described in Refs. \cite{Critelli:2017oub,Grefa:2021qvt}, i.e. the initial conditions required for integrating the EOM for the perturbation, $P(r_{\textrm{start}},\omega)$ and $P'(r_{\textrm{start}},\omega)$, are obtained by Taylor expanding $P(r,\omega)$ to second order around $r_{\textrm{start}}$. The holographic Kubo formula for the baryon conductivity in physical units reads as follows \cite{DeWolfe:2011ts,Rougemont:2015ona,Rougemont:2017tlu}
\begin{equation}\label{eq:conduct}
    \sigma_{B}(T,\mu_{B})=-\frac{\Lambda}{2\kappa_{5}^{2}\phi_{A}^{1/\nu}}\lim_{\omega\rightarrow0}\frac{1}{\omega}\left(e^{2A}hf(\phi)\textrm{Im}[a^{*}a']\right).
\end{equation}
The term $e^{2A}hf(\phi)\textrm{Im}[a^{*}a']$ in Eq.\ (\ref{eq:conduct}) is a radially conserved flux which, consequently, may be evaluated at any value of the holographic coordinate $r$. We also remark that the strict dc limit of vanishing frequency, $\omega\to 0$, which should be implemented in the Kubo formula \eqref{eq:conduct}, may be numerically problematic. Therefore, we actually approximate the dc limit in the numerical evaluation of Eq.\ \eqref{eq:conduct} by taking a small but nonzero evaluation frequency $\omega=\omega_{\textrm{eval}}\equiv 10^{-5}$. One consistency check that must be performed to verify that this is indeed a good approximation to the dc limit is to test for several selected values of $(T,\mu_B)$ whether $\sigma_B(T,\mu_B)$ remains approximately unchanged when evaluated using different small values of $\omega$ around $\omega=\omega_{\textrm{eval}}\equiv 10^{-5}$. Typically, the results remain approximately unchanged for $\omega \sim 10^{-7}$ --- $10^{-2}$, while for very small frequencies some numerical problems with artificial divergences occur, and for $\omega\gtrsim 10^{-1}$ the frequency is no longer small enough to be a good approximation to the dc limit.

\begin{figure}
\centering

   \includegraphics[width=0.49\textwidth]{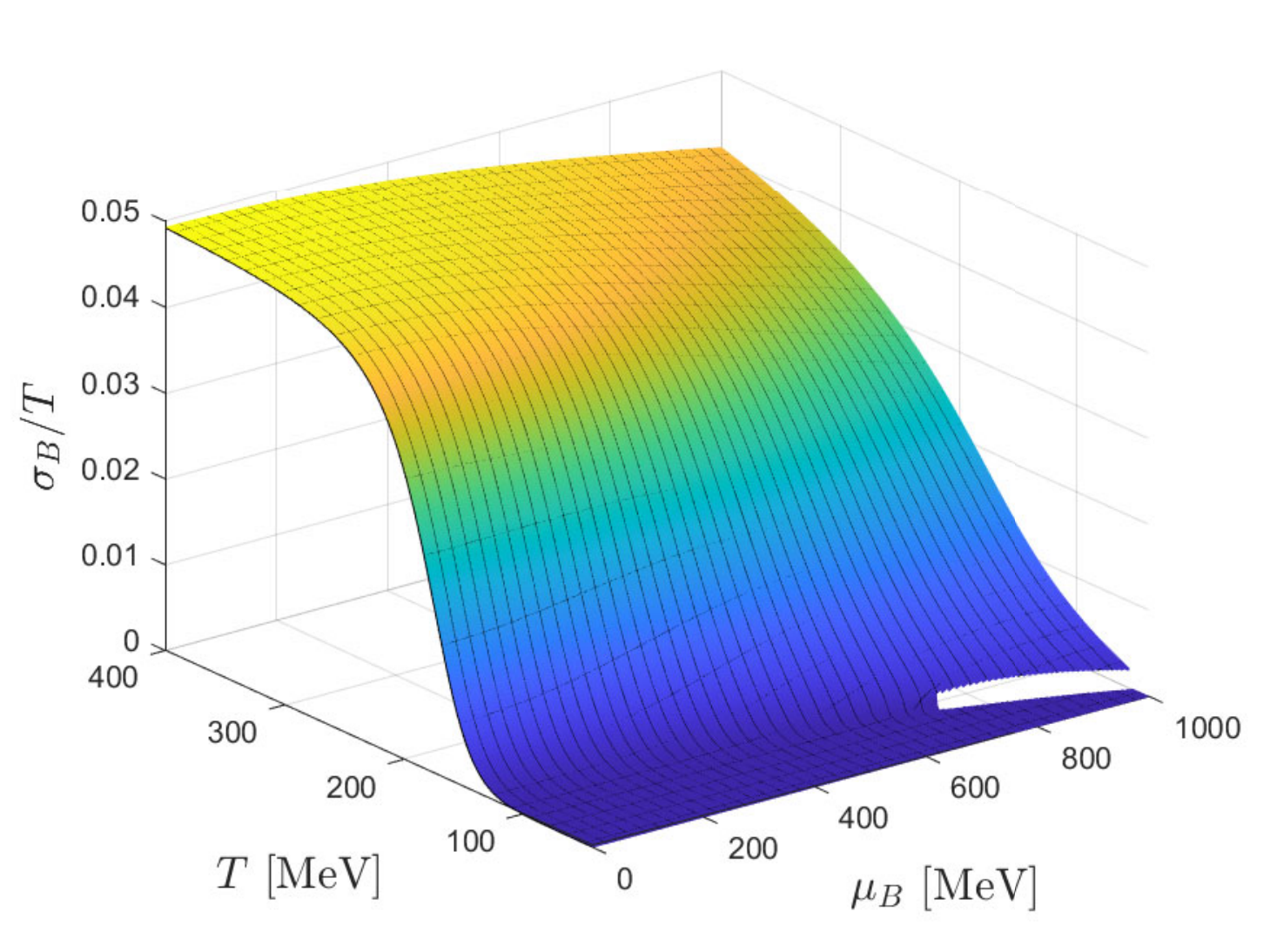}
   \includegraphics[width=0.49\textwidth]{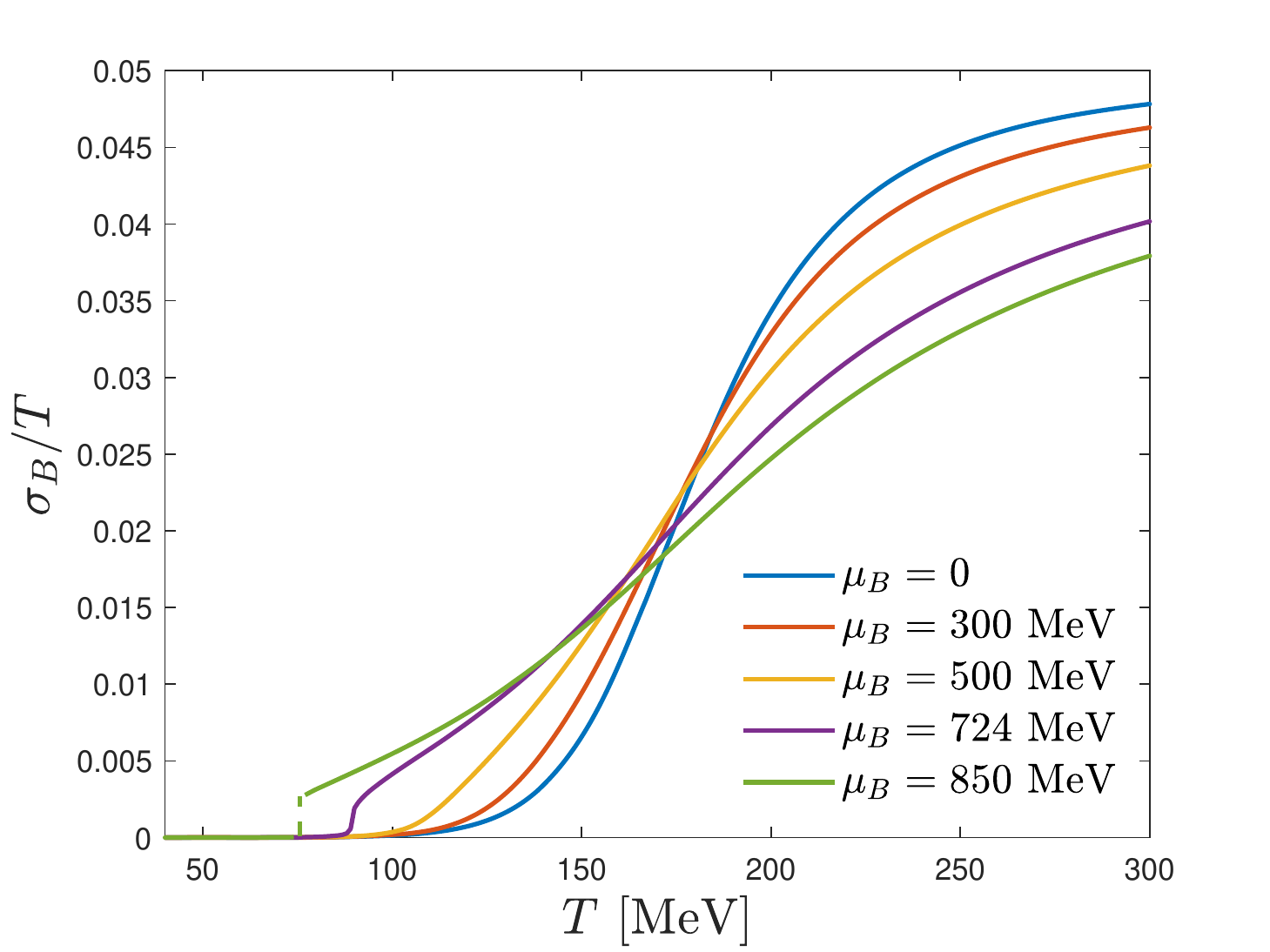} 
    \caption{Upper panel: scaled baryon conductivity $\sigma_B/T$ as a function of temperature and baryon chemical potential. Lower panel: scaled baryon conductivity as a function of the temperature, for several values of the chemical potential.}
    \label{fig:barConduct}
\end{figure}

The results for the baryon conductivity as a function of $(T,\mu_{B})$ are shown in Fig.\ \ref{fig:barConduct}: the upper panel shows the full surface plot as a function of $T$ and $\mu_B$, while the bottom panel shows slices at constant $\mu_B$ as functions of the temperature. The overall dependence of the baryon conductivity on the baryon chemical potential is relatively small, and it remains finite at the CEP, where it develops an infinite slope. The fact that $\sigma_B$ is finite at the critical point indicates that this approach is in the model B dynamical universality class \cite{Hohenberg:1977ym}. Beyond the critical point and over the line of first order phase transition, $\sigma_{B}/T$ exhibits a small discontinuity gap represented by a dashed line in Fig.\ \ref{fig:barConduct}. The discontinuity gap remains relatively small up to $\mu_B\sim 1000$ MeV. Another prominent feature of the baryon conductivity which can be clearly seen in the lower panel of Fig.\ \ref{fig:barConduct}, is the sharp crossing region between the different curves with fixed values of $\mu_B$. This feature has been also observed in the older EMD model of Ref.\ \cite{Rougemont:2015ona}, and seems to be a robust feature of this observable. Interestingly, in Ref.\ \cite{Rougemont:2018ivt}, the characteristic equilibration times of the baryon current estimated from the imaginary part of the lowest nonhydrodynamic quasinormal modes in the $SO(3)$ triplet channel of the present EMD model have been also shown to present a sharp crossing region very similar to the one observed for the baryon conductivity, although at slightly higher values of temperature.

\subsection{Baryon diffusion}
\label{sec:baryondiff}

The fact that the baryon conductivity remains finite at the CEP has a consequence on the baryon diffusion, which controls the fluid response to inhomogeneities in the baryon density. As shown in Ref.\ \cite{Iqbal:2008by}, the baryon diffusion coefficient $D_{B}$ can be holographically evaluated using the Nernst-Einstein's relation
\begin{equation}\label{eq:diffus}
    D_{B}=\frac{\sigma_{B}}{\chi_{2}^{B}},
\end{equation}
where $\chi_{2}^{B}$ is the second order baryon susceptibility. The $n$th order baryon susceptibility is defined as
\begin{equation}\label{eq:chi2}
    \chi_{n}^{B}=\frac{\partial^{n}P}{\partial\mu_{B}^{n}}=\frac{\partial^{n-1}\rho_{B}}{\partial\mu_{B}^{n-1}}.
\end{equation}

\begin{figure}[ht!]
    \centering
    \includegraphics[width=0.49\textwidth]{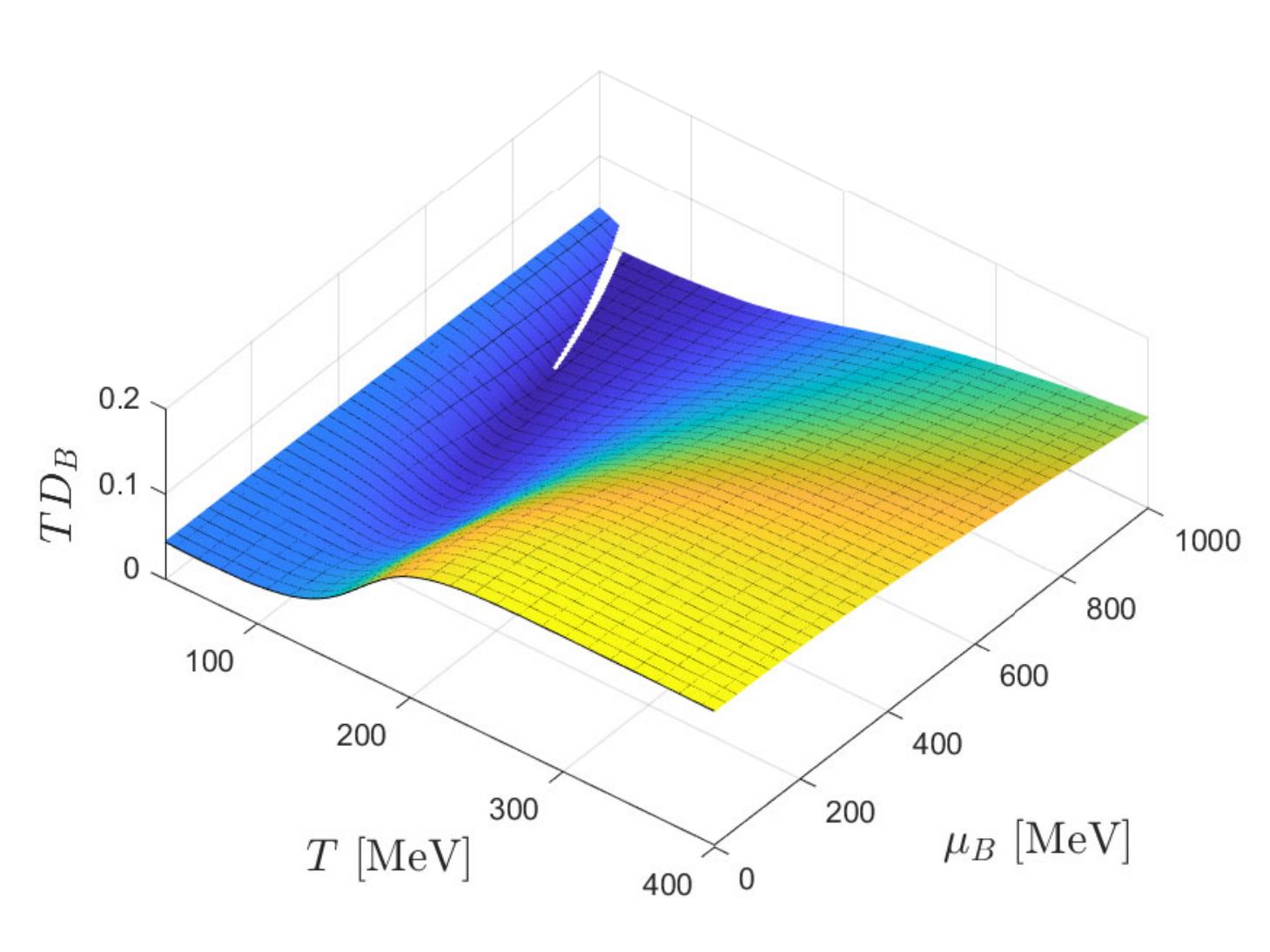}
    \includegraphics[width=0.49\textwidth]{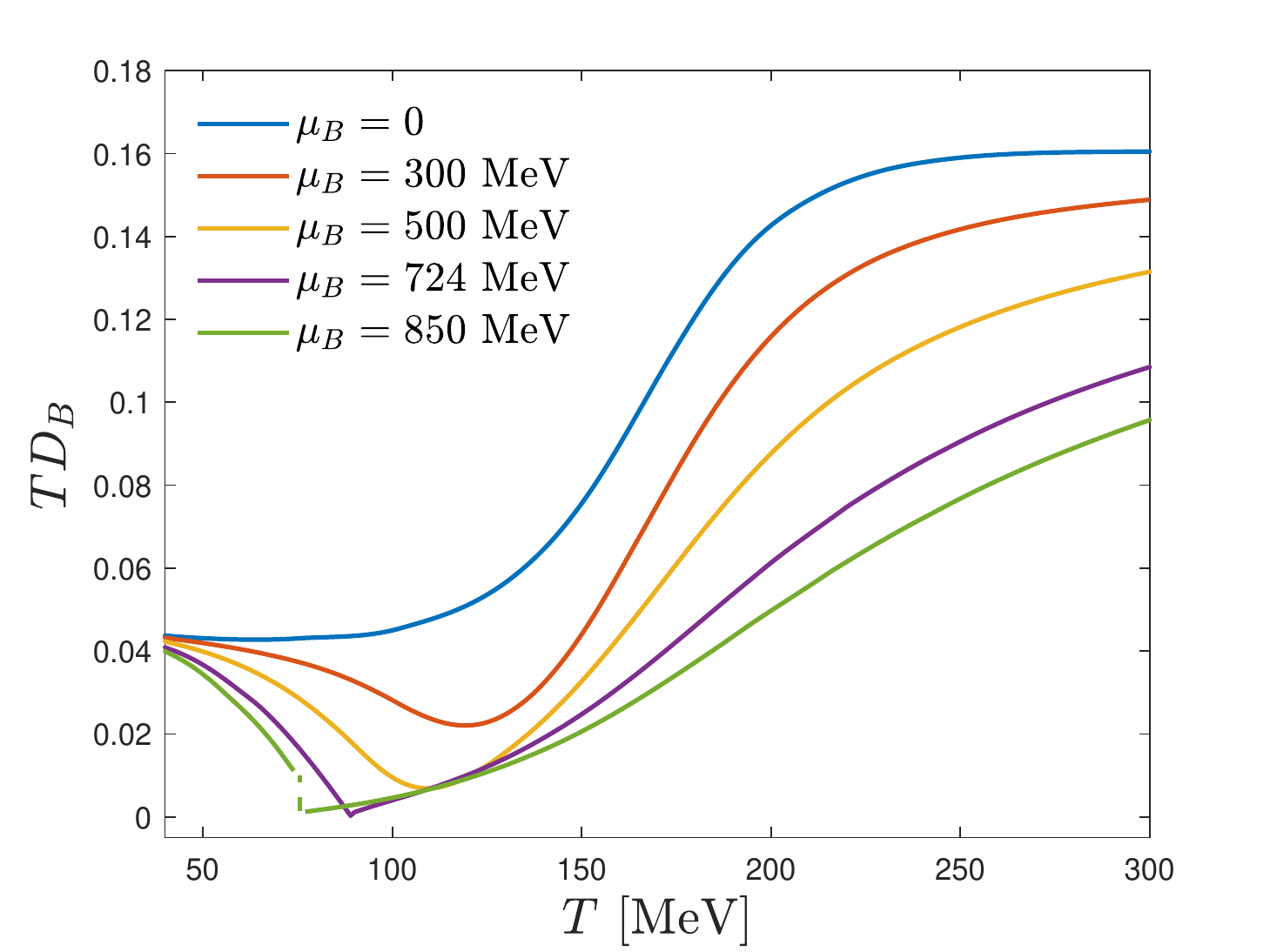}
    \caption{Upper panel: scaled baryon diffusion coefficient $TD_B$ as a function of temperature and baryon chemical potential. Lower panel: scaled baryon diffusion as a function of temperature, for several values of the baryon chemical potential.}
    \label{fig:barDiffu}
\end{figure}

The holographic results for the baryon diffusion are displayed in Fig.\ \ref{fig:barDiffu} and clearly show the suppression of baryon charge diffusion as the baryon chemical potential increases. One also notices the formation of a dip in the baryon diffusion at finite $\mu_B$, which moves toward the CEP. As reported in Refs. \cite{Critelli:2017oub,Grefa:2021qvt}, the location of the CEP was identified by the numerical divergence of the second order baryon susceptibility, which together with the finite behavior of the baryon conductivity results in a vanishing baryon diffusion at the CEP. Beyond the critical point and across the line of first-order phase transition, the baryon diffusion exhibits a small discontinuity gap, which grows with increasing $\mu_B$. Given that $D_B$ has an effect on the rapidity distribution of net-protons \cite{Denicol:2018wdp}, we anticipate that this behavior may lead to some interesting consequences in hydrodynamic simulations.

\subsection{Thermal conductivity}
\label{sec:thermalcond}

The thermal conductivity at finite baryon chemical potential can be obtained from the following relation \cite{Kapusta:2012zb}
\begin{equation}\label{eq:therCond}
    \sigma_{T}=\frac{D_{B}}{T}\chi_{2}^{B}\left(\frac{\epsilon+P}{\rho_{B}}\right)^{2}=T\sigma_{B}\left(\frac{s}{\rho_{B}}+\frac{\mu_{B}}{T}\right)^{2} .
\end{equation}

Eq.\ \eqref{eq:therCond} could be problematic in the limit of vanishing chemical potential since $\rho_{B}\rightarrow 0$ in this limit. However, because $\rho_{B}(T,\mu_{B}\rightarrow0)\rightarrow\mu_{B}\chi_{2}^{B}(T,\mu_{B}\rightarrow0)$, one can define the following dimensionless combination as done e.g. in Ref.\ \cite{Jain:2009pw}, which is well behaved in the limit of zero baryon chemical potential\footnote{We employed in the first line of Eq.\ \eqref{eq:therCondApprox} the holographic relation $\eta/s=1/4\pi$.}

\begin{align}\label{eq:therCondApprox}
    \frac{\mu_{B}^{2}\sigma_{T}}{\eta T}(T,\mu_{B})&=4\pi\frac{\sigma_{B}}{T}\frac{\mu_B^2}{Ts}\left(\frac{Ts}{\rho_{B}}+\mu_B\right)^{2}\nonumber\\
    &\to 4\pi\frac{\sigma_{B}}{T}\frac{s}{T^{3}}\left(\frac{T^{2}}{\chi_{2}^{B}}\right)^{2},\,\,\, \textrm{as}\,\,\, \mu_B\to 0.
\end{align}

\noindent Since it is the combination $\mu_B^2\sigma_T$ that remains finite in the $\mu_B\to 0$ limit, another possible dimensionless normalization, employed e.g. in Ref.\ \cite{Rougemont:2015ona}, and which is perhaps a better representative of the behavior of the thermal conductivity (because it does not mix with the effects of the shear viscosity, $\eta$), is given by
\begin{align}\label{eq:thermcond2}
\frac{\mu_B^2\sigma_T}{T^4}(T,\mu_B) &= \frac{\sigma_B}{T}\left(\frac{\mu_B}{\rho_B}\,\frac{s}{T}+\frac{\mu_B^2}{T^2}\right)^2 \nonumber\\
&\to \frac{\sigma_B}{T}\left(\frac{s}{T^3}\,\frac{T^2}{\chi_2^B}\right)^2, \,\,\, \textrm{as} \,\,\, \mu_B\to 0.
\end{align}

\begin{figure*}[ht!]
    \centering
    \includegraphics[width=0.49\textwidth]{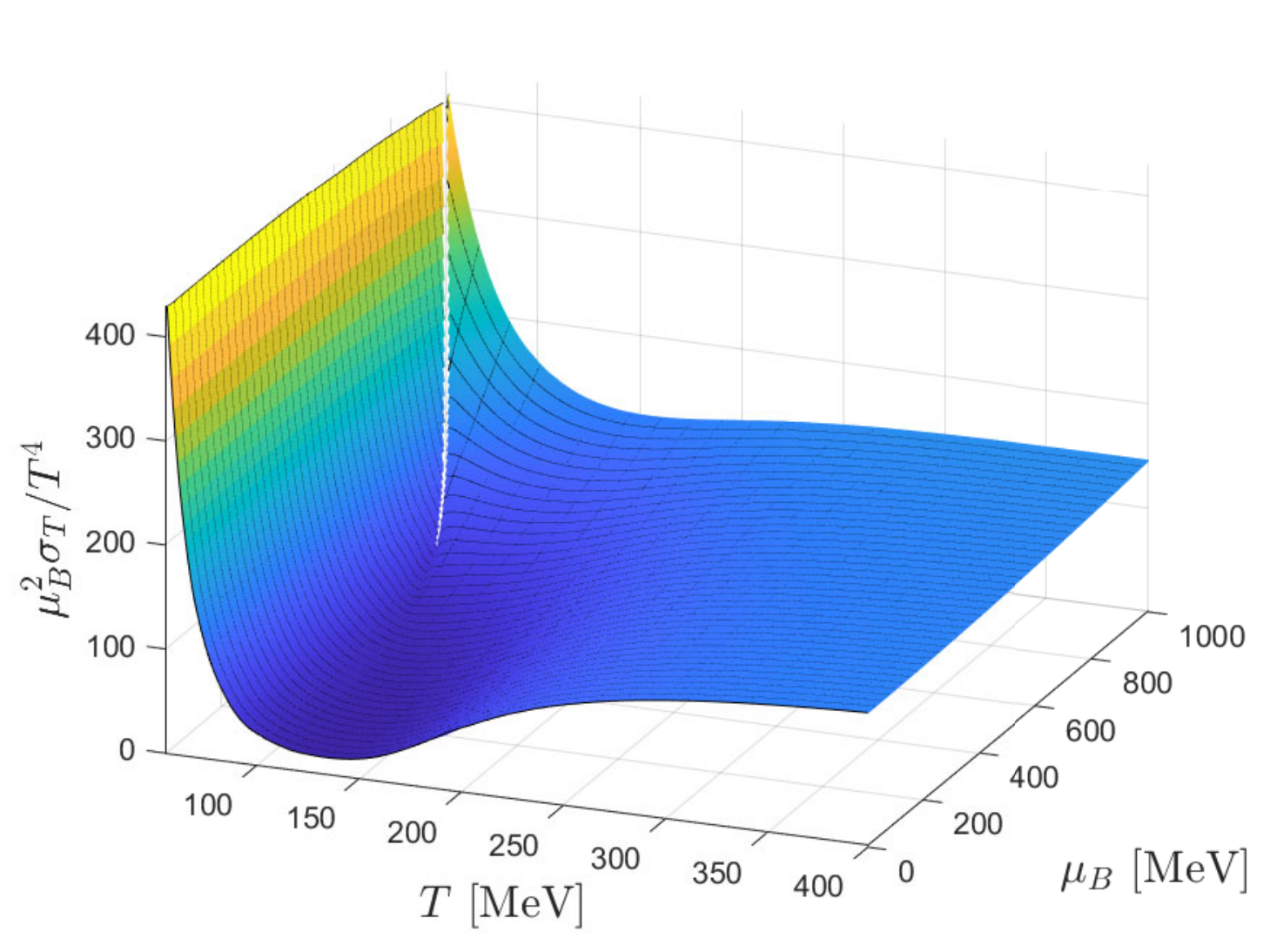}
    \includegraphics[width=0.49\textwidth]{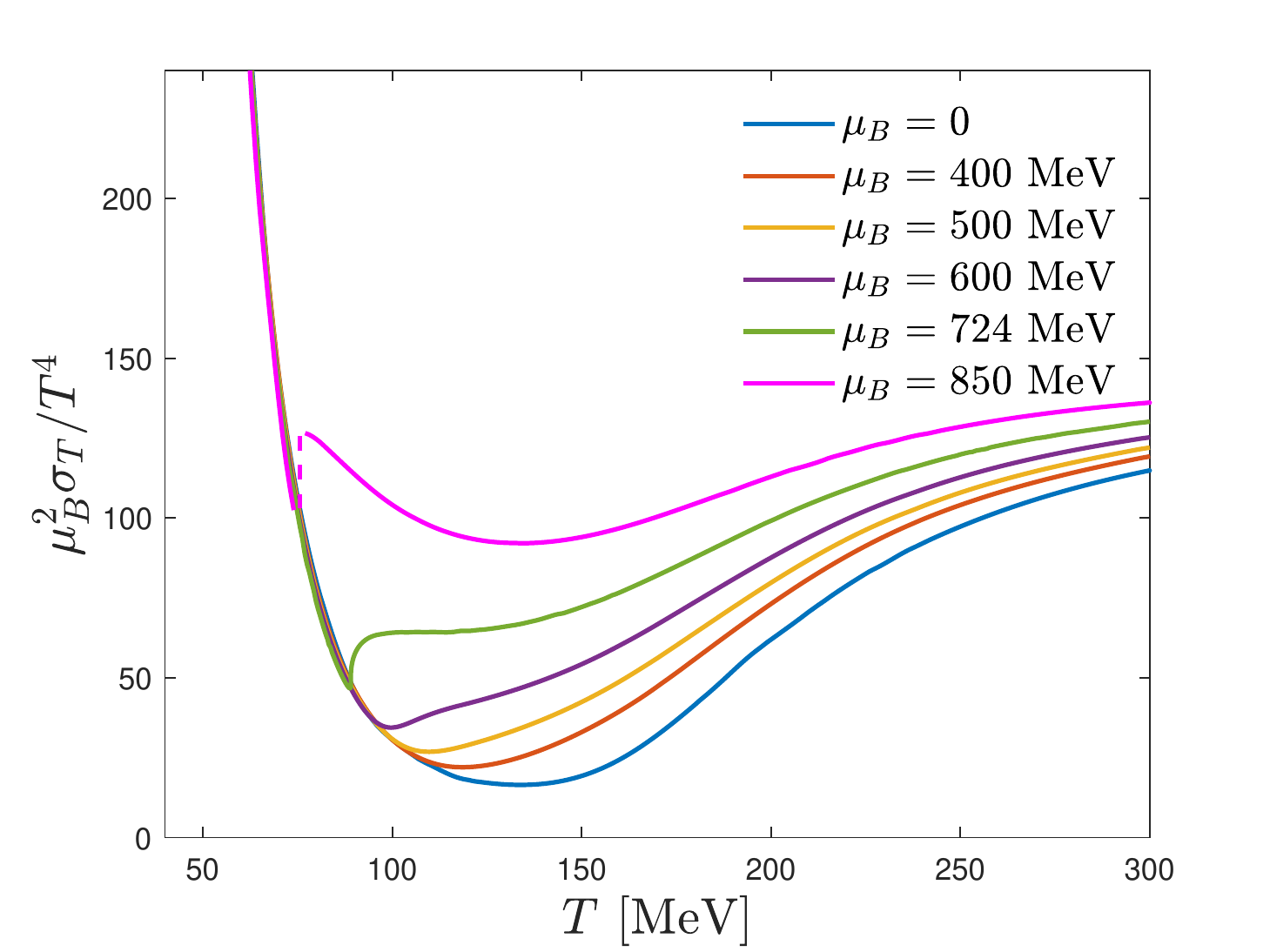}
    \caption{Holographic thermal conductivity as a dimensionless combination given by Eq.\ \eqref{eq:thermcond2}. The left panel shows this combination as a function of temperature and chemical potential, while the right panel shows it as a function of the temperature, for different values of $\mu_B$.}
    \label{fig:thermalConduc}
\end{figure*}

\noindent Our results obtained using Eq.\ \eqref{eq:thermcond2} are shown in Fig.\ \ref{fig:thermalConduc}. The minimum that the thermal conductivity exhibits at $\mu_{B}=0$ moves toward the critical point and becomes narrower to finally turn into a cusp at the critical point. However, at some value of the baryon chemical potential, the thermal conductivity starts to develop another local minimum, which does not follow the critical point. It starts as a change in concavity and the minimum really appears beyond the critical chemical potential as one may observe in Fig.\ \ref{fig:thermalConduc}.

\section{Energy Loss}
\label{sec:energyloss}

In this section we present the predictions from the EMD model for the energy loss of heavy and light partons in the strongly coupled, hot and baryon dense QGP. From the theoretical point of view, it is interesting to study how a hot, baryon dense and strongly interacting medium affects the energy loss experienced by fast moving probes in particular in the vicinity of the CEP and along the line of first order phase transition predicted by this model. At low beam energies (large baryon densities) the high $p_T$ spectra drops off rapidly \cite{STAR:2017sal} so it is clear that fewer jets exist at low $\sqrt{s}$.  However, studies by STAR have found hints of jet quenching effects across the beam energy scan using comparisons between central and peripheral collisions \cite{STAR:2017ieb}. Thus, it is interesting to study heavy and hard probes across the phase diagram, with specific interest on a critical point/first-order phase transition. Initial studies have attempted to calculate $\hat{q}/T^3$ at large densities and generally found that it should increase with increasing $\mu_B$ \cite{Rougemont:2015wca,Rougemont:2017tlu,McLaughlin:2021dlq}.

\subsection{Heavy quark drag force}
\label{sec:dragforce}

In the holographic trailing string approach \cite{Gubser:2006bz,Herzog:2006gh} (see also \cite{Gursoy:2009kk} for the generalization of this method including nonconformal effects due to a dilaton field), a heavy quark moving through the strongly coupled medium with a constant velocity $v$ in some direction, $x$ for example, is represented by the endpoint of an open string attached to the boundary, while the remainder of the string trails behind it, having its other endpoint attached to a new 2D black hole horizon developed over the string world sheet within the bulk \cite{Gubser:2006nz,Casalderrey-Solana:2007ahi,Gursoy:2010aa}. As the quark moves at the boundary, it loses energy and momentum through the drag force $F_\textrm{drag}=dp_{x}/dt$, which can be computed through the energy flow $dE/dx$ from the string endpoint at the boundary toward the other string endpoint located at the world sheet horizon within the bulk.

For the EMD model, it has been shown in Ref.\ \cite{Rougemont:2015wca} that the heavy quark drag force is given by
\begin{equation}\label{eq:Fdrag}
    \frac{F_{\textrm{drag}}}{\sqrt{\lambda_{t}}T^{2}}(T,\mu_{B};v)=-8\pi vh_{0}^{\textrm{far}}e^{\sqrt{2/3}\phi(r_{*})+2A(r_{*})},
\end{equation}
where $\lambda_{t}=1/\alpha'\,^2=1/l_{s}^{4}$ is the `t Hooft coupling,\footnote{The `t Hooft coupling is expected to be related to the coupling and the number of colors of the dual gauge theory through the holographic dictionary \cite{Maldacena:1997re,Gubser:1998bc,Witten:1998qj,Witten:1998zw}. However, the precise relation is only known for top-down holographic constructions.} $l_{s}$ is the fundamental string length, and $r_{*}$ is the radial location of the string world sheet horizon, which is obtained as the numerical solution of the following equation \cite{Rougemont:2015wca}
\begin{equation}\label{eq:r_star}
    h(r_{*})=h_{0}^{\textrm{far}}v^{2}.
\end{equation}

In the conformal limit, corresponding to values of temperature much larger than any other dimensionful scale of the system, the EMD backgrounds approach the AdS$_{5}$-Schwarzschild metric, and correspondingly, the conformal limit of Eq.\ \eqref{eq:Fdrag} should approach the well-known $\mathcal{N}=4$ Super-Yang Mills (SYM) value \cite{Gubser:2006bz,Herzog:2006gh}
\begin{equation}\label{eq:FdragCFT}
    \lim_{T\rightarrow\infty} \frac{F_{\textrm{drag}}}{\sqrt{\lambda_{t}}T^{2}}(T,\mu_{B};v)\to-\frac{\pi\gamma(v)v}{2}=-\frac{\pi v}{2\sqrt{1-v^{2}}}.
\end{equation}

\begin{figure}[h]
    \centering
    \includegraphics[width=0.49\textwidth]{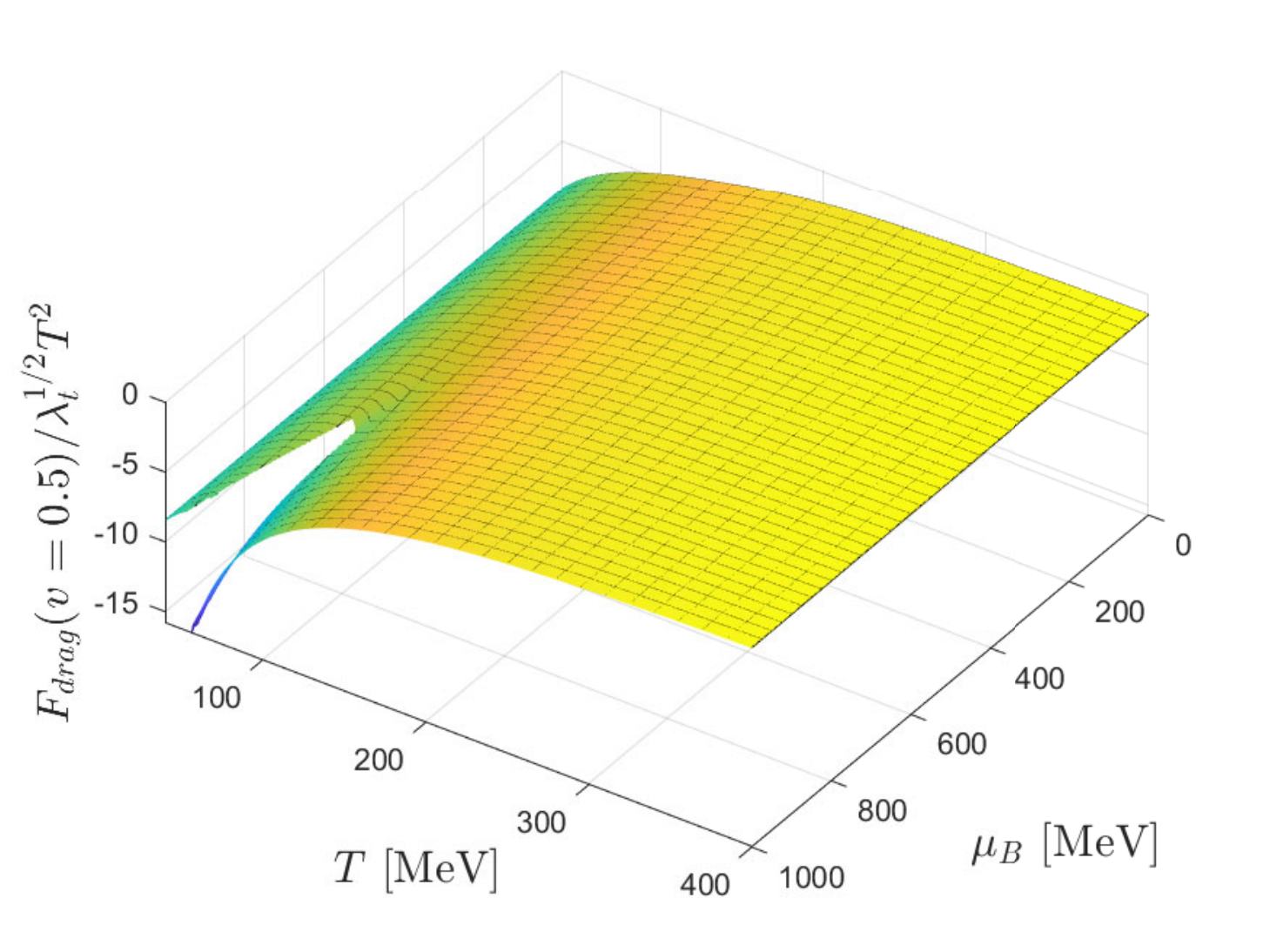}
    \includegraphics[width=0.49\textwidth]{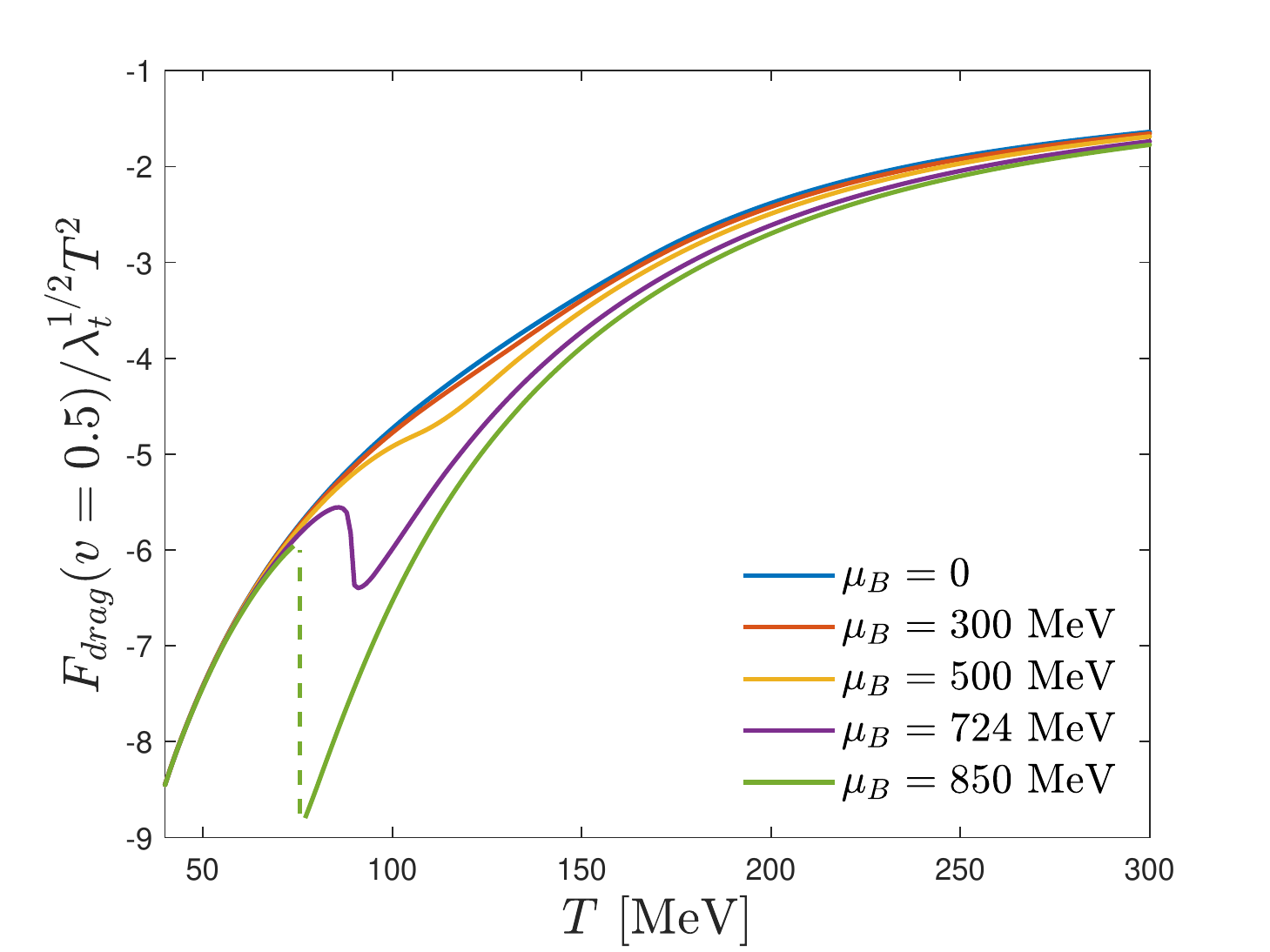}
     \caption{Heavy quark drag force at $v=0.5$ as a function of temperature and baryon chemical potential (top) and slices at constant $\mu_B$ (bottom).}
    \label{fig:Fdrag05}
\end{figure}

\begin{figure}[h]
    \centering
    \includegraphics[width=0.49\textwidth]{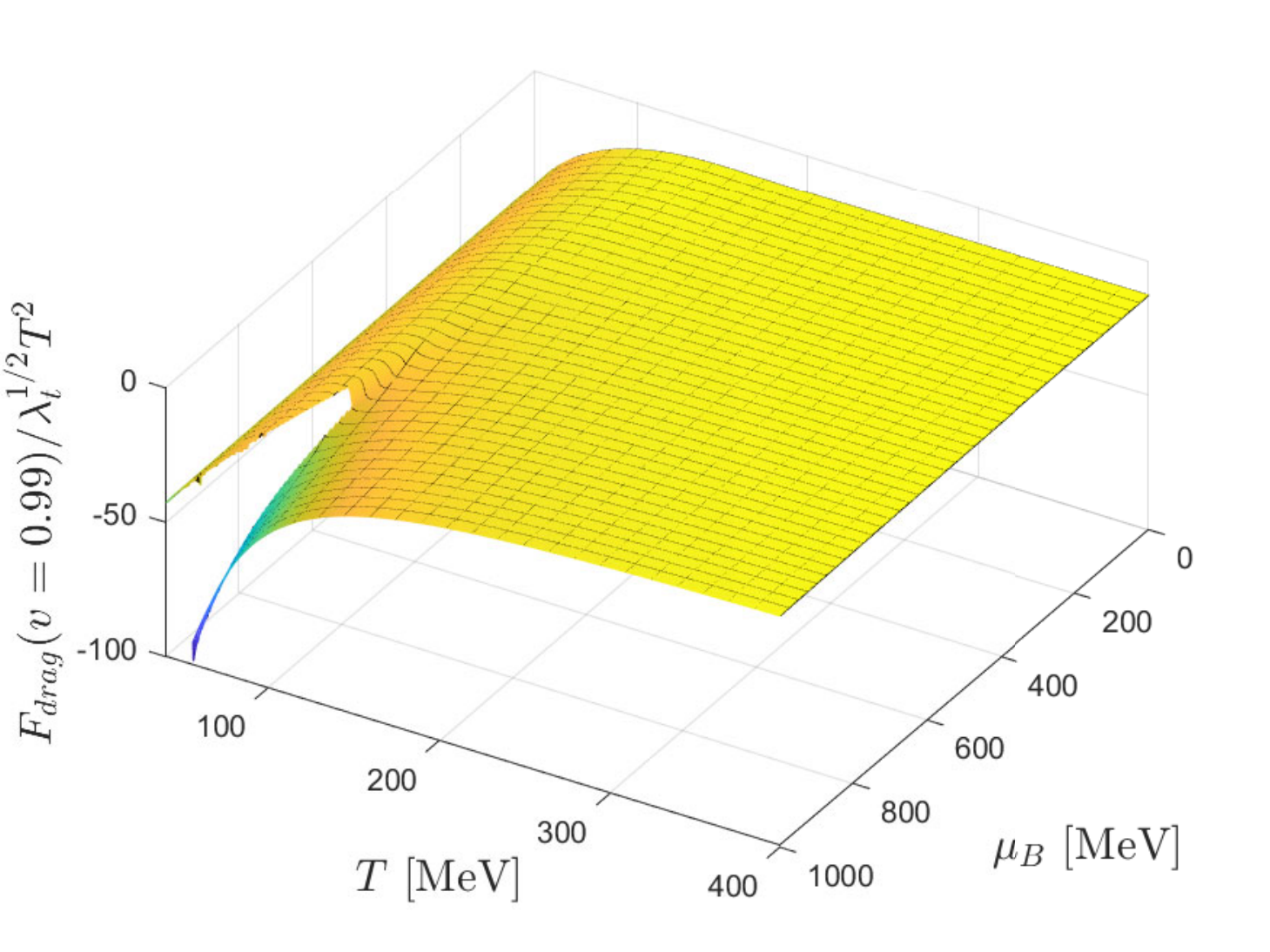}
    \includegraphics[width=0.49\textwidth]{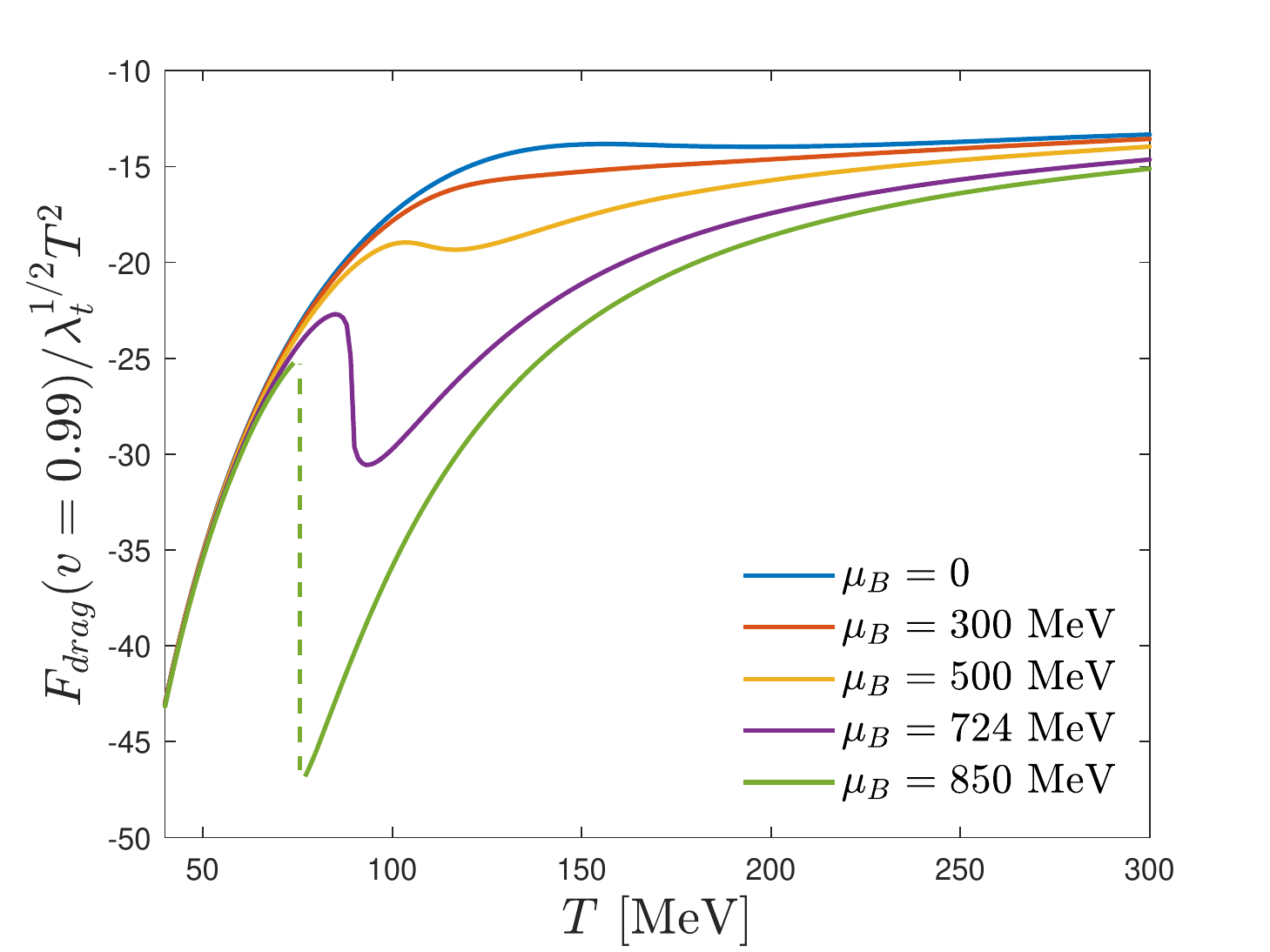}
     \caption{Heavy quark drag force at $v=0.99$ as a function of temperature and baryon chemical potential (top) and slices at constant $\mu_B$ (bottom).}
    \label{fig:Fdrag09}
\end{figure}

The EMD numerical results for the heavy quark drag force are shown in Figs.\ \ref{fig:Fdrag05} (for $v=0.5$) and \ref{fig:Fdrag09} (for $v=0.99$). The plots for small and large $v$ display some qualitative differences. Indeed, for $v=0.5$ it is only possible to observe an appreciable splitting between the curves with fixed values of $\mu_B$ at high values of the baryon chemical potential, while for $v=0.99$ the splitting is clear already at lower values of $\mu_B$, showing that the drag force is more sensitive to the baryon density of the medium at higher quark velocities. One can also see that in both cases ($v=0.5$ and $v=0.99$), the heavy quark energy loss associated with the drag force increases in magnitude by lowering the temperature and/or by increasing the baryon chemical potential of the medium, developing a dip and an inflection point at high $\mu_B$, with the latter moving toward the CEP as the baryon density of the fluid is enhanced. On top of the first-order line the heavy quark drag force presents a large discontinuity gap, which significantly increases with the magnitude of the quark velocity.
The aforementioned observations also suggest that a very heavy quark (e.g. the bottom), which might not achieve a very high velocity within the plasma, is less sensitive to the in-medium effects in comparison with a less massive quark (e.g. the charm), which could attain higher velocities within the fluid.

\subsection{Heavy quark Langevin diffusion coefficients}
\label{sec:langevin}

A holographic treatment of Langevin diffusion processes was originally proposed in Refs.\ \cite{Gubser:2006nz,Casalderrey-Solana:2007ahi}, and further generalized to include nonconformal effects associated to a dilaton field in Ref.\ \cite{Gursoy:2010aa}. The Brownian motion of a heavy quark moving through a medium can be approximately modeled by a linearized local Langevin equation that describes the thermal fluctuations of a heavy quark trajectory with constant velocity. Attempts have been made to extract the diffusion coefficient from experimental data using a Bayesian analysis in \cite{Xu:2017obm} at $\mu_B=0$, although that work used the simplifying assumption that the perpendicular and parallel diffusion coefficients are equal (even though their microscopic formulas are different), so their results are not directly comparable to ours. Here we make use of the nonconformal holographic formulas derived in Ref.\ \cite{Gursoy:2010aa}, and properly adapted to the EMD setup in Ref.\ \cite{Rougemont:2015wca}, to evaluate the heavy quark Langevin diffusion coefficients in the limit of zero frequency, which corresponds to the long time behavior of the stochastic diffusion process in the hot and baryon dense medium described by our holographic model.

The EMD holographic formulas for the heavy quark Langevin diffusion coefficients, perpendicular and parallel to the quark velocity are given by, respectively,
\begin{eqnarray}\label{eq:kappaPerp}
    &\frac{\kappa_{\perp}}{\sqrt{\lambda_{t}}T^{3}}(T,\mu_{B};v)=16\pi v(h_{0}^{\textrm{far}})^{3/2}e^{\sqrt{2/3}\phi(r_{*})+3A(r_{*})}\times \nonumber \\
    &\times\sqrt{h'(r_{*})\left[4A'(r_{*})+\sqrt{\frac{8}{3}}\phi'(r_{*})+\frac{h'(r_{*})}{h(r_{*})}\right]},
\end{eqnarray}
    
\begin{eqnarray}\label{eq:kappaPar}
    &\frac{\kappa_{\parallel}}{\sqrt{\lambda_{t}}T^{3}}(T,\mu_{B};v)=16\pi v^{3}(h_{0}^{\textrm{far}})^{5/2}\frac{e^{\sqrt{2/3}\phi(r_{*})+3A(r_{*})}}{h'(r_{*})^2}\times \nonumber \\
    &\times\left(h'(r_{*})\left[4A'(r_{*})+\sqrt{\frac{8}{3}}\phi'(r_{*})+\frac{h'(r_{*})}{h(r_{*})}\right]\right)^{3/2}.
\end{eqnarray}

\begin{figure*}
    \centering
    \includegraphics[width=0.49\textwidth]{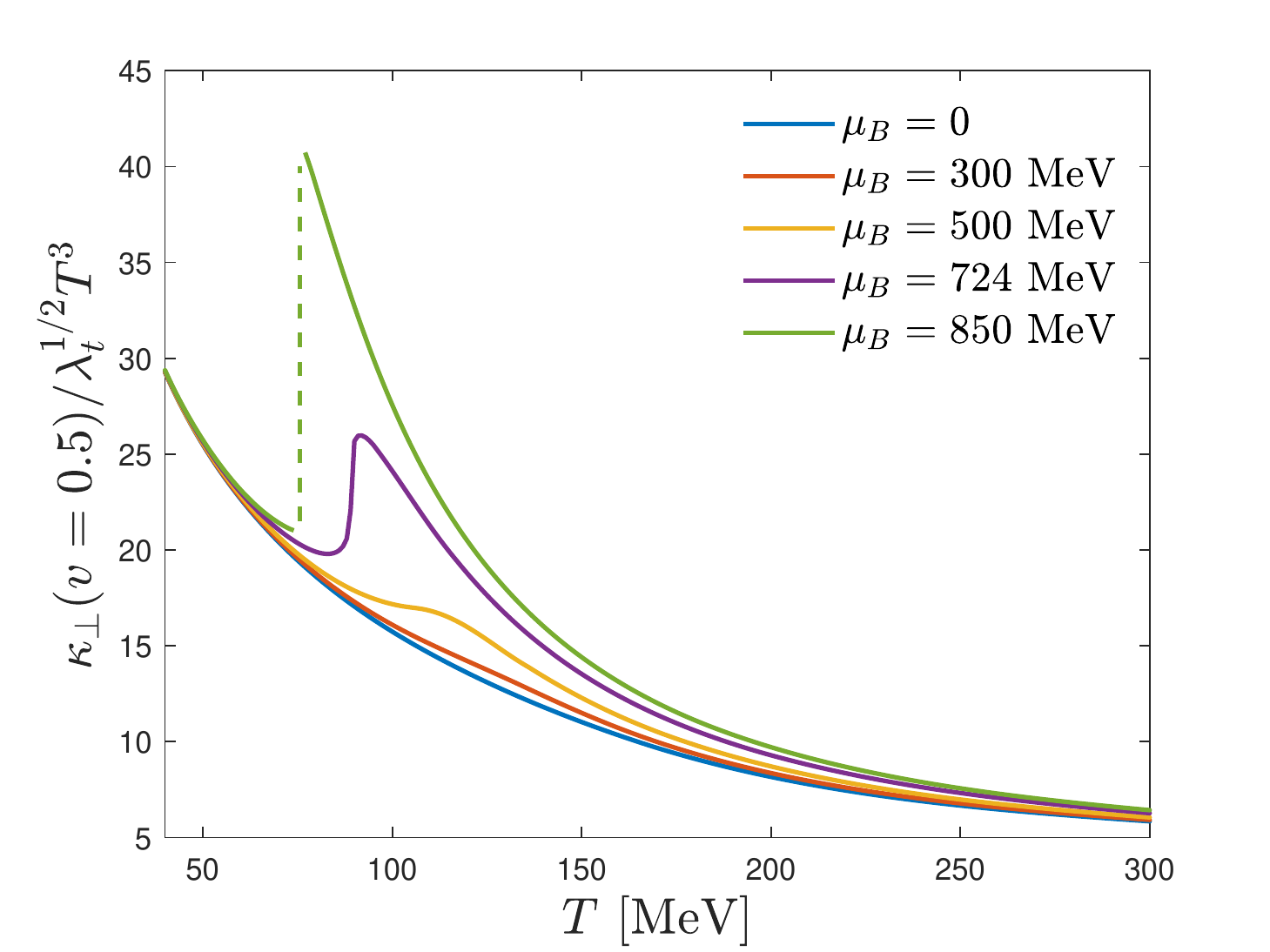}
    \includegraphics[width=0.49\textwidth]{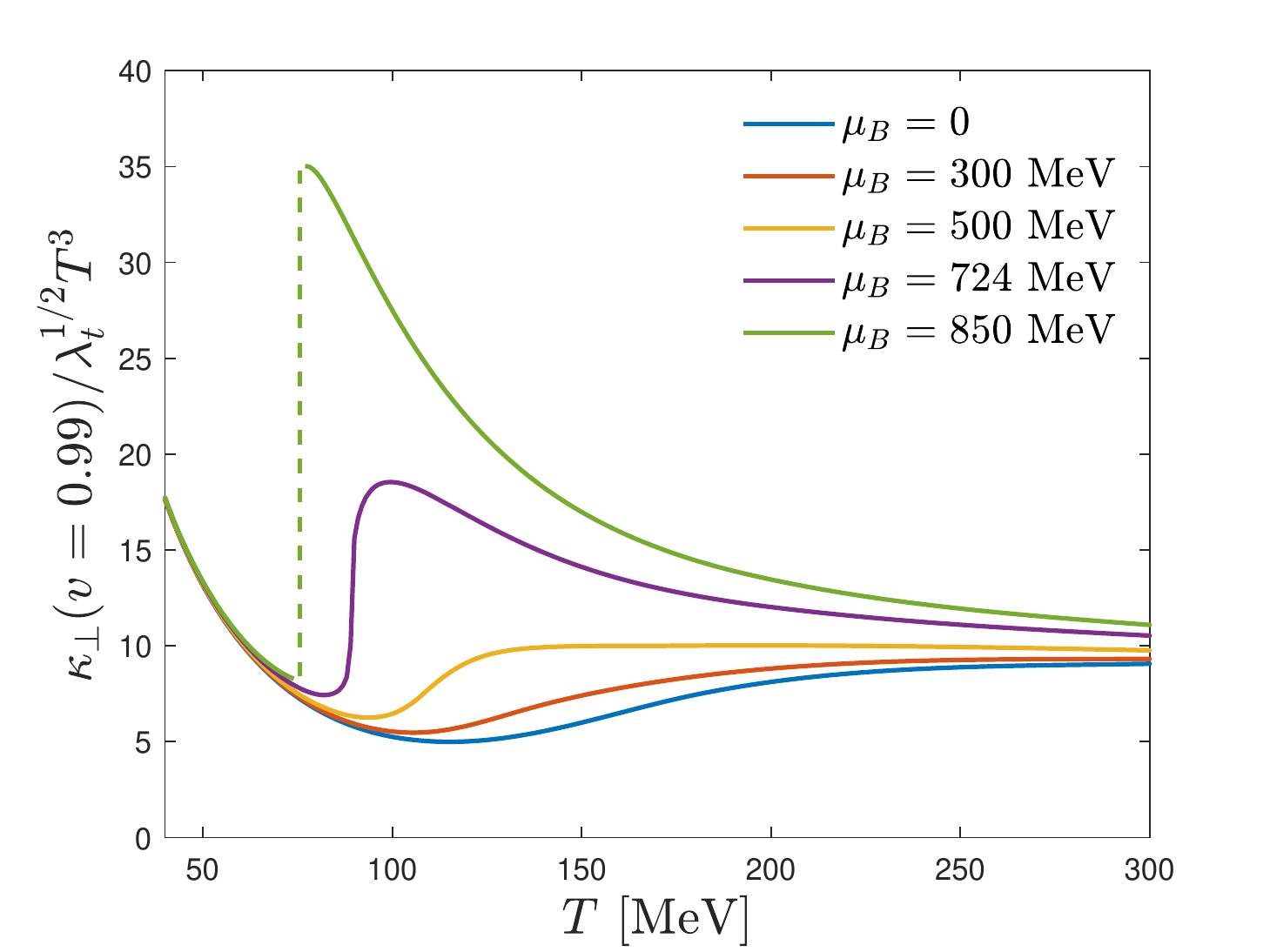}
    \newline
    \centering
    \includegraphics[width=0.49\textwidth]{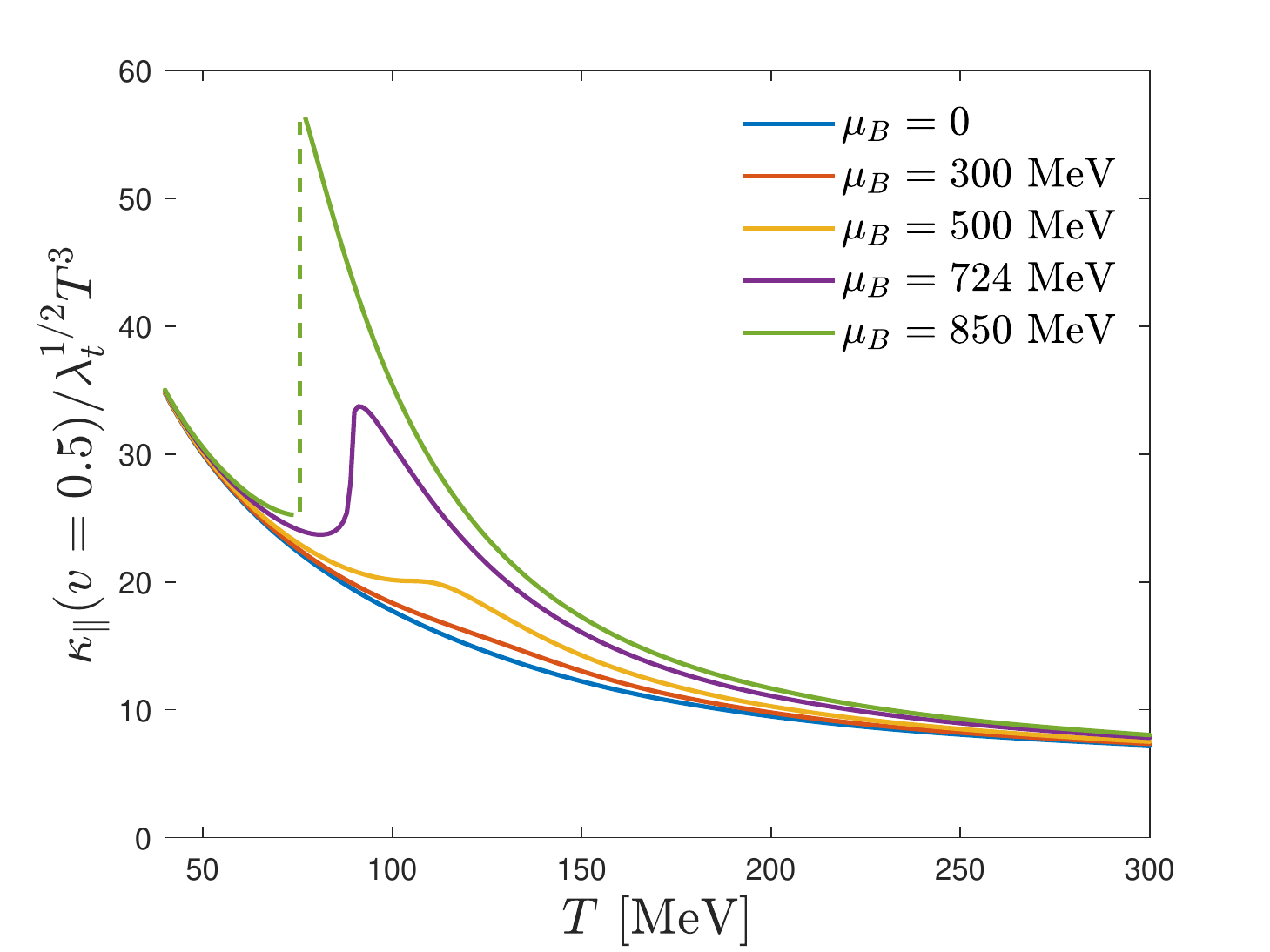}
    \includegraphics[width=0.49\textwidth]{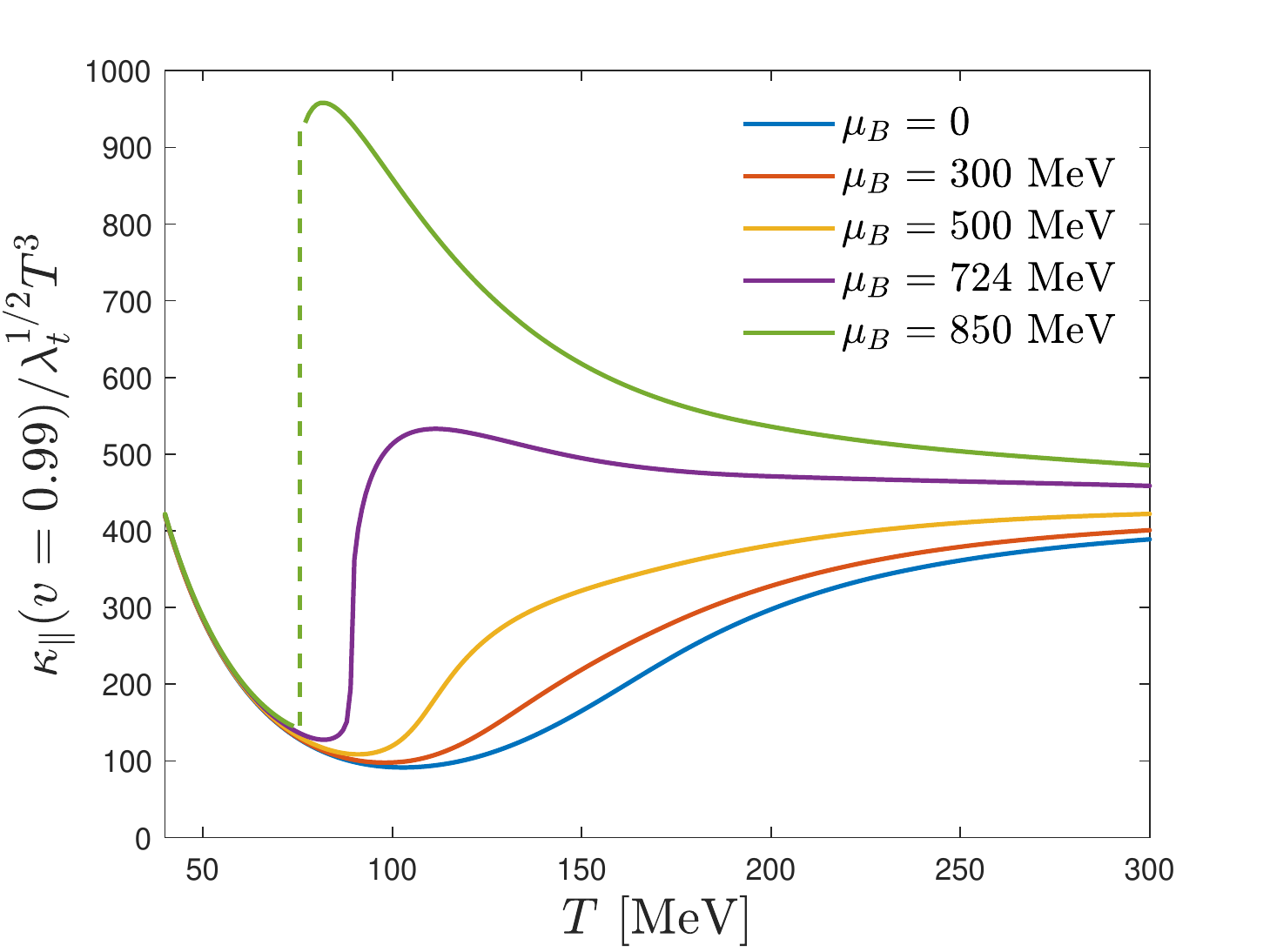}
    \caption{Perpendicular (upper panels) and parallel (lower panels) Langevin diffusion coefficients as functions of the temperature for several values of baryon chemical potential and two quark velocities: $v=0.5$ (left panels), and $v=0.99$ (right panels).}
    \label{fig:langevin}
\end{figure*}

\noindent The corresponding conformal limits are given by \cite{Gubser:2006nz,Casalderrey-Solana:2007ahi}
\begin{align}
    \lim_{T\to\infty}\frac{\kappa_{\perp}}{\sqrt{\lambda_{t}}T^{3}}(T,\mu_B;v)&\to\pi\gamma(v)^{1/2}=\frac{\pi}{(1-v^{2})^{1/4}},\\
      \lim_{T\to\infty}\frac{\kappa_{\parallel}}{\sqrt{\lambda_{t}}T^{3}}(T,\mu_B;v)&\to\pi\gamma(v)^{5/2}=\frac{\pi}{(1-v^{2})^{5/4}}.
\end{align}

We remark that, as discussed in detail in Ref.\ \cite{Gursoy:2010aa}, it is possible to define two velocity-dependent jet quenching parameters associated with the heavy quark Langevin diffusion coefficients as follows
\begin{align}\label{eq:heavyqhats}
\hat{q}_\perp=\frac{\langle p_\perp^2 \rangle}{vt}=\frac{2\kappa_\perp}{v}, \qquad \hat{q}_\parallel=\frac{\langle \Delta p_\parallel^2 \rangle}{vt}=\frac{\kappa_\parallel}{v},
\end{align}
where $\langle p_\perp^2 \rangle$ and $\langle \Delta p_\parallel^2 \rangle$ are, respectively, linear approximations for the noise-averaged transverse and longitudinal momentum fluctuations of a heavy quark after traveling a distance $vt$ within the medium. The velocity-dependent coefficient $\hat{q}_\perp(T,\mu_B;v)$ gives the transverse momentum broadening of a heavy quark moving with velocity $v$ within the fluid \cite{Gursoy:2010aa,Gubser:2006nz}.

Our numerical results are shown in Fig.\ \ref{fig:langevin} for the perpendicular (upper panels) and parallel (lower panels) Langevin diffusion coefficients, where we considered again two different velocities: the left panels correspond to $v=0.5$, while the right panels refer to $v=0.99$.

Analogously to the heavy quark drag force, the heavy quark Langevin diffusion coefficients are also enhanced with increasing the baryon density of the medium, and they also display a more sensitive dependence on the temperature and the baryon chemical potential of the fluid at large velocities. As the baryon density of the medium is enhanced, these coefficients develop a peak and an inflection point, with the latter moving toward the CEP, where it acquires an infinite slope. On top of the first order line these quantities develop a large discontinuity gap and in particular, the gap in the parallel diffusion coefficient at $v=0.99$ is extremely large. One also notices that, at fixed velocities, the parallel diffusion coefficient is always larger than the perpendicular one, in consonance with the universal inequality $\kappa_\parallel \ge \kappa_\perp$ (the equality is saturated in the limit $v\to 0$) obtained in Ref.\ \cite{Gursoy:2010aa}. We note that, due to the large difference between $\kappa_\parallel $ and $ \kappa_\perp$ in our framework even at $\mu_B=0$, it would be interesting to revisit the Bayesian analysis in   \cite{Xu:2017obm} relaxing the simplifying assumption that $\kappa_\parallel = \kappa_\perp$.

\subsection{The jet quenching parameter}
\label{sec:jet}

The energy loss from collisional and radiative processes of high energy partons produced by the interaction with the hot and dense medium they travel through can be characterized by the jet quenching parameter $\hat{q}$, defined as the rate for transverse momentum broadening \cite{Baier:1996sk}.  At vanishing $\mu_B$, there have been a number of studies of $\hat{q}$ across both the QGP and HRG phases.  It is generally thought that $\hat{q}/T^3$ increases with increasing temperature in the HRG phase \cite{Dorau:2019ozd,McLaughlin:2021dlq}, while the QGP phase comparatively has larger values of $\hat{q}/T^3$. It is still an open question how the two phases join together, where some work appears to indicate a jump/peak in $\hat{q}/T^3$ at the onset of the QGP phase \cite{JET:2013cls,Shi:2018izg} whereas others argue for a smooth matching across the phase transition \cite{Baier:2002tc,Rougemont:2015wca,Rougemont:2017tlu,McLaughlin:2021dlq}. Additionally, there are known tensions when comparing extractions of $\hat{q}$ at RHIC and the LHC \cite{Andres:2016iys}.

The holographic formalism proposed in Refs.\ \cite{Liu:2006ug,Liu:2006he} to describe the jet quenching parameter $\hat{q}(T,\mu_B)$ associated with the transverse momentum broadening of light partons moving at the speed of light \cite{Baier:1996sk} was employed in Ref.\ \cite{Rougemont:2015wca} in order to obtain the following holographic formula in the context of the EMD model
\begin{equation}\label{eq:qhat}
    \frac{\hat{q}}{\sqrt{\lambda_{t}}T^{3}}(T,\mu_{B})=\frac{64\pi^{2}h_{0}^{\textrm{far}}}{\int_{r_{\textrm{start}}}^{r_{\textrm{max}}}dr\frac{e^{-\sqrt{2/3}\phi(r)-3A(r)}}{\sqrt{h(r)\left[h_{0}^{far}-h(r)\right]}}},
\end{equation}
with the associated conformal limit given by \cite{Liu:2006ug}
\begin{equation}
    \lim_{T\to\infty}\frac{\hat{q}}{\sqrt{\lambda_{t}}T^{3}}(T,\mu_B)=\frac{\pi^{3/2}\Gamma(3/4)}{\Gamma(5/4)}\approx7.52814.
\end{equation}

\begin{figure}[h]
    \centering
    \includegraphics[width=0.49\textwidth]{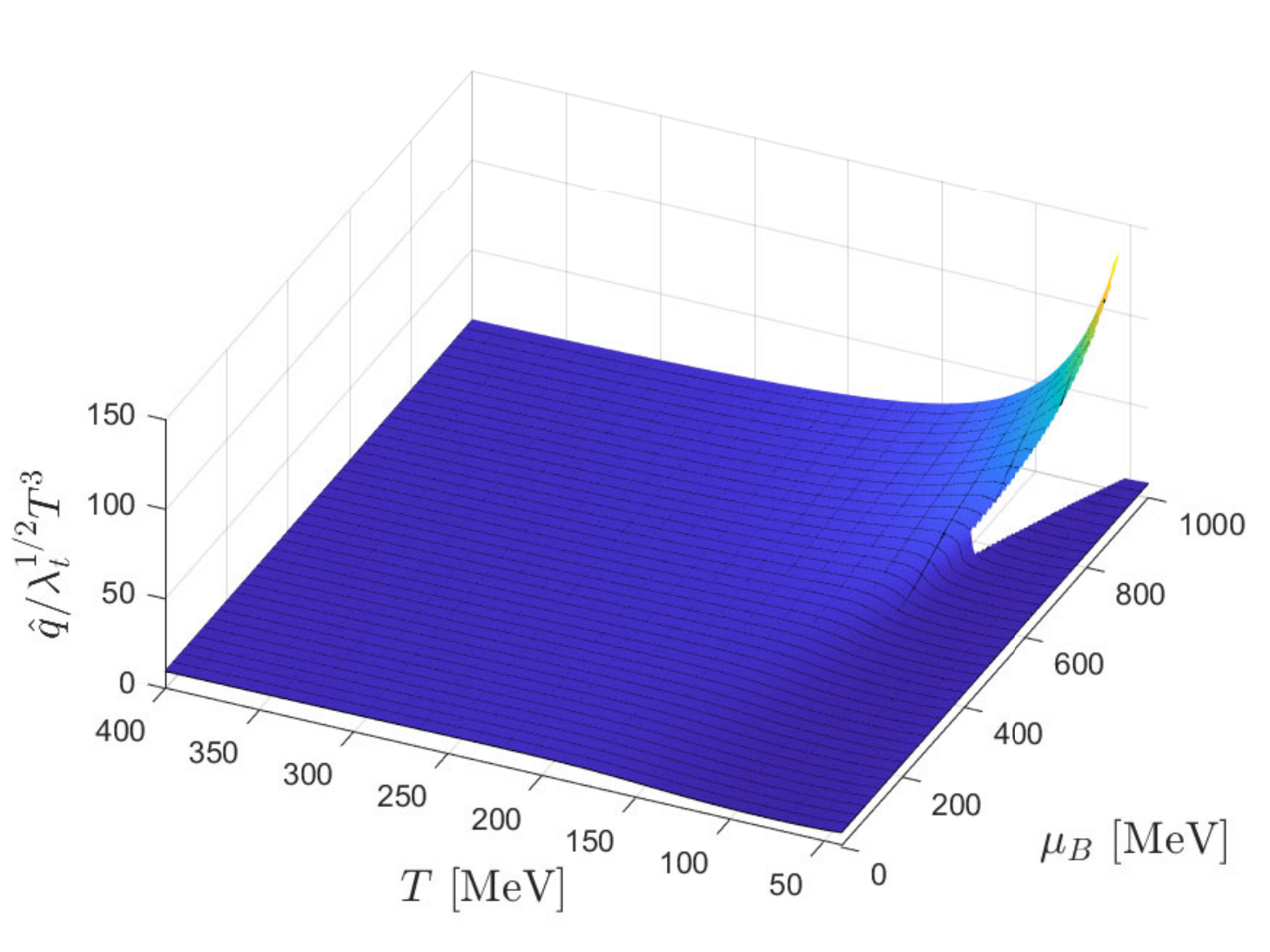}
    \includegraphics[width=0.49\textwidth]{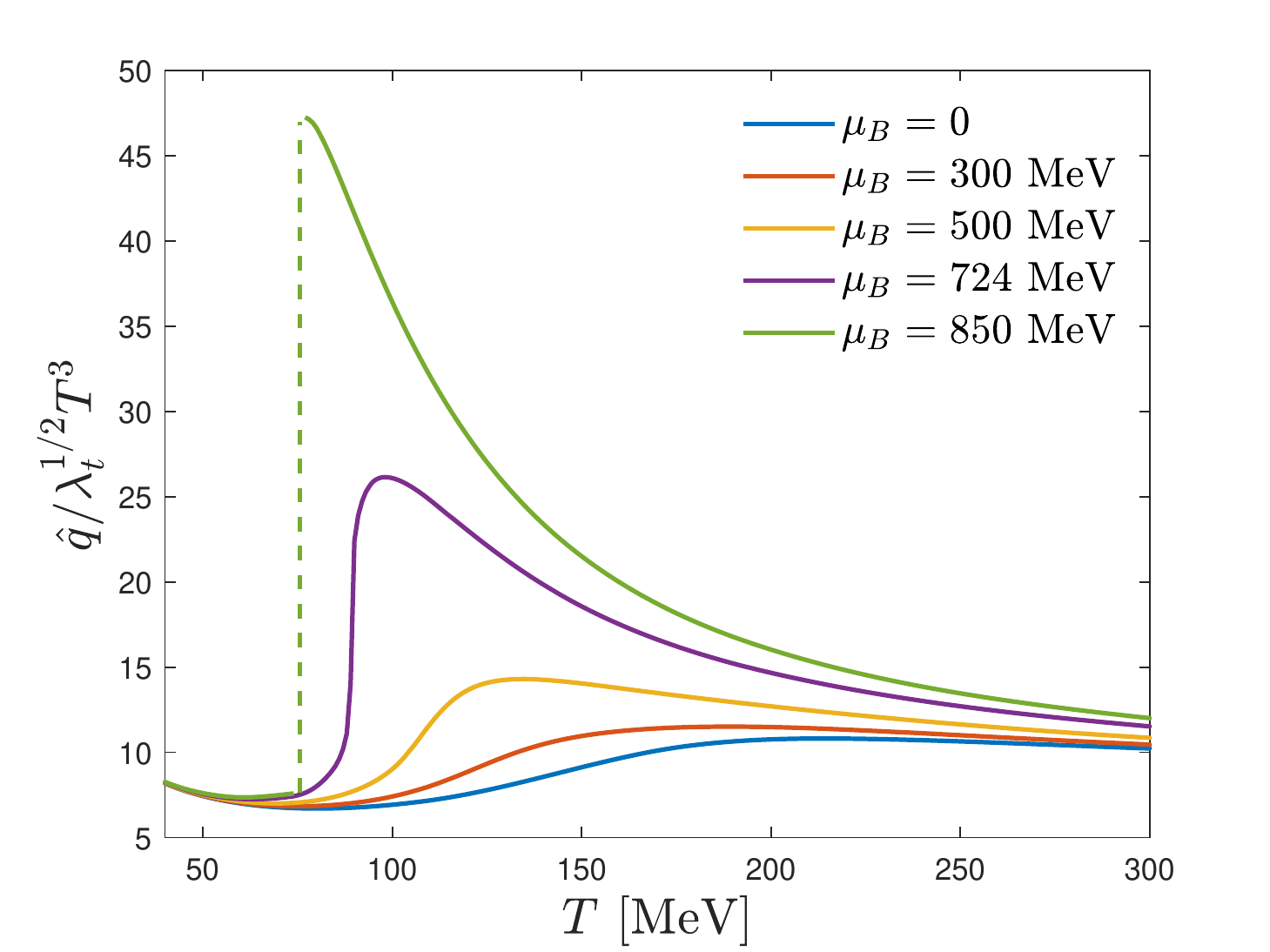}
     \caption{Upper panel: scaled jet quenching parameter $\hat{q}/\lambda_{t}^{1/2}T^{3}$ obtained from Eq. (\ref{eq:qhat}) as a function of temperature and baryon chemical potential. Lower panel: scaled jet quenching parameter as a function of the temperature, for several values of $\mu_B$.}
    \label{fig:qhat}
\end{figure}

The results for the holographic jet quenching parameter $\hat{q}$ given by Eq.\ \eqref{eq:qhat} are shown in Fig.\ \ref{fig:qhat}. The jet quenching parameter, akin to the heavy quark drag force and the Langevin diffusion coefficients, is also enhanced as one increases the  baryon chemical potential of the medium, indicating more jet suppression and parton energy loss in the baryon rich regime. This quantity displays a peak around the crossover region which largely increases in magnitude, while also becoming sharper as the chemical potential increases; however, it is the inflection point that moves toward the CEP as the baryon density of the fluid is increased, acquiring an infinite slope in the critical region. On top of the first-order line the jet quenching parameter presents a large discontinuity gap, as shown in Fig.\ \ref{fig:qhat}.

\section{Shear and Bulk Viscosities}
\label{sec:viscosities}

In this section we present our results for the shear and bulk viscosities of the hot and baryon dense medium described by our EMD model.

\subsection{Shear viscosity}
\label{sec:shear}

The nearly perfect fluidity of the QGP, characterized by a small value of its shear viscosity to entropy density ratio, $\eta/s$, is one of the most striking features of the deconfined QCD medium produced in relativistic heavy ion collisions. The calculation of the shear viscosity \cite{Moore:2020pfu}, which measures the medium's resistance to sheared flow in the presence of a velocity gradient of the fluid, as well as other transport observables, has faced great difficulties from \textit{ab initio} lattice calculations \cite{Meyer:2011gj}. However, the (almost) universal holographic result for the shear viscosity to entropy density ratio, $\eta/s=1/4\pi$ \cite{Kovtun:2004de}, which is valid for a broad class of strongly coupled fluids with holographic duals,\footnote{The result $\eta/s=1/4\pi$ holds for any holographic model which is isotropic, translationally invariant, and has at most two derivatives of the metric field in the bulk gravity action. Our EMD model fits in such restrictions, and therefore has $\eta/s=1/4\pi$ for any value of $T>0$ and $\mu_B\ge 0$.} has been successful in obtaining compatibility with computed bounds for $\eta/s$ extracted from the Bayesian analysis of several experimental data of heavy ion collisions \cite{Bernhard:2016tnd,JETSCAPE:2020shq,Nijs:2021clz}.

Although for the present EMD model $\eta/s=1/4\pi$ for any value of temperature and chemical potential, we note that the actual measure of fluidity in a baryon dense medium is given instead by the following dimensionless combination \cite{Liao:2009gb}
\begin{equation}\label{eq:etaHol}
    \frac{\eta T}{\epsilon+P}(T,\mu_{B})=\frac{1}{4\pi\left(1+\frac{\mu_{B}\rho_{B}}{Ts}\right)},
\end{equation}
which normalizes the product between the temperature and the shear viscosity by the enthalpy density of the medium, and where the right-hand side of Eq.\ \eqref{eq:etaHol}, which depends solely on thermodynamical observables, holds for holographic models with $\eta/s=1/4\pi$.

\begin{figure}[h]
    \centering
    \includegraphics[width=0.49\textwidth]{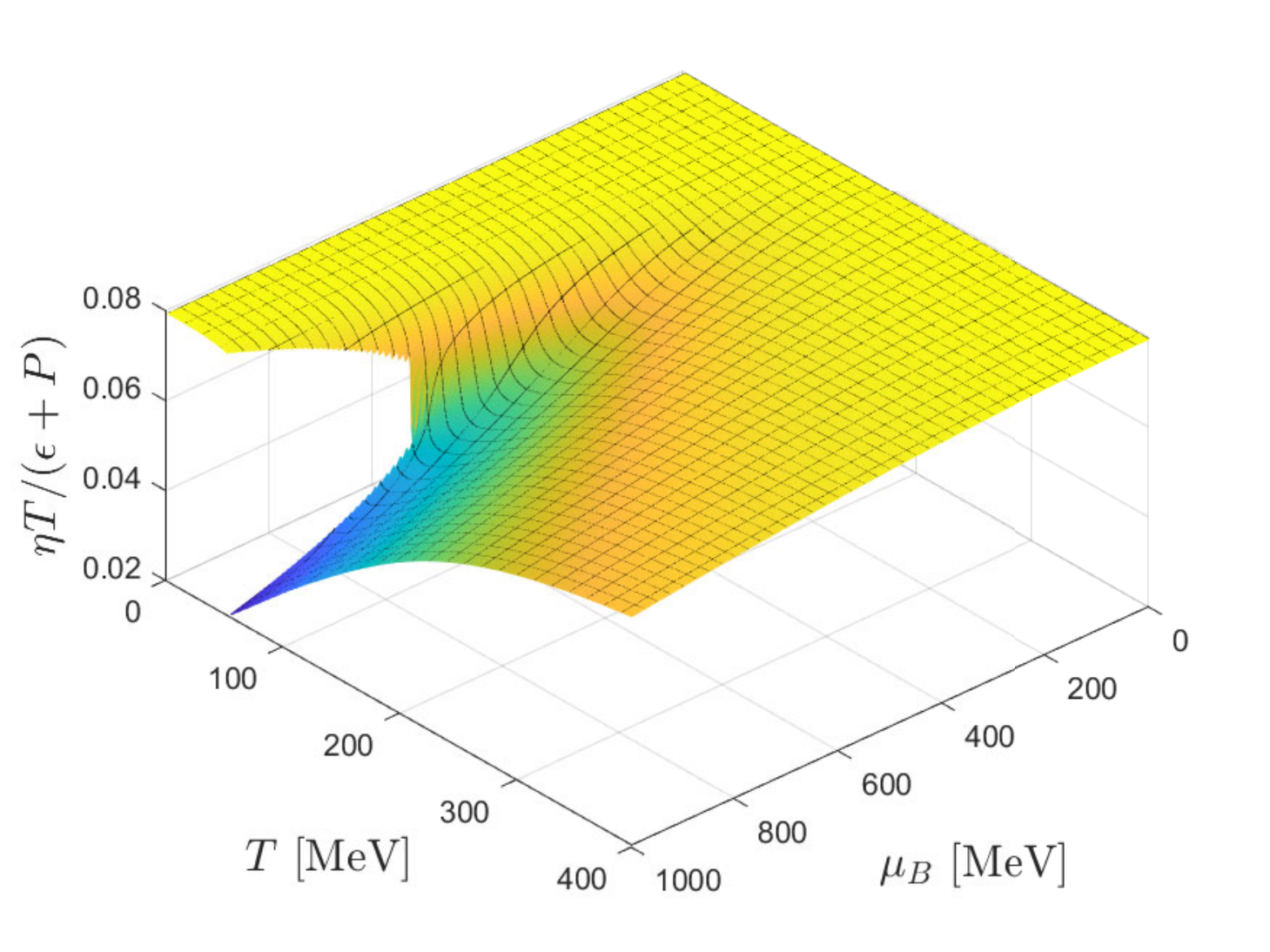}
    \includegraphics[width=0.49\textwidth]{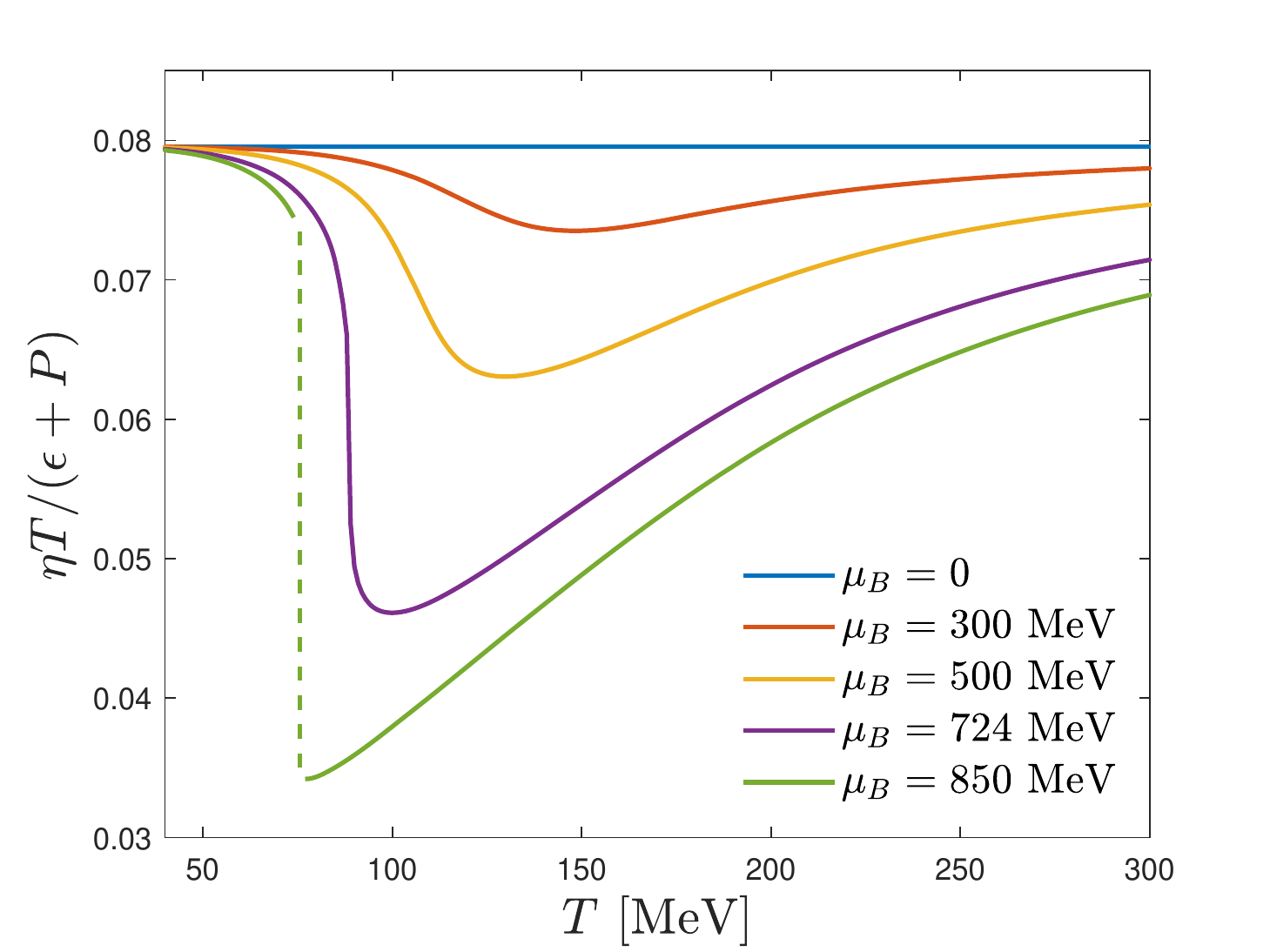}
     \caption{Holographic shear viscosity times temperature over enthalpy density obtained from Eq.\ \eqref{eq:etaHol} as a function of $T$ and $\mu_B$ (top), and the same observable as a function of the temperature for several values of $\mu_B$ (bottom). This dimensionless combination reduces to $\eta/s=1/4\pi$ at $\mu_{B}=0$.}
    \label{fig:etaHolographic}
\end{figure}

Our results for the normalized shear viscosity, defined in Eq. \eqref{eq:etaHol}, are presented in Fig.\ \ref{fig:etaHolographic}. At $\mu_{B}=0$, Eq.\eqref{eq:etaHol} reduces to $\eta/s=1/4\pi$, whereas at finite baryon chemical potential, the normalized shear viscosity develops a nontrivial dependence on the temperature, and a minimum and an inflection point are produced, with the latter moving toward the CEP as the baryon density is increased. At the CEP, an infinite slope is observed\footnote{The shear viscosity is actually finite at the critical point in our model. This is again compatible with model B dynamical universality class \cite{Hohenberg:1977ym}, in contrast to what is expected to hold in QCD where $\eta$ (very slowly) diverges at the critical point \cite{Son:2004iv}.}, while for values of $\mu_B$ beyond the CEP, the normalized shear viscosity develops a discontinuity gap at the first-order line, as a consequence of the corresponding discontinuity gaps in the baryon density $\rho_{B}$ and the entropy density $s$ in this region of the phase diagram \cite{Grefa:2021qvt}. The fact that the normalized shear viscosity given by Eq.\ \eqref{eq:etaHol} decreases with increasing values of $\mu_B$ indicates that the QGP becomes even closer to the perfect fluid limit in the baryon dense regime. As far as we are aware, no theoretical models have studied a discontinuity across the first-order phase transition line within a realistic relativistic viscous hydrodynamics framework, so we do not know what the consequences of this effect would be in hydrodynamic  simulations. However, the presence of any shear viscosity at large baryon densities can significantly affect the passage through the QCD phase diagram \cite{Feng:2018anl,Dore:2020jye,Du:2021zqz}

\subsection{Bulk viscosity}
\label{sec:bulk}

The bulk viscosity measures the medium's resistance to deformations associated with a compression or an expansion of the fluid, and has been shown to also play an important role in relativistic heavy ion collisions. It has a complicated interplay with shear viscosity in the QGP \cite{Noronha-Hostler:2014dqa} and it can also affect the transverse momentum spectra, the azimuthal momentum anisotropy, and the multiplicity of charged hadrons produced in heavy ion collisions \cite{Ryu:2015vwa}. Large uncertainties remain on the extraction of bulk viscosity due to uncertainties in the correct out-of-equilibrium description at the point where a fluid switches to particles \cite{Monnai:2009ad,Dusling:2011fd,Noronha-Hostler:2013gga,Noronha-Hostler:2014dqa,JETSCAPE:2020mzn}. However, it is generally believed that, at vanishing baryon densities, a peak exists around the crossover region \cite{Noronha-Hostler:2008kkf,Ryu:2015vwa,Bernhard:2016tnd,Bernhard:2019bmu}. There have been arguments that the bulk viscosity should diverge at the QCD critical point such that $\zeta \propto \xi^3$ \cite{Monnai:2016kud} (where $\xi$ is the correlation length), compatible with the expectation that QCD is in the model H dynamical universality class \cite{Son:2004iv}. The presence of such divergence has strong implications for the search for the critical point and the applicability of hydrodynamics \cite{Monnai:2016kud,Dore:2020jye,Dore:2021sbl}.

In this section we compute the bulk viscosity for the EMD model and obtain its dependence on $T$ and $\mu_{B}$, including the region of the phase diagram comprising the CEP and the line of first-order phase transition.
As discussed in Refs.\ \cite{DeWolfe:2011ts,Rougemont:2017tlu}, the EOM for the relevant linearized homogeneous perturbation in the $SO(3)$ singlet channel of the EMD model, $\mathcal{H}(r,\omega)$, is given by

\begin{align}\label{eq:bulkEOM}
     \mathcal{H}''&+\left(4A'+\frac{h'}{h}+\frac{2\phi''}{\phi}-\frac{2A''}{A'}\right)\mathcal{H}'\nonumber\\
    &+\left[\frac{e^{-2A}\omega^{2}}{h^{2}}+\frac{h'}{h}\left(\frac{A''}{A'}-\frac{\phi''}{\phi'}\right)\right.\nonumber\\
    &+\left.\frac{e^{-2A}}{h\phi'}(3A'f'(\phi)-f(\phi)\phi')\Phi'^{2}\right]\mathcal{H}=0,
\end{align}
which must be solved with in-falling wave condition at the black hole horizon, and normalized to unity at the boundary, which may be done by setting
\begin{equation}\label{eq:bulkWaveCond}
    \mathcal{H}(r,\omega)\equiv\frac{r^{-i\omega}F(r,\omega)}{r_\textrm{max}^{-i\omega}F(r_\textrm{max},\omega)},
\end{equation}
where the EOM for $F(r,\omega)$ and the initial conditions $F(r_{\textrm{start}},\omega)$ and $F'(r_{\textrm{start}},\omega)$ are obtained in an analogous way to what was previously discussed in the case of the baryon conductivity, below Eq.\ \eqref{eq:condWaveCond}.

The ratio between the bulk viscosity and the entropy density in the EMD model is then calculated by making use of the following holographic Kubo formula \cite{DeWolfe:2011ts,Rougemont:2017tlu}
\begin{equation} \label{eq:zeta}
    \frac{\zeta}{s}(T,\mu_{B})=-\frac{1}{36\pi}\lim_{\omega\rightarrow0}\frac{1}{\omega}\left(\frac{e^{4A}h\phi'^{2}\textrm{Im}[\mathcal{H}^{*}\mathcal{H}']}{A'^{2}}\right),
\end{equation}
where analogous observations to what was discussed below Eq.\ \eqref{eq:conduct} also apply here.

\begin{figure}[h!]
    \centering
    \includegraphics[width=0.49\textwidth]{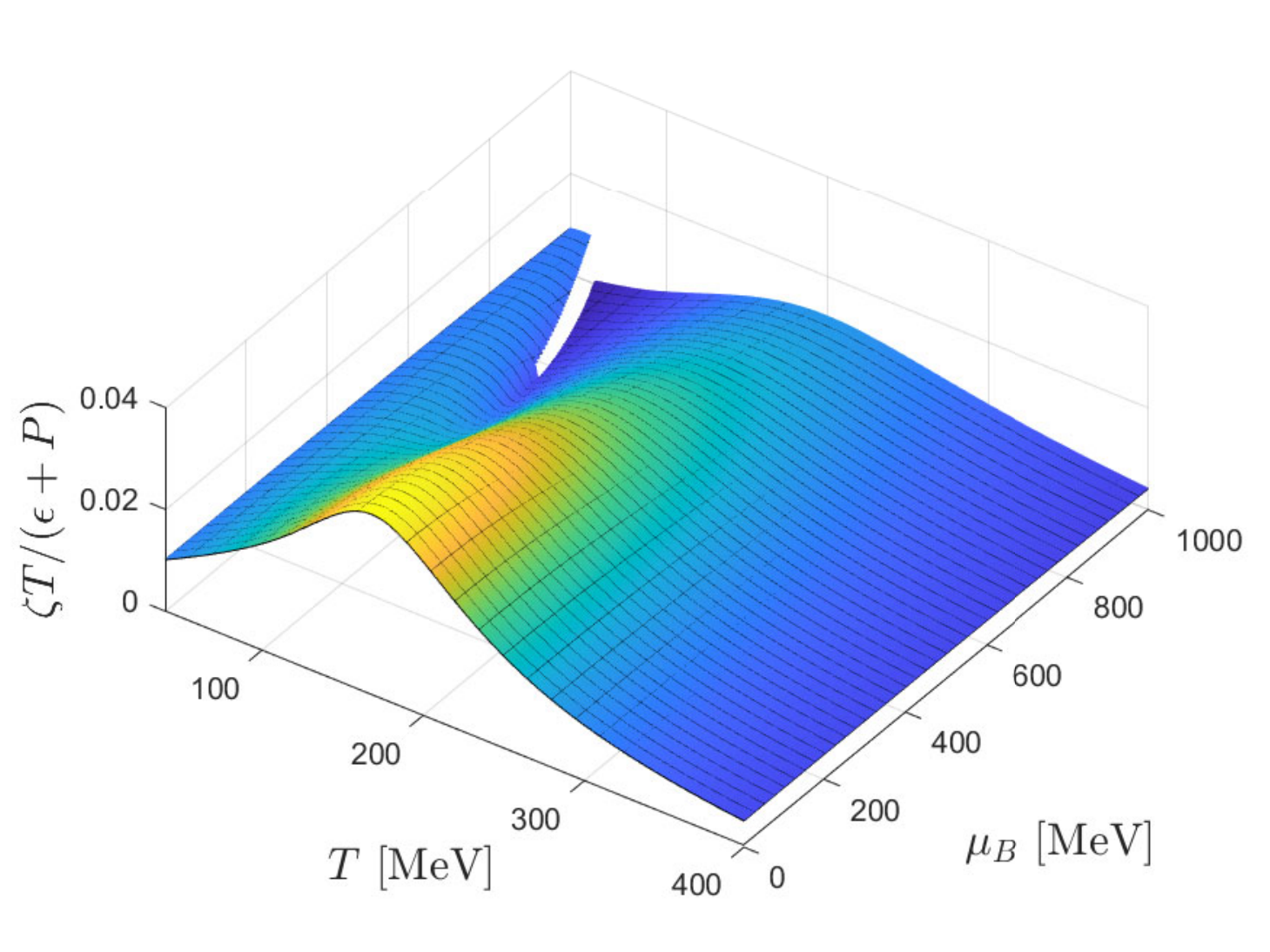}
    \includegraphics[width=0.49\textwidth]{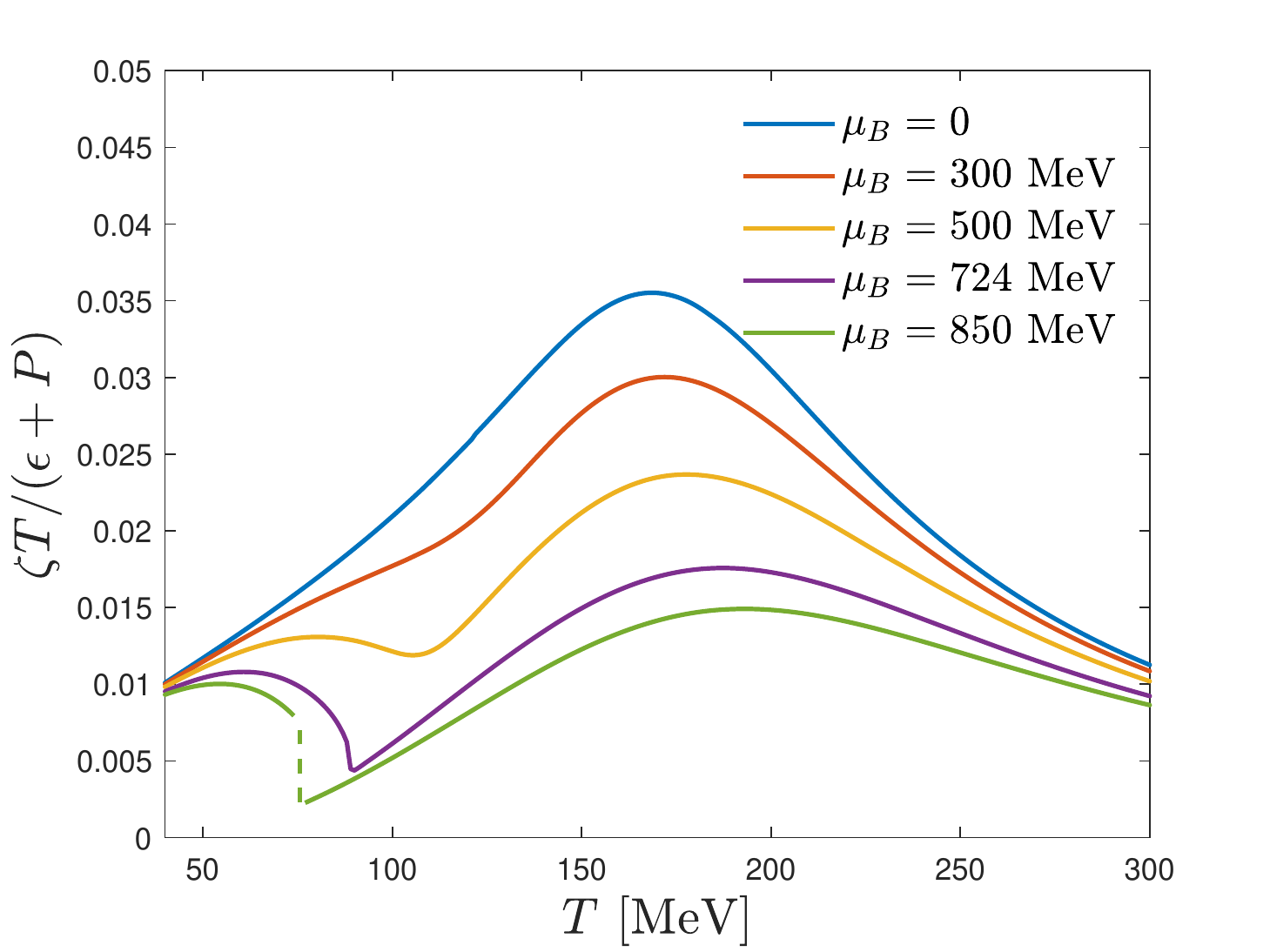}
     \caption{Holographic bulk viscosity $\zeta T/(\epsilon+P)$ as a function of $T$ and $\mu_B$ (top), and the same observable as a function of the temperature for several values of $\mu_B$ (bottom).}
     \label{fig:BulkVisc}
\end{figure}

From Eq.\ \eqref{eq:zeta} and from the thermodynamic state variables given by Eqs.\ \eqref{eq:thermodynamics}, one can also obtain the following dimensionless combination, which naturally appears in the hydrodynamic expression for the bulk viscous pressure of charged fluids, like the baryon dense QGP,
\begin{equation}\label{eq:BulkViscosity}
    \frac{\zeta T}{\epsilon+P}(T,\mu_{B})=\frac{\zeta}{s}(T,\mu_{B})\frac{1}{1+\frac{\mu_{B}\rho_{B}}{Ts}},
\end{equation}
and reduces to $\zeta/s$ when $\mu_{B}=0$.

The numerical results obtained using Eq.\ \eqref{eq:BulkViscosity} are presented in Fig.\ \ref{fig:BulkVisc}. The holographic bulk viscosity presents a peak in the crossover region at $\mu_{B}=0$. However, this peak does not evolve toward the CEP as $\mu_{B}$ is enhanced. Instead, its location slightly shifts toward higher temperatures as the baryon chemical potential is increased. This is an important qualitative difference in comparison with some earlier versions of the EMD model as e.g. Refs.\ \cite{DeWolfe:2011ts,Rougemont:2017tlu}. Additionally, whereas in Ref.\ \cite{DeWolfe:2011ts} the magnitude of the peak remains about the same as the chemical potential is increased up to the critical region, our present results show that the height of the peak of the bulk viscosity decreases as $\mu_{B}$ increases, similarly to what is observed in Ref.\ \cite{Rougemont:2017tlu}. Consequently, one concludes that the behavior of the peak observed in the holographic bulk viscosity as the baryon density of the medium is increased is a model-dependent feature.

In the present EMD model, the normalized bulk viscosity given by Eq.\ \eqref{eq:BulkViscosity} develops a dip at finite $\mu_B$, which moves toward the CEP as the baryon density is increased. At the CEP, the bulk viscosity acquires an infinite slope, similarly to what is observed for the shear viscosity in Fig.\ \ref{fig:etaHolographic}. In fact, the quantity $\zeta/s$ computed from Eq.\ \eqref{eq:zeta} also exhibits a divergent slope at the CEP as shown in Fig.\ \ref{fig:bulk_loop}. For values of $\mu_B$ beyond the CEP, and on top of the line of first order phase transition, the bulk viscosity develops a discontinuity gap.

As observed for the shear viscosity, also the bulk viscosity is overall suppressed as the baryon chemical potential of the medium is increased, indicating that viscous effects become smaller in the baryon dense regime of the QGP. Such a prediction seems to be a robust feature of the EMD setup, since it is also observed in earlier versions of the model discussed e.g. in Refs.\ \cite{DeWolfe:2011ts,Rougemont:2017tlu}.

Another important observation extracted from the behavior of the holographic bulk viscosity is the fact that this transport observable remains finite at the CEP in EMD holography \cite{DeWolfe:2011ts}, contrary to the prediction obtained in some other effective models regarding a divergent bulk viscosity at the CEP \cite{Monnai:2016kud,PhysRevE.55.403,Buchel:2009mf}. This is possibly reminiscent of the different dynamical universality classes \cite{Hohenberg:1977ym} expected to hold for QCD (type-H) \cite{Son:2004iv} and holographic / large-$N_c$ approaches (type-B) \cite{Natsuume:2010bs}. Also, there are differences in the static critical exponents, since only the corresponding mean field values for the 3D Ising universality were found in the EMD holographic model of Ref.\ \cite{DeWolfe:2010he}.

\begin{figure}[h!]
    \centering
    \includegraphics[width=0.49\textwidth]{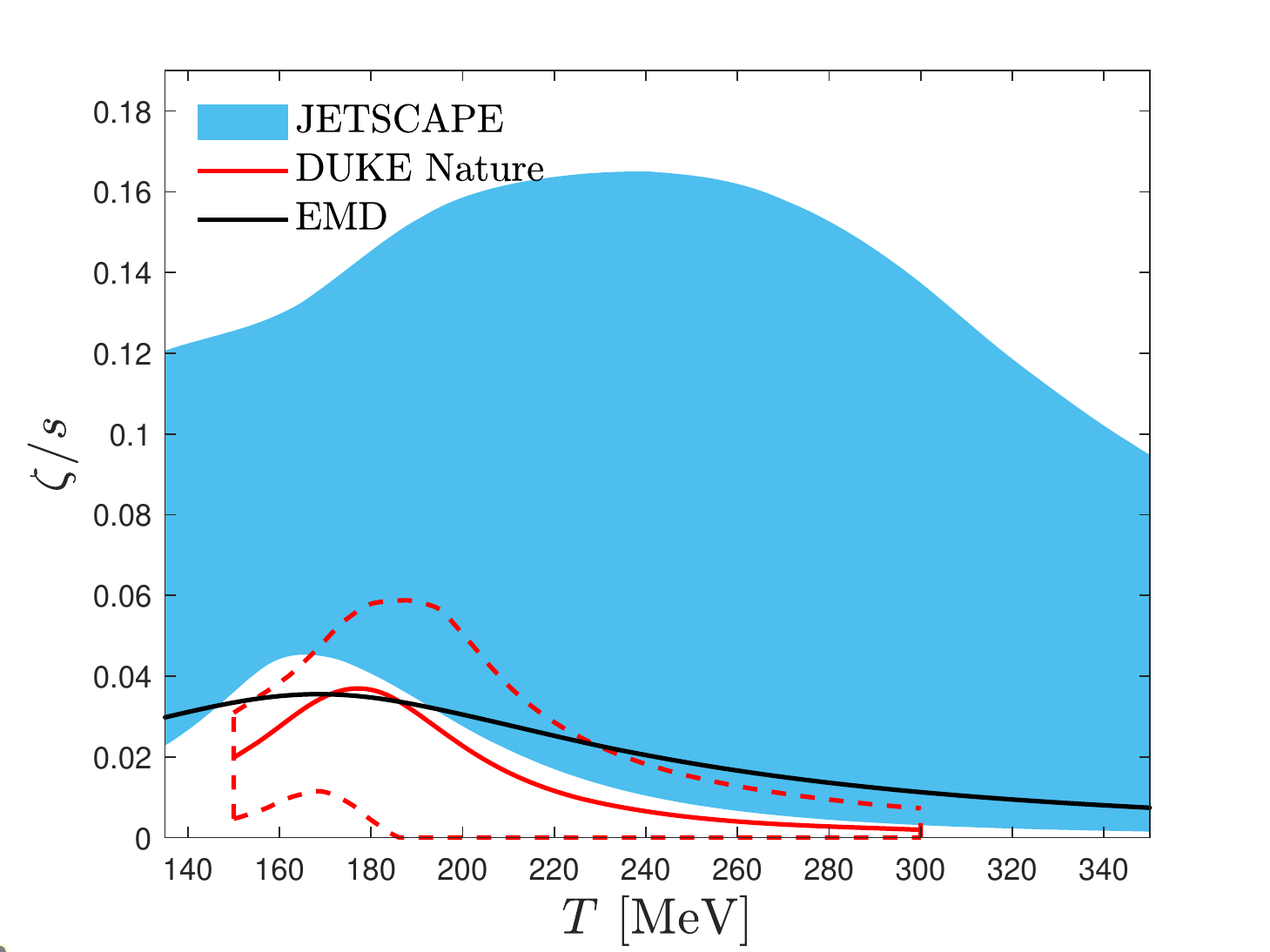}
    \caption{Holographic bulk viscosity $\zeta/s$ at $\mu_{B}=0$ compared with the 90\% credible intervals of the JETSCAPE Bayesian model from \cite{JETSCAPE:2020shq}, and with the results of the Duke group \cite{Bernhard:2019bmu}.}
    \label{fig:zeta_hydro}
\end{figure}

The peak of the bulk viscosity at zero density, which in the current holographic model is located at $T\sim 168$ MeV, seems to be a feature of this dynamical observable since a similar profile is found in Bayesian analyses where $\zeta/s$ is extracted from comparisons of relativistic hydrodynamics calculations to experimental data. A comparison between the holographic bulk viscosity at $\mu_{B}=0$ and the recent Bayesian analyses from two different groups \cite{Bernhard:2019bmu,JETSCAPE:2020shq} is shown in Fig.\ \ref{fig:zeta_hydro}.

\section{Out-of-equilibrium phase diagram}
\label{out}

\begin{figure}
    \centering
    \includegraphics[width=0.49\textwidth]{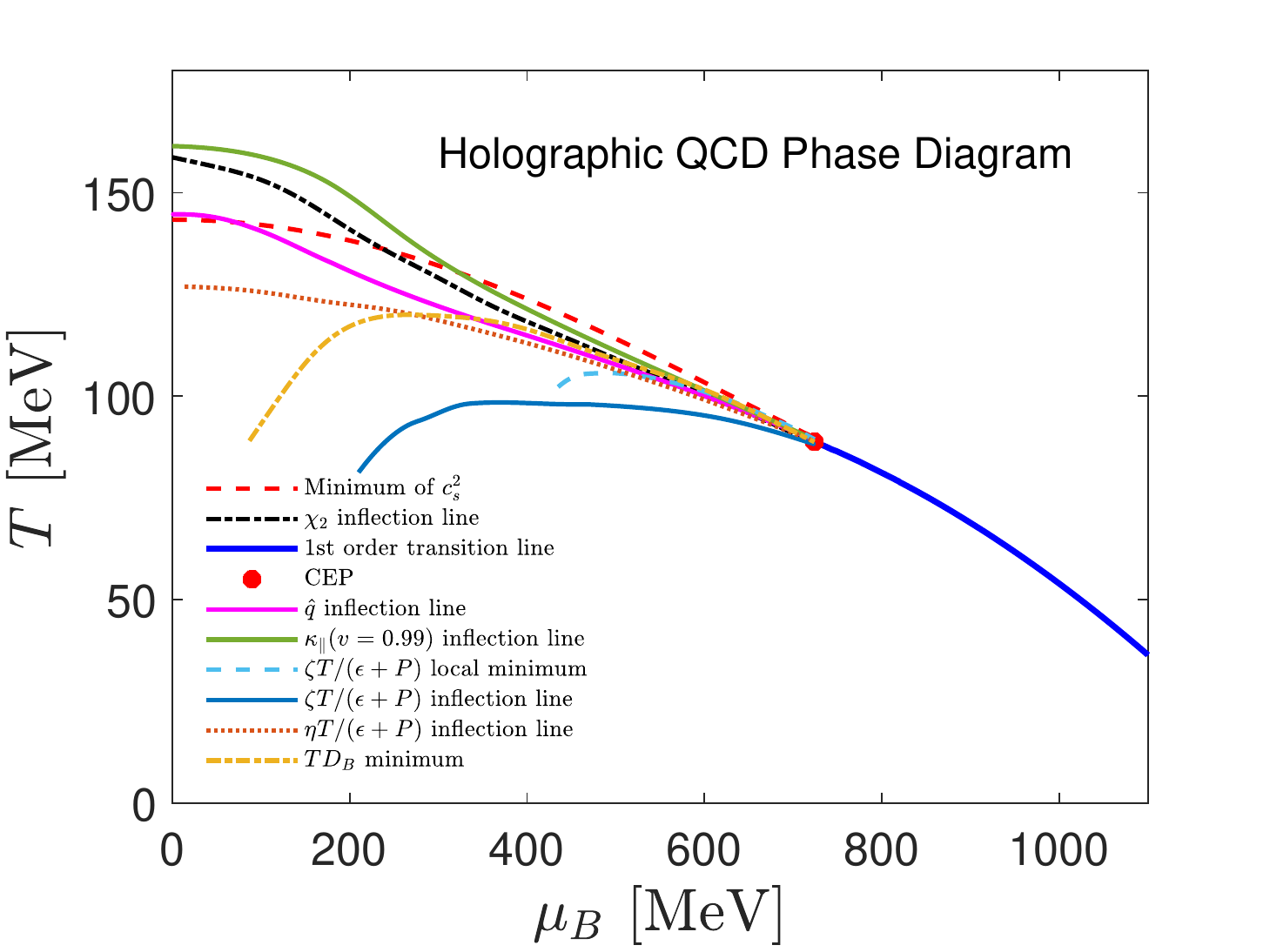}
    \caption{Holographic phase diagram which also includes the inflection lines and extrema from transport observables to characterize the crossover region. The inflection lines of the jet quenching parameter $\hat{q}$, the parallel Langevin coefficient $\kappa_{\parallel}$ at $v=0.99$, and the minimum and inflection point of $\zeta T/(\epsilon+P)$, the inflection of $\eta T/(\epsilon+P)$, and minimum of $D_{B}$ are also displayed in the phase diagram in addition to the lines obtained from equilibrium variables. Notice that some of these characteristic points are only produced at nonzero values of the baryon chemical potential.} 
    \label{fig:phaseDiagramDynam}
\end{figure}

For a crossover transition it is possible to have a wide range of pseudocritical temperatures  that depend on what observable one is studying.  However, at a critical point one expects that all pseudocritical temperatures should converge. Then, exactly at the critical point  the (properly normalized, hence, dimensionless) transport coefficients should display an infinite slope. Furthermore, one expects that across the first-order phase transition line a discontinuous gap should open up and become wider with increasing $\mu_B$. This behavior is found in all transport quantities shown here.  

However, one surprising effect that we find is that sometimes the most interesting characteristic points at $\mu_B\rightarrow 0$ (such as the peak in the bulk viscosity) are not the relevant points for criticality. In the case of the bulk viscosity, a minimum is formed at intermediate $\mu_B$ that eventually moves to the location of the critical point, whereas the peak in bulk viscosity remains at nearly the same temperature even at large $\mu_B$ (contrary to what happens in some older EMD models considered e.g. in Refs.\ \cite{DeWolfe:2011ts} and \cite{Rougemont:2017tlu}). A similar effect is seen for the drag force  at $v=0.5$ in Fig.\ \ref{fig:Fdrag05} and for the Langevin diffusion coefficients also at $v=0.5$ in Fig.\ \ref{fig:langevin}.

Thus, while some characteristic points of transport coefficients can be also employed to identify the crossover region in addition to the equilibrium state variables used in Ref.\ \cite{Grefa:2021qvt}, one must be careful in correctly identifying the actual sequence of inflection points which evolve toward the CEP as the baryon density increases. In particular, due to the aforementioned nonuniversal behavior of the peak in the bulk viscosity in different models, caution is needed in order to avoid being misled by false ``transition lines'' in the crossover region associated with points that do not evolve toward the CEP.

We summarize the characteristic points that are relevant to the critical point in Fig.\ \ref{fig:phaseDiagramDynam} and compare them to the equilibrium points found previously in \cite{Grefa:2021qvt}. The inflection lines of the jet quenching parameter $\hat{q}$, the parallel Langevin coefficient $\kappa_{\parallel}$ at $v=0.99$, and the minimum and inflection point of $\zeta T/(\epsilon+P)$, the inflection of $\eta T/(\epsilon+P)$, and minimum of $D_B$ have been used as proxies for a transition line in the crossover region. Interestingly enough, within the crossover region and far from the CEP, none of the characteristic curves follow the same behavior, even though they all converge at the CEP. Additionally, it also appears that the spread in temperature in the equilibrium lines is smaller than the out-of-equilibrium ones (at a fixed point in $\mu_B$). This suggest that out-of-equilibrium quantities demonstrate an even wider cross-over than equilibrium ones.

\section{Conclusions}
\label{sec:conclusion}

In this work, we studied several transport properties of the hot and dense QGP across the phase diagram using the holographic EMD model of Refs.\ \cite{Critelli:2017oub,Grefa:2021qvt}. This EMD model is in quantitative agreement with state-of-the-art lattice QCD thermodynamics with $2+1$ flavors at the physical point, both at zero and finite baryon density \cite{Borsanyi:2013bia,Bellwied:2015lba,Borsanyi:2021sxv}. We found that the diffusion of baryon charge and also the hydrodynamic shear and bulk viscosities are suppressed with increasing baryon density, indicating that the medium becomes even closer to perfect fluidity at large densities. Moreover, we also found that the jet quenching parameter, the heavy quark drag force, and the momentum diffusion are enhanced as one increases the baryon density of the medium toward the critical region of the phase diagram.

Overall, the different physical quantities display a discontinuity gap at the line of first order phase transition, while developing an infinite slope at the critical point, which further degenerates into different kinds of extrema in the crossover region depending on the observable considered. The baryon conductivity is not greatly affected by the increasing baryon chemical potential and remains finite at the critical point, although it develops an infinite slope. The baryon diffusion coefficient develops a minimum as $\mu_{B}$ increases, and vanishes exactly at the critical point, where this minimum becomes a cusp. The minimum exhibited by the thermal conductivity at vanishing chemical potential also extrapolates to the critical point, where it too becomes a cusp. Furthermore, the characteristic points displayed by the heavy quark drag force, the Langevin diffusion coefficients, the jet quenching parameter and the shear viscosity all move toward the CEP. It is noteworthy that most of these inflection points are only present at finite values of the baryon chemical potential, i.e. are not found at $\mu_{B}=0$. The same is true for the minimum displayed by the normalized bulk viscosity, which only appears at $\mu_{B}\gtrsim 435$ MeV, and also coincides with the CEP at larger densities.

The inflection lines and extrema displayed by some of the transport observables can be used as markers for the crossover region in the phase diagram, as shown in Fig. \ref{fig:phaseDiagramDynam}, where they have been used as proxies for a crossover line.

Finally, it is important to try to address the phenomenological reliability, and also discuss the limitations of the present holographic EMD approach. As aforementioned, the present EMD model has the merits of being able to quantitatively describe state-of-the-art lattice data on QCD thermodynamics, both at zero and finite baryon density, moreover it naturally encompasses the almost perfect fluidity of the strongly coupled QGP produced in heavy ion collisions, besides also predicting a bulk viscosity at zero density which is in the ballpark of values favored in state-of-the-art Bayesian analyses of phenomenological models simultaneously describing several sets of heavy-ion data. However, the present EMD model also has relevant limitations. In fact, for instance, it does not describe the chiral condensate and chiral symmetry breaking, and it is also unable to describe hadron thermodynamics (a limitation which is shared with all other holographic gauge/gravity models which we know---the reason being ultimately tied to the large $N_c^{-2}$ suppression of the pressure of the medium in the hadronic phase relatively to the deconfined QGP phase in the large $N_c$ limit). Furthermore, the fact that the present EMD model is in good quantitative agreement with the latest lattice QCD data at finite baryon density does not automatically guarantee that the predictions made for regions of the QCD phase diagram well beyond the reach of current lattice simulations are phenomenologically reliable. Indeed, the fact that the EMD model of Ref. \cite{Cai:2022omk} is also able to obtain a good quantitative agreement with lattice QCD thermodynamics at zero and finite baryon density, while still predicting the QCD CEP at a significantly different location than in our model, shows that the available lattice data is not enough to strongly constraint such a prediction in the EMD class of holographic models. It may also be that the freedom of choosing the functional forms for the free functions $V(\phi)$ and $f(\phi)$ of the EMD model plays a relevant role in this issue. This important question is something which is currently under investigation in our research group and we hope to report new findings in this regard soon.
 
We also hope that the novel behavior displayed by the multitude of transport coefficients in the vicinity of the critical point and across the first-order line, computed for the first time in this paper, will motivate new studies and simulations of the out-of-equilibrium dynamics of the hot and baryon-rich quark-gluon plasma.

\acknowledgments
This material is based upon work supported by the National Science Foundation under grants No. PHY-1654219 and No. PHY-2116686. This work was supported in part by the National Science Foundation (NSF) within the framework of the MUSES collaboration, under grant number No. OAC-2103680, the US-DOE Nuclear Science Grant No. DE-SC0020633, U.S.-DOE Office of Science, Office of Nuclear Physics, within the framework of the Beam Energy Scan Topical (BEST) Collaboration. J.N. is partially supported by the U.S. Department of Energy, Office of Science, Office for Nuclear Physics under Award No. DE-SC0021301.

\appendix
\section{NUMERICAL PROCEDURE}
\label{sec:numerical}

The EMD black hole background fields are obtained by numerically solving the EOMs resulting from the gravitational action in Eq.\ \eqref{eq:action}, with the pair of initial conditions $(\phi_{0},\Phi_{1})$, which are, respectively, the value of the dilaton field and the value of the radial derivative of the Maxwell field both evaluated at the black hole horizon \cite{DeWolfe:2010he,Critelli:2017oub}. Each pair of values chosen for these initial conditions translates, through the numerical solutions for the EMD fields and the holographic dictionary given by Eq.\ \eqref{eq:thermodynamics}, into different thermal states of the dual gauge theory living at the boundary. In order to obtain the numerical solutions for the transport coefficients over a finite region of the QCD diagram, we use the same set of initial conditions that gave rise to the mapping into the QCD phase diagram region as reported in our previous work regarding the equilibrium state variables of this EMD model \cite{Grefa:2021qvt}. The numerical procedure used to generate the EMD backgrounds can be summarized as follows:
\begin{itemize}
    \item At $\Phi_{1}=0$, which corresponds to vanishing chemical potential, we choose the values for $\phi_{0}$ such that the mapping to the solutions in the temperature axis ranging from $T=2$ MeV to $T=550$ MeV (at $\mu_{B}=0$) is equally spaced.
    \item For each chosen value of $\phi_{0}$, $\Phi_{1}$ is varied to map the QCD phase diagram
completely up to $\mu_{B}=1100$ MeV producing lines of constant $\phi_{0}$. These lines run into the phase diagram where they eventually cross each other starting at the location of the CEP and producing a region of multiple solutions that corresponds to the coexistence region of thermodynamically stable, metastable, and unstable extrema of the free energy. This is the region where the equilibrium state variables $s$ and $\rho_B$ obtained directly from the solutions of the black hole fields exhibit a multivalued \textit{S}-shape, and the line of first-order phase transition was obtained as described in \cite{Grefa:2021qvt}. The resulting first-order line is shown in Fig.\ \ref{fig:phaseDiagramDynam}. Interestingly enough, the numerical results for the transport coefficients obtained by solving the corresponding perturbation equations on top of the EMD background black hole solutions exhibit the same characteristic \textit{S}-shape for $\mu_{B}>\mu_{B}^{c}$ near the first-order line, with the exception of the bulk viscosity $\zeta/s$ and the thermal conductivity $\mu_B^2\sigma_T/T^4$, which develop instead a loop for $\mu_{B}>\mu_{B}^{c}$ over the coexistence region. This is illustrated in Fig.\ \ref{fig:bulk_loop} for the bulk viscosity.

\begin{figure}[h!]
    \centering
    \includegraphics[width=0.49\textwidth]{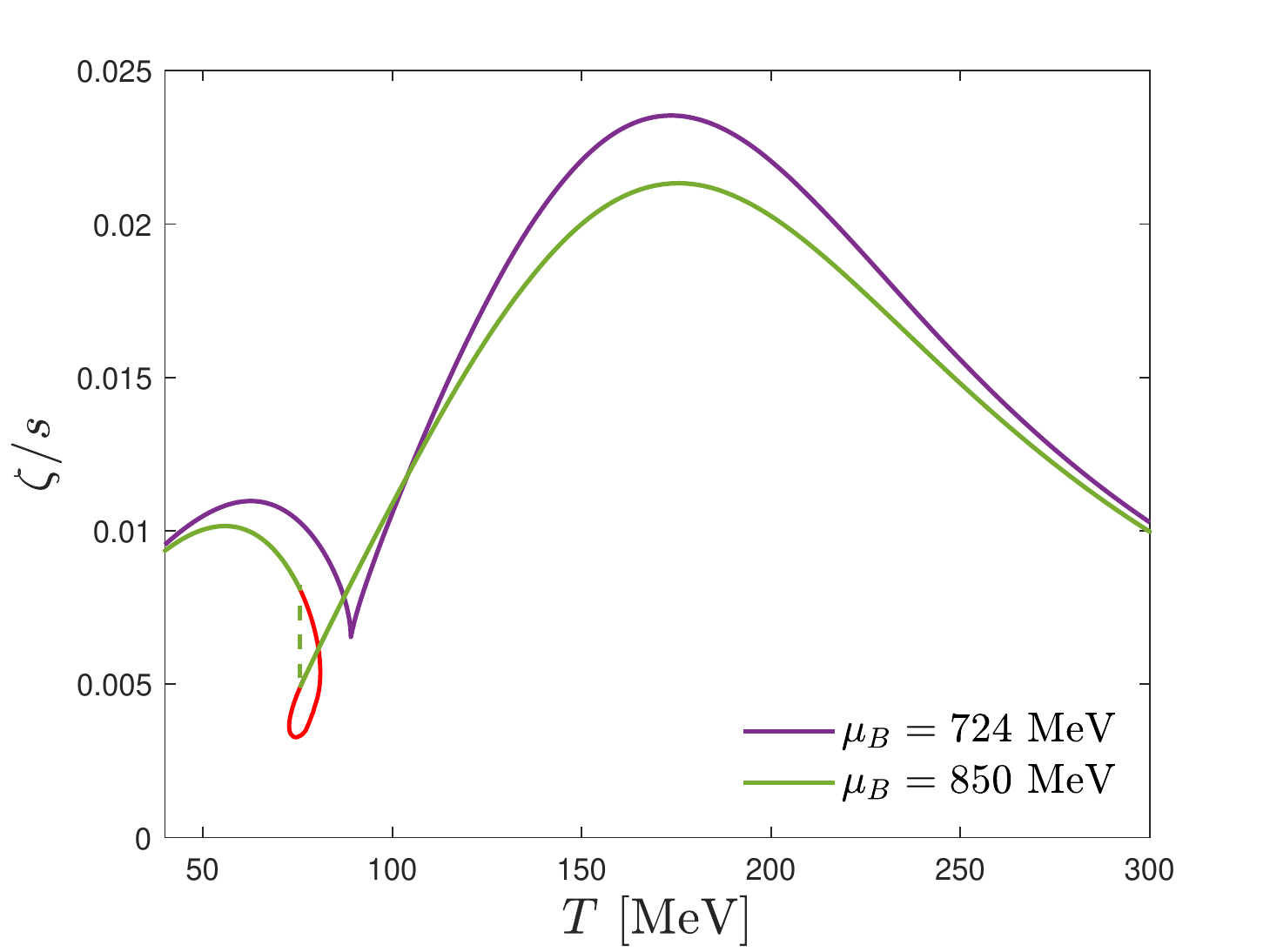}
    \caption{Numerical results from Eq.\ \eqref{eq:zeta}. Bulk viscosity over entropy density $\zeta/s$ as a function of the temperature at the critical baryon chemical potential $\mu_{B}^{c}$, and at $\mu_{B}>\mu_{B}^{c}$ before considering the line of first-order phase transition. The dashed line corresponds to the discontinuity gap when taking into account the first-order phase transition, while the line in red represents the unstable and metastable branches of solutions.}
    \label{fig:bulk_loop}
\end{figure}

\item The lines of constant $\phi_{0}$ that are mapped into the QCD phase diagram, which are the trajectories where the equilibrium state variables and transport coefficients are computed, are significantly affected by the numerical noise associated with the fitting of the ultraviolet coefficients that appear in the holographic formulas in Eq.\ \eqref{eq:thermodynamics} and the formulas regarding the dynamical variables. In particular, the coefficient $\phi_{A}$ raised to the powers $-1/\nu$ and $-3/\nu$ is the most affected by the noise. In order to obtain smooth mapping over the phase diagram a filtering process is applied. These lines of constant $\phi_{0}$ are formed by 3000 black hole solutions. This number of solutions allows us to treat the lines with a cubic smoothing
spline (CSS) filter that gets rid of big bumps, and then a Savitzky-Golay (SG) filter that preserves the shape and features of the signal. Once the lines of constant $\phi_{0}$ are fixed, they are fitted with a cubic spline to obtain lines of constant $\mu_{B}$, and lines of constant $T$.

\item The final step is to obtain the transport coefficients as single-valued functions over trajectories of constant $T$ or $\mu_{B}$. We use the information about the first-order line to differentiate the thermodynamically stable minima of the free energy from the thermodynamically metastable and unstable saddle points or maxima that are also solutions to the black hole EOMs as done for the equilibrium variables in \cite{Grefa:2021qvt}. This procedure reveals the discontinuity present in the transport coefficients that corresponds to the line of first-order phase transition, as exemplified in Fig.\ \ref{fig:bulk_loop}. For a value of $\mu_{B}=850$ MeV, the green dashed line represents the discontinuity gap in the bulk viscosity over entropy density, $\zeta/s$, when the first-order phase transition line is taken into account.
\end{itemize}

We close this section with some remarks related to the numerical calculation of the transport coefficients considered in the present work. In the EMD model, the baryon conductivity and the bulk viscosity over entropy density are obtained from the holographic Kubo formulas \eqref{eq:conduct} and \eqref{eq:zeta}, respectively, and the main features and basic consistency tests regarding their numerical calculation have been already discussed before in the text [see around and below Eqs.\ \eqref{eq:condWaveCond} and \eqref{eq:conduct}, and also around and below Eqs.\ \eqref{eq:bulkWaveCond} and \eqref{eq:zeta}].

Regarding the numerical calculation of the jet quenching parameter given by the integral in Eq.\ \eqref{eq:qhat}, due to small numerical oscillations in the value of $h(r)$ for $r\sim r_{\text{max}}$, the factor $[h_{0}^{\text{far}}-h(r)]$ may eventually evaluate to small negative values due to numerical roundoff errors, thus rendering a small spurious imaginary part for the jet quenching parameter and an inadequate oscillatory behavior which makes the numerical integration difficult to perform in a reliable way in the near-boundary region. In order to circumvent this issue, a possible approach is to cut the numerical integration at some $r_{\text{stop}}$ chosen just before the radial position where the change of sign in the factor $[h_{0}^{\text{far}}-h(r)]$ happens for a given pair of initial conditions \cite{Rougemont:2020had}. One can check that, in general, even by cutting the numerical integration at $r$ considerably less than $(r_\text{stop} - 0.02)$, our numerical results do not change by a significant amount.

\bibliography{references,NOTinspire}

\begin{thebibliography}{143}%
\makeatletter
\providecommand \@ifxundefined [1]{%
 \@ifx{#1\undefined}
}%
\providecommand \@ifnum [1]{%
 \ifnum #1\expandafter \@firstoftwo
 \else \expandafter \@secondoftwo
 \fi
}%
\providecommand \@ifx [1]{%
 \ifx #1\expandafter \@firstoftwo
 \else \expandafter \@secondoftwo
 \fi
}%
\providecommand \natexlab [1]{#1}%
\providecommand \enquote  [1]{``#1''}%
\providecommand \bibnamefont  [1]{#1}%
\providecommand \bibfnamefont [1]{#1}%
\providecommand \citenamefont [1]{#1}%
\providecommand \href@noop [0]{\@secondoftwo}%
\providecommand \href [0]{\begingroup \@sanitize@url \@href}%
\providecommand \@href[1]{\@@startlink{#1}\@@href}%
\providecommand \@@href[1]{\endgroup#1\@@endlink}%
\providecommand \@sanitize@url [0]{\catcode `\\12\catcode `\$12\catcode
  `\&12\catcode `\#12\catcode `\^12\catcode `\_12\catcode `\%12\relax}%
\providecommand \@@startlink[1]{}%
\providecommand \@@endlink[0]{}%
\providecommand \url  [0]{\begingroup\@sanitize@url \@url }%
\providecommand \@url [1]{\endgroup\@href {#1}{\urlprefix }}%
\providecommand \urlprefix  [0]{URL }%
\providecommand \Eprint [0]{\href }%
\providecommand \doibase [0]{http://dx.doi.org/}%
\providecommand \selectlanguage [0]{\@gobble}%
\providecommand \bibinfo  [0]{\@secondoftwo}%
\providecommand \bibfield  [0]{\@secondoftwo}%
\providecommand \translation [1]{[#1]}%
\providecommand \BibitemOpen [0]{}%
\providecommand \bibitemStop [0]{}%
\providecommand \bibitemNoStop [0]{.\EOS\space}%
\providecommand \EOS [0]{\spacefactor3000\relax}%
\providecommand \BibitemShut  [1]{\csname bibitem#1\endcsname}%
\let\auto@bib@innerbib\@empty
\bibitem [{\citenamefont {Adcox}\ \emph {et~al.}(2005)\citenamefont {Adcox}
  \emph {et~al.}}]{PHENIX:2004vcz}%
  \BibitemOpen
  \bibfield  {author} {\bibinfo {author} {\bibfnamefont {K.}~\bibnamefont
  {Adcox}} \emph {et~al.} (\bibinfo {collaboration} {PHENIX}),\ }\href
  {\doibase 10.1016/j.nuclphysa.2005.03.086} {\bibfield  {journal} {\bibinfo
  {journal} {Nucl. Phys. A}\ }\textbf {\bibinfo {volume} {757}},\ \bibinfo
  {pages} {184} (\bibinfo {year} {2005})},\ \Eprint
  {http://arxiv.org/abs/nucl-ex/0410003} {arXiv:nucl-ex/0410003} \BibitemShut
  {NoStop}%
\bibitem [{\citenamefont {Arsene}\ \emph {et~al.}(2005)\citenamefont {Arsene}
  \emph {et~al.}}]{BRAHMS:2004adc}%
  \BibitemOpen
  \bibfield  {author} {\bibinfo {author} {\bibfnamefont {I.}~\bibnamefont
  {Arsene}} \emph {et~al.} (\bibinfo {collaboration} {BRAHMS}),\ }\href
  {\doibase 10.1016/j.nuclphysa.2005.02.130} {\bibfield  {journal} {\bibinfo
  {journal} {Nucl. Phys. A}\ }\textbf {\bibinfo {volume} {757}},\ \bibinfo
  {pages} {1} (\bibinfo {year} {2005})},\ \Eprint
  {http://arxiv.org/abs/nucl-ex/0410020} {arXiv:nucl-ex/0410020} \BibitemShut
  {NoStop}%
\bibitem [{\citenamefont {Back}\ \emph {et~al.}(2005)\citenamefont {Back} \emph
  {et~al.}}]{PHOBOS:2004zne}%
  \BibitemOpen
  \bibfield  {author} {\bibinfo {author} {\bibfnamefont {B.~B.}\ \bibnamefont
  {Back}} \emph {et~al.} (\bibinfo {collaboration} {PHOBOS}),\ }\href {\doibase
  10.1016/j.nuclphysa.2005.03.084} {\bibfield  {journal} {\bibinfo  {journal}
  {Nucl. Phys. A}\ }\textbf {\bibinfo {volume} {757}},\ \bibinfo {pages} {28}
  (\bibinfo {year} {2005})},\ \Eprint {http://arxiv.org/abs/nucl-ex/0410022}
  {arXiv:nucl-ex/0410022} \BibitemShut {NoStop}%
\bibitem [{\citenamefont {Adams}\ \emph {et~al.}(2005)\citenamefont {Adams}
  \emph {et~al.}}]{STAR:2005gfr}%
  \BibitemOpen
  \bibfield  {author} {\bibinfo {author} {\bibfnamefont {J.}~\bibnamefont
  {Adams}} \emph {et~al.} (\bibinfo {collaboration} {STAR}),\ }\href {\doibase
  10.1016/j.nuclphysa.2005.03.085} {\bibfield  {journal} {\bibinfo  {journal}
  {Nucl. Phys. A}\ }\textbf {\bibinfo {volume} {757}},\ \bibinfo {pages} {102}
  (\bibinfo {year} {2005})},\ \Eprint {http://arxiv.org/abs/nucl-ex/0501009}
  {arXiv:nucl-ex/0501009} \BibitemShut {NoStop}%
\bibitem [{\citenamefont {Aad}\ \emph {et~al.}(2013)\citenamefont {Aad} \emph
  {et~al.}}]{ATLAS:2013xzf}%
  \BibitemOpen
  \bibfield  {author} {\bibinfo {author} {\bibfnamefont {G.}~\bibnamefont
  {Aad}} \emph {et~al.} (\bibinfo {collaboration} {ATLAS}),\ }\href {\doibase
  10.1007/JHEP11(2013)183} {\bibfield  {journal} {\bibinfo  {journal} {JHEP}\
  }\textbf {\bibinfo {volume} {11}},\ \bibinfo {pages} {183} (\bibinfo {year}
  {2013})},\ \Eprint {http://arxiv.org/abs/1305.2942} {arXiv:1305.2942
  [hep-ex]} \BibitemShut {NoStop}%
\bibitem [{\citenamefont {Cebra}\ \emph {et~al.}(2014)\citenamefont {Cebra},
  \citenamefont {Brovko}, \citenamefont {Flores}, \citenamefont {Haag},\ and\
  \citenamefont {Klay}}]{Cebra:2014sxa}%
  \BibitemOpen
  \bibfield  {author} {\bibinfo {author} {\bibfnamefont {D.}~\bibnamefont
  {Cebra}}, \bibinfo {author} {\bibfnamefont {S.~G.}\ \bibnamefont {Brovko}},
  \bibinfo {author} {\bibfnamefont {C.~E.}\ \bibnamefont {Flores}}, \bibinfo
  {author} {\bibfnamefont {B.~A.}\ \bibnamefont {Haag}}, \ and\ \bibinfo
  {author} {\bibfnamefont {J.~L.}\ \bibnamefont {Klay}},\ }\href@noop {} {\
  (\bibinfo {year} {2014})},\ \Eprint {http://arxiv.org/abs/1408.1369}
  {arXiv:1408.1369 [nucl-ex]} \BibitemShut {NoStop}%
\bibitem [{\citenamefont {Adamczewski-Musch}\ \emph {et~al.}(2019)\citenamefont
  {Adamczewski-Musch} \emph {et~al.}}]{HADES:2019auv}%
  \BibitemOpen
  \bibfield  {author} {\bibinfo {author} {\bibfnamefont {J.}~\bibnamefont
  {Adamczewski-Musch}} \emph {et~al.} (\bibinfo {collaboration} {HADES}),\
  }\href {\doibase 10.1038/s41567-019-0583-8} {\bibfield  {journal} {\bibinfo
  {journal} {Nature Phys.}\ }\textbf {\bibinfo {volume} {15}},\ \bibinfo
  {pages} {1040} (\bibinfo {year} {2019})}\BibitemShut {NoStop}%
\bibitem [{\citenamefont {Friese}(2006)}]{Friese:2006dj}%
  \BibitemOpen
  \bibfield  {author} {\bibinfo {author} {\bibfnamefont {V.}~\bibnamefont
  {Friese}},\ }\href {\doibase 10.1016/j.nuclphysa.2006.06.018} {\bibfield
  {journal} {\bibinfo  {journal} {Nucl. Phys. A}\ }\textbf {\bibinfo {volume}
  {774}},\ \bibinfo {pages} {377} (\bibinfo {year} {2006})}\BibitemShut
  {NoStop}%
\bibitem [{\citenamefont {Tahir}\ \emph {et~al.}(2005)\citenamefont {Tahir}
  \emph {et~al.}}]{Tahir:2005zz}%
  \BibitemOpen
  \bibfield  {author} {\bibinfo {author} {\bibfnamefont {N.~A.}\ \bibnamefont
  {Tahir}} \emph {et~al.},\ }\href {\doibase 10.1103/PhysRevLett.95.035001}
  {\bibfield  {journal} {\bibinfo  {journal} {Phys. Rev. Lett.}\ }\textbf
  {\bibinfo {volume} {95}},\ \bibinfo {pages} {035001} (\bibinfo {year}
  {2005})}\BibitemShut {NoStop}%
\bibitem [{\citenamefont {Lutz}\ \emph {et~al.}(2009)\citenamefont {Lutz} \emph
  {et~al.}}]{Lutz:2009ff}%
  \BibitemOpen
  \bibfield  {author} {\bibinfo {author} {\bibfnamefont {M.~F.~M.}\
  \bibnamefont {Lutz}} \emph {et~al.} (\bibinfo {collaboration} {PANDA}),\
  }\href@noop {} {\  (\bibinfo {year} {2009})},\ \Eprint
  {http://arxiv.org/abs/0903.3905} {arXiv:0903.3905 [hep-ex]} \BibitemShut
  {NoStop}%
\bibitem [{\citenamefont {Durante}\ \emph {et~al.}(2019)\citenamefont {Durante}
  \emph {et~al.}}]{Durante:2019hzd}%
  \BibitemOpen
  \bibfield  {author} {\bibinfo {author} {\bibfnamefont {M.}~\bibnamefont
  {Durante}} \emph {et~al.},\ }\href {\doibase 10.1088/1402-4896/aaf93f}
  {\bibfield  {journal} {\bibinfo  {journal} {Phys. Scripta}\ }\textbf
  {\bibinfo {volume} {94}},\ \bibinfo {pages} {033001} (\bibinfo {year}
  {2019})},\ \Eprint {http://arxiv.org/abs/1903.05693} {arXiv:1903.05693
  [nucl-th]} \BibitemShut {NoStop}%
\bibitem [{\citenamefont {Kekelidze}\ \emph {et~al.}(2017)\citenamefont
  {Kekelidze}, \citenamefont {Kovalenko}, \citenamefont {Lednicky},
  \citenamefont {Matveev}, \citenamefont {Meshkov}, \citenamefont {Sorin},\
  and\ \citenamefont {Trubnikov}}]{Kekelidze:2017tgp}%
  \BibitemOpen
  \bibfield  {author} {\bibinfo {author} {\bibfnamefont {V.}~\bibnamefont
  {Kekelidze}}, \bibinfo {author} {\bibfnamefont {A.}~\bibnamefont
  {Kovalenko}}, \bibinfo {author} {\bibfnamefont {R.}~\bibnamefont {Lednicky}},
  \bibinfo {author} {\bibfnamefont {V.}~\bibnamefont {Matveev}}, \bibinfo
  {author} {\bibfnamefont {I.}~\bibnamefont {Meshkov}}, \bibinfo {author}
  {\bibfnamefont {A.}~\bibnamefont {Sorin}}, \ and\ \bibinfo {author}
  {\bibfnamefont {G.}~\bibnamefont {Trubnikov}},\ }\href {\doibase
  10.1016/j.nuclphysa.2017.06.031} {\bibfield  {journal} {\bibinfo  {journal}
  {Nucl. Phys. A}\ }\textbf {\bibinfo {volume} {967}},\ \bibinfo {pages} {884}
  (\bibinfo {year} {2017})}\BibitemShut {NoStop}%
\bibitem [{\citenamefont {Kekelidze}\ \emph {et~al.}(2016)\citenamefont
  {Kekelidze}, \citenamefont {Kovalenko}, \citenamefont {Lednicky},
  \citenamefont {Matveev}, \citenamefont {Meshkov}, \citenamefont {Sorin},\
  and\ \citenamefont {Trubnikov}}]{Kekelidze:2016wkp}%
  \BibitemOpen
  \bibfield  {author} {\bibinfo {author} {\bibfnamefont {V.}~\bibnamefont
  {Kekelidze}}, \bibinfo {author} {\bibfnamefont {A.}~\bibnamefont
  {Kovalenko}}, \bibinfo {author} {\bibfnamefont {R.}~\bibnamefont {Lednicky}},
  \bibinfo {author} {\bibfnamefont {V.}~\bibnamefont {Matveev}}, \bibinfo
  {author} {\bibfnamefont {I.}~\bibnamefont {Meshkov}}, \bibinfo {author}
  {\bibfnamefont {A.}~\bibnamefont {Sorin}}, \ and\ \bibinfo {author}
  {\bibfnamefont {G.}~\bibnamefont {Trubnikov}},\ }\href {\doibase
  10.1016/j.nuclphysa.2016.03.019} {\bibfield  {journal} {\bibinfo  {journal}
  {Nucl. Phys. A}\ }\textbf {\bibinfo {volume} {956}},\ \bibinfo {pages} {846}
  (\bibinfo {year} {2016})}\BibitemShut {NoStop}%
\bibitem [{\citenamefont {Dexheimer}\ \emph {et~al.}(2021)\citenamefont
  {Dexheimer}, \citenamefont {Noronha}, \citenamefont {Noronha-Hostler},
  \citenamefont {Ratti},\ and\ \citenamefont {Yunes}}]{Dexheimer:2020zzs}%
  \BibitemOpen
  \bibfield  {author} {\bibinfo {author} {\bibfnamefont {V.}~\bibnamefont
  {Dexheimer}}, \bibinfo {author} {\bibfnamefont {J.}~\bibnamefont {Noronha}},
  \bibinfo {author} {\bibfnamefont {J.}~\bibnamefont {Noronha-Hostler}},
  \bibinfo {author} {\bibfnamefont {C.}~\bibnamefont {Ratti}}, \ and\ \bibinfo
  {author} {\bibfnamefont {N.}~\bibnamefont {Yunes}},\ }\href {\doibase
  10.1088/1361-6471/abe104} {\bibfield  {journal} {\bibinfo  {journal} {J.
  Phys. G}\ }\textbf {\bibinfo {volume} {48}},\ \bibinfo {pages} {073001}
  (\bibinfo {year} {2021})},\ \Eprint {http://arxiv.org/abs/2010.08834}
  {arXiv:2010.08834 [nucl-th]} \BibitemShut {NoStop}%
\bibitem [{\citenamefont {Heinz}\ and\ \citenamefont
  {Snellings}(2013)}]{Heinz:2013th}%
  \BibitemOpen
  \bibfield  {author} {\bibinfo {author} {\bibfnamefont {U.}~\bibnamefont
  {Heinz}}\ and\ \bibinfo {author} {\bibfnamefont {R.}~\bibnamefont
  {Snellings}},\ }\href {\doibase 10.1146/annurev-nucl-102212-170540}
  {\bibfield  {journal} {\bibinfo  {journal} {Ann. Rev. Nucl. Part. Sci.}\
  }\textbf {\bibinfo {volume} {63}},\ \bibinfo {pages} {123} (\bibinfo {year}
  {2013})},\ \Eprint {http://arxiv.org/abs/1301.2826} {arXiv:1301.2826
  [nucl-th]} \BibitemShut {NoStop}%
\bibitem [{\citenamefont {Luzum}\ and\ \citenamefont
  {Petersen}(2014)}]{Luzum:2013yya}%
  \BibitemOpen
  \bibfield  {author} {\bibinfo {author} {\bibfnamefont {M.}~\bibnamefont
  {Luzum}}\ and\ \bibinfo {author} {\bibfnamefont {H.}~\bibnamefont
  {Petersen}},\ }\href {\doibase 10.1088/0954-3899/41/6/063102} {\bibfield
  {journal} {\bibinfo  {journal} {J. Phys. G}\ }\textbf {\bibinfo {volume}
  {41}},\ \bibinfo {pages} {063102} (\bibinfo {year} {2014})},\ \Eprint
  {http://arxiv.org/abs/1312.5503} {arXiv:1312.5503 [nucl-th]} \BibitemShut
  {NoStop}%
\bibitem [{\citenamefont {Meyer}(2011)}]{Meyer:2011gj}%
  \BibitemOpen
  \bibfield  {author} {\bibinfo {author} {\bibfnamefont {H.~B.}\ \bibnamefont
  {Meyer}},\ }\href {\doibase 10.1140/epja/i2011-11086-3} {\bibfield  {journal}
  {\bibinfo  {journal} {Eur. Phys. J. A}\ }\textbf {\bibinfo {volume} {47}},\
  \bibinfo {pages} {86} (\bibinfo {year} {2011})},\ \Eprint
  {http://arxiv.org/abs/1104.3708} {arXiv:1104.3708 [hep-lat]} \BibitemShut
  {NoStop}%
\bibitem [{\citenamefont {Ding}\ \emph {et~al.}(2015)\citenamefont {Ding},
  \citenamefont {Karsch},\ and\ \citenamefont {Mukherjee}}]{Ding:2015ona}%
  \BibitemOpen
  \bibfield  {author} {\bibinfo {author} {\bibfnamefont {H.-T.}\ \bibnamefont
  {Ding}}, \bibinfo {author} {\bibfnamefont {F.}~\bibnamefont {Karsch}}, \ and\
  \bibinfo {author} {\bibfnamefont {S.}~\bibnamefont {Mukherjee}},\ }\href
  {\doibase 10.1142/S0218301315300076} {\bibfield  {journal} {\bibinfo
  {journal} {Int. J. Mod. Phys. E}\ }\textbf {\bibinfo {volume} {24}},\
  \bibinfo {pages} {1530007} (\bibinfo {year} {2015})},\ \Eprint
  {http://arxiv.org/abs/1504.05274} {arXiv:1504.05274 [hep-lat]} \BibitemShut
  {NoStop}%
\bibitem [{\citenamefont {Ratti}(2018)}]{Ratti:2018ksb}%
  \BibitemOpen
  \bibfield  {author} {\bibinfo {author} {\bibfnamefont {C.}~\bibnamefont
  {Ratti}},\ }\href {\doibase 10.1088/1361-6633/aabb97} {\bibfield  {journal}
  {\bibinfo  {journal} {Rept. Prog. Phys.}\ }\textbf {\bibinfo {volume} {81}},\
  \bibinfo {pages} {084301} (\bibinfo {year} {2018})},\ \Eprint
  {http://arxiv.org/abs/1804.07810} {arXiv:1804.07810 [hep-lat]} \BibitemShut
  {NoStop}%
\bibitem [{\citenamefont {Rothkopf}(2020)}]{Rothkopf:2019ipj}%
  \BibitemOpen
  \bibfield  {author} {\bibinfo {author} {\bibfnamefont {A.}~\bibnamefont
  {Rothkopf}},\ }\href {\doibase 10.1016/j.physrep.2020.02.006} {\bibfield
  {journal} {\bibinfo  {journal} {Phys. Rept.}\ }\textbf {\bibinfo {volume}
  {858}},\ \bibinfo {pages} {1} (\bibinfo {year} {2020})},\ \Eprint
  {http://arxiv.org/abs/1912.02253} {arXiv:1912.02253 [hep-ph]} \BibitemShut
  {NoStop}%
\bibitem [{\citenamefont {Auvinen}\ \emph {et~al.}(2018)\citenamefont
  {Auvinen}, \citenamefont {Bernhard}, \citenamefont {Bass},\ and\
  \citenamefont {Karpenko}}]{Auvinen:2017fjw}%
  \BibitemOpen
  \bibfield  {author} {\bibinfo {author} {\bibfnamefont {J.}~\bibnamefont
  {Auvinen}}, \bibinfo {author} {\bibfnamefont {J.~E.}\ \bibnamefont
  {Bernhard}}, \bibinfo {author} {\bibfnamefont {S.~A.}\ \bibnamefont {Bass}},
  \ and\ \bibinfo {author} {\bibfnamefont {I.}~\bibnamefont {Karpenko}},\
  }\href {\doibase 10.1103/PhysRevC.97.044905} {\bibfield  {journal} {\bibinfo
  {journal} {Phys. Rev. C}\ }\textbf {\bibinfo {volume} {97}},\ \bibinfo
  {pages} {044905} (\bibinfo {year} {2018})},\ \Eprint
  {http://arxiv.org/abs/1706.03666} {arXiv:1706.03666 [hep-ph]} \BibitemShut
  {NoStop}%
\bibitem [{\citenamefont {Dore}\ \emph {et~al.}(2020)\citenamefont {Dore},
  \citenamefont {Noronha-Hostler},\ and\ \citenamefont
  {McLaughlin}}]{Dore:2020jye}%
  \BibitemOpen
  \bibfield  {author} {\bibinfo {author} {\bibfnamefont {T.}~\bibnamefont
  {Dore}}, \bibinfo {author} {\bibfnamefont {J.}~\bibnamefont
  {Noronha-Hostler}}, \ and\ \bibinfo {author} {\bibfnamefont {E.}~\bibnamefont
  {McLaughlin}},\ }\href {\doibase 10.1103/PhysRevD.102.074017} {\bibfield
  {journal} {\bibinfo  {journal} {Phys. Rev. D}\ }\textbf {\bibinfo {volume}
  {102}},\ \bibinfo {pages} {074017} (\bibinfo {year} {2020})},\ \Eprint
  {http://arxiv.org/abs/2007.15083} {arXiv:2007.15083 [nucl-th]} \BibitemShut
  {NoStop}%
\bibitem [{\citenamefont {Sch\"afer}\ \emph {et~al.}(2021)\citenamefont
  {Sch\"afer}, \citenamefont {Karpenko}, \citenamefont {Wu}, \citenamefont
  {Hammelmann},\ and\ \citenamefont {Elfner}}]{Schafer:2021csj}%
  \BibitemOpen
  \bibfield  {author} {\bibinfo {author} {\bibfnamefont {A.}~\bibnamefont
  {Sch\"afer}}, \bibinfo {author} {\bibfnamefont {I.}~\bibnamefont {Karpenko}},
  \bibinfo {author} {\bibfnamefont {X.-Y.}\ \bibnamefont {Wu}}, \bibinfo
  {author} {\bibfnamefont {J.}~\bibnamefont {Hammelmann}}, \ and\ \bibinfo
  {author} {\bibfnamefont {H.}~\bibnamefont {Elfner}},\ }\href@noop {} {\
  (\bibinfo {year} {2021})},\ \Eprint {http://arxiv.org/abs/2112.08724}
  {arXiv:2112.08724 [hep-ph]} \BibitemShut {NoStop}%
\bibitem [{\citenamefont {Denicol}\ \emph {et~al.}(2012)\citenamefont
  {Denicol}, \citenamefont {Niemi}, \citenamefont {Molnar},\ and\ \citenamefont
  {Rischke}}]{Denicol:2012cn}%
  \BibitemOpen
  \bibfield  {author} {\bibinfo {author} {\bibfnamefont {G.~S.}\ \bibnamefont
  {Denicol}}, \bibinfo {author} {\bibfnamefont {H.}~\bibnamefont {Niemi}},
  \bibinfo {author} {\bibfnamefont {E.}~\bibnamefont {Molnar}}, \ and\ \bibinfo
  {author} {\bibfnamefont {D.~H.}\ \bibnamefont {Rischke}},\ }\href {\doibase
  10.1103/PhysRevD.85.114047} {\bibfield  {journal} {\bibinfo  {journal} {Phys.
  Rev. D}\ }\textbf {\bibinfo {volume} {85}},\ \bibinfo {pages} {114047}
  (\bibinfo {year} {2012})},\ \bibinfo {note} {[Erratum: Phys.Rev.D 91, 039902
  (2015)]},\ \Eprint {http://arxiv.org/abs/1202.4551} {arXiv:1202.4551
  [nucl-th]} \BibitemShut {NoStop}%
\bibitem [{\citenamefont {Denicol}\ \emph {et~al.}(2018)\citenamefont
  {Denicol}, \citenamefont {Gale}, \citenamefont {Jeon}, \citenamefont
  {Monnai}, \citenamefont {Schenke},\ and\ \citenamefont
  {Shen}}]{Denicol:2018wdp}%
  \BibitemOpen
  \bibfield  {author} {\bibinfo {author} {\bibfnamefont {G.~S.}\ \bibnamefont
  {Denicol}}, \bibinfo {author} {\bibfnamefont {C.}~\bibnamefont {Gale}},
  \bibinfo {author} {\bibfnamefont {S.}~\bibnamefont {Jeon}}, \bibinfo {author}
  {\bibfnamefont {A.}~\bibnamefont {Monnai}}, \bibinfo {author} {\bibfnamefont
  {B.}~\bibnamefont {Schenke}}, \ and\ \bibinfo {author} {\bibfnamefont
  {C.}~\bibnamefont {Shen}},\ }\href {\doibase 10.1103/PhysRevC.98.034916}
  {\bibfield  {journal} {\bibinfo  {journal} {Phys. Rev. C}\ }\textbf {\bibinfo
  {volume} {98}},\ \bibinfo {pages} {034916} (\bibinfo {year} {2018})},\
  \Eprint {http://arxiv.org/abs/1804.10557} {arXiv:1804.10557 [nucl-th]}
  \BibitemShut {NoStop}%
\bibitem [{\citenamefont {An}\ \emph {et~al.}(2022)\citenamefont {An} \emph
  {et~al.}}]{An:2021wof}%
  \BibitemOpen
  \bibfield  {author} {\bibinfo {author} {\bibfnamefont {X.}~\bibnamefont {An}}
  \emph {et~al.},\ }\href {\doibase 10.1016/j.nuclphysa.2021.122343} {\bibfield
   {journal} {\bibinfo  {journal} {Nucl. Phys. A}\ }\textbf {\bibinfo {volume}
  {1017}},\ \bibinfo {pages} {122343} (\bibinfo {year} {2022})},\ \Eprint
  {http://arxiv.org/abs/2108.13867} {arXiv:2108.13867 [nucl-th]} \BibitemShut
  {NoStop}%
\bibitem [{\citenamefont {Du}\ \emph {et~al.}(2021)\citenamefont {Du},
  \citenamefont {An},\ and\ \citenamefont {Heinz}}]{Du:2021zqz}%
  \BibitemOpen
  \bibfield  {author} {\bibinfo {author} {\bibfnamefont {L.}~\bibnamefont
  {Du}}, \bibinfo {author} {\bibfnamefont {X.}~\bibnamefont {An}}, \ and\
  \bibinfo {author} {\bibfnamefont {U.}~\bibnamefont {Heinz}},\ }\href
  {\doibase 10.1103/PhysRevC.104.064904} {\bibfield  {journal} {\bibinfo
  {journal} {Phys. Rev. C}\ }\textbf {\bibinfo {volume} {104}},\ \bibinfo
  {pages} {064904} (\bibinfo {year} {2021})},\ \Eprint
  {http://arxiv.org/abs/2107.02302} {arXiv:2107.02302 [hep-ph]} \BibitemShut
  {NoStop}%
\bibitem [{\citenamefont {Moore}\ and\ \citenamefont
  {Teaney}(2005)}]{Moore:2004tg}%
  \BibitemOpen
  \bibfield  {author} {\bibinfo {author} {\bibfnamefont {G.~D.}\ \bibnamefont
  {Moore}}\ and\ \bibinfo {author} {\bibfnamefont {D.}~\bibnamefont {Teaney}},\
  }\href {\doibase 10.1103/PhysRevC.71.064904} {\bibfield  {journal} {\bibinfo
  {journal} {Phys. Rev. C}\ }\textbf {\bibinfo {volume} {71}},\ \bibinfo
  {pages} {064904} (\bibinfo {year} {2005})},\ \Eprint
  {http://arxiv.org/abs/hep-ph/0412346} {arXiv:hep-ph/0412346} \BibitemShut
  {NoStop}%
\bibitem [{\citenamefont {Baier}\ \emph {et~al.}(1997)\citenamefont {Baier},
  \citenamefont {Dokshitzer}, \citenamefont {Mueller}, \citenamefont {Peigne},\
  and\ \citenamefont {Schiff}}]{Baier:1996sk}%
  \BibitemOpen
  \bibfield  {author} {\bibinfo {author} {\bibfnamefont {R.}~\bibnamefont
  {Baier}}, \bibinfo {author} {\bibfnamefont {Y.~L.}\ \bibnamefont
  {Dokshitzer}}, \bibinfo {author} {\bibfnamefont {A.~H.}\ \bibnamefont
  {Mueller}}, \bibinfo {author} {\bibfnamefont {S.}~\bibnamefont {Peigne}}, \
  and\ \bibinfo {author} {\bibfnamefont {D.}~\bibnamefont {Schiff}},\ }\href
  {\doibase 10.1016/S0550-3213(96)00581-0} {\bibfield  {journal} {\bibinfo
  {journal} {Nucl. Phys. B}\ }\textbf {\bibinfo {volume} {484}},\ \bibinfo
  {pages} {265} (\bibinfo {year} {1997})},\ \Eprint
  {http://arxiv.org/abs/hep-ph/9608322} {arXiv:hep-ph/9608322} \BibitemShut
  {NoStop}%
\bibitem [{\citenamefont {Maldacena}(1998)}]{Maldacena:1997re}%
  \BibitemOpen
  \bibfield  {author} {\bibinfo {author} {\bibfnamefont {J.~M.}\ \bibnamefont
  {Maldacena}},\ }\href {\doibase 10.1023/A:1026654312961} {\bibfield
  {journal} {\bibinfo  {journal} {Adv. Theor. Math. Phys.}\ }\textbf {\bibinfo
  {volume} {2}},\ \bibinfo {pages} {231} (\bibinfo {year} {1998})},\ \Eprint
  {http://arxiv.org/abs/hep-th/9711200} {arXiv:hep-th/9711200} \BibitemShut
  {NoStop}%
\bibitem [{\citenamefont {Gubser}\ \emph {et~al.}(1998)\citenamefont {Gubser},
  \citenamefont {Klebanov},\ and\ \citenamefont {Polyakov}}]{Gubser:1998bc}%
  \BibitemOpen
  \bibfield  {author} {\bibinfo {author} {\bibfnamefont {S.~S.}\ \bibnamefont
  {Gubser}}, \bibinfo {author} {\bibfnamefont {I.~R.}\ \bibnamefont
  {Klebanov}}, \ and\ \bibinfo {author} {\bibfnamefont {A.~M.}\ \bibnamefont
  {Polyakov}},\ }\href {\doibase 10.1016/S0370-2693(98)00377-3} {\bibfield
  {journal} {\bibinfo  {journal} {Phys. Lett. B}\ }\textbf {\bibinfo {volume}
  {428}},\ \bibinfo {pages} {105} (\bibinfo {year} {1998})},\ \Eprint
  {http://arxiv.org/abs/hep-th/9802109} {arXiv:hep-th/9802109} \BibitemShut
  {NoStop}%
\bibitem [{\citenamefont {Witten}(1998{\natexlab{a}})}]{Witten:1998qj}%
  \BibitemOpen
  \bibfield  {author} {\bibinfo {author} {\bibfnamefont {E.}~\bibnamefont
  {Witten}},\ }\href {\doibase 10.4310/ATMP.1998.v2.n2.a2} {\bibfield
  {journal} {\bibinfo  {journal} {Adv. Theor. Math. Phys.}\ }\textbf {\bibinfo
  {volume} {2}},\ \bibinfo {pages} {253} (\bibinfo {year}
  {1998}{\natexlab{a}})},\ \Eprint {http://arxiv.org/abs/hep-th/9802150}
  {arXiv:hep-th/9802150} \BibitemShut {NoStop}%
\bibitem [{\citenamefont {Witten}(1998{\natexlab{b}})}]{Witten:1998zw}%
  \BibitemOpen
  \bibfield  {author} {\bibinfo {author} {\bibfnamefont {E.}~\bibnamefont
  {Witten}},\ }\href {\doibase 10.4310/ATMP.1998.v2.n3.a3} {\bibfield
  {journal} {\bibinfo  {journal} {Adv. Theor. Math. Phys.}\ }\textbf {\bibinfo
  {volume} {2}},\ \bibinfo {pages} {505} (\bibinfo {year}
  {1998}{\natexlab{b}})},\ \Eprint {http://arxiv.org/abs/hep-th/9803131}
  {arXiv:hep-th/9803131} \BibitemShut {NoStop}%
\bibitem [{\citenamefont {Critelli}\ \emph {et~al.}(2017)\citenamefont
  {Critelli}, \citenamefont {Noronha}, \citenamefont {Noronha-Hostler},
  \citenamefont {Portillo}, \citenamefont {Ratti},\ and\ \citenamefont
  {Rougemont}}]{Critelli:2017oub}%
  \BibitemOpen
  \bibfield  {author} {\bibinfo {author} {\bibfnamefont {R.}~\bibnamefont
  {Critelli}}, \bibinfo {author} {\bibfnamefont {J.}~\bibnamefont {Noronha}},
  \bibinfo {author} {\bibfnamefont {J.}~\bibnamefont {Noronha-Hostler}},
  \bibinfo {author} {\bibfnamefont {I.}~\bibnamefont {Portillo}}, \bibinfo
  {author} {\bibfnamefont {C.}~\bibnamefont {Ratti}}, \ and\ \bibinfo {author}
  {\bibfnamefont {R.}~\bibnamefont {Rougemont}},\ }\href {\doibase
  10.1103/PhysRevD.96.096026} {\bibfield  {journal} {\bibinfo  {journal} {Phys.
  Rev. D}\ }\textbf {\bibinfo {volume} {96}},\ \bibinfo {pages} {096026}
  (\bibinfo {year} {2017})},\ \Eprint {http://arxiv.org/abs/1706.00455}
  {arXiv:1706.00455 [nucl-th]} \BibitemShut {NoStop}%
\bibitem [{\citenamefont {Gubser}\ \emph {et~al.}(2008)\citenamefont {Gubser},
  \citenamefont {Nellore}, \citenamefont {Pufu},\ and\ \citenamefont
  {Rocha}}]{Gubser:2008yx}%
  \BibitemOpen
  \bibfield  {author} {\bibinfo {author} {\bibfnamefont {S.~S.}\ \bibnamefont
  {Gubser}}, \bibinfo {author} {\bibfnamefont {A.}~\bibnamefont {Nellore}},
  \bibinfo {author} {\bibfnamefont {S.~S.}\ \bibnamefont {Pufu}}, \ and\
  \bibinfo {author} {\bibfnamefont {F.~D.}\ \bibnamefont {Rocha}},\ }\href
  {\doibase 10.1103/PhysRevLett.101.131601} {\bibfield  {journal} {\bibinfo
  {journal} {Phys. Rev. Lett.}\ }\textbf {\bibinfo {volume} {101}},\ \bibinfo
  {pages} {131601} (\bibinfo {year} {2008})},\ \Eprint
  {http://arxiv.org/abs/0804.1950} {arXiv:0804.1950 [hep-th]} \BibitemShut
  {NoStop}%
\bibitem [{\citenamefont {DeWolfe}\ \emph
  {et~al.}(2011{\natexlab{a}})\citenamefont {DeWolfe}, \citenamefont {Gubser},\
  and\ \citenamefont {Rosen}}]{DeWolfe:2010he}%
  \BibitemOpen
  \bibfield  {author} {\bibinfo {author} {\bibfnamefont {O.}~\bibnamefont
  {DeWolfe}}, \bibinfo {author} {\bibfnamefont {S.~S.}\ \bibnamefont {Gubser}},
  \ and\ \bibinfo {author} {\bibfnamefont {C.}~\bibnamefont {Rosen}},\ }\href
  {\doibase 10.1103/PhysRevD.83.086005} {\bibfield  {journal} {\bibinfo
  {journal} {Phys. Rev. D}\ }\textbf {\bibinfo {volume} {83}},\ \bibinfo
  {pages} {086005} (\bibinfo {year} {2011}{\natexlab{a}})},\ \Eprint
  {http://arxiv.org/abs/1012.1864} {arXiv:1012.1864 [hep-th]} \BibitemShut
  {NoStop}%
\bibitem [{\citenamefont {DeWolfe}\ \emph
  {et~al.}(2011{\natexlab{b}})\citenamefont {DeWolfe}, \citenamefont {Gubser},\
  and\ \citenamefont {Rosen}}]{DeWolfe:2011ts}%
  \BibitemOpen
  \bibfield  {author} {\bibinfo {author} {\bibfnamefont {O.}~\bibnamefont
  {DeWolfe}}, \bibinfo {author} {\bibfnamefont {S.~S.}\ \bibnamefont {Gubser}},
  \ and\ \bibinfo {author} {\bibfnamefont {C.}~\bibnamefont {Rosen}},\ }\href
  {\doibase 10.1103/PhysRevD.84.126014} {\bibfield  {journal} {\bibinfo
  {journal} {Phys. Rev. D}\ }\textbf {\bibinfo {volume} {84}},\ \bibinfo
  {pages} {126014} (\bibinfo {year} {2011}{\natexlab{b}})},\ \Eprint
  {http://arxiv.org/abs/1108.2029} {arXiv:1108.2029 [hep-th]} \BibitemShut
  {NoStop}%
\bibitem [{\citenamefont {Borsanyi}\ \emph {et~al.}(2010)\citenamefont
  {Borsanyi}, \citenamefont {Endrodi}, \citenamefont {Fodor}, \citenamefont
  {Jakovac}, \citenamefont {Katz}, \citenamefont {Krieg}, \citenamefont
  {Ratti},\ and\ \citenamefont {Szabo}}]{Borsanyi:2010cj}%
  \BibitemOpen
  \bibfield  {author} {\bibinfo {author} {\bibfnamefont {S.}~\bibnamefont
  {Borsanyi}}, \bibinfo {author} {\bibfnamefont {G.}~\bibnamefont {Endrodi}},
  \bibinfo {author} {\bibfnamefont {Z.}~\bibnamefont {Fodor}}, \bibinfo
  {author} {\bibfnamefont {A.}~\bibnamefont {Jakovac}}, \bibinfo {author}
  {\bibfnamefont {S.~D.}\ \bibnamefont {Katz}}, \bibinfo {author}
  {\bibfnamefont {S.}~\bibnamefont {Krieg}}, \bibinfo {author} {\bibfnamefont
  {C.}~\bibnamefont {Ratti}}, \ and\ \bibinfo {author} {\bibfnamefont {K.~K.}\
  \bibnamefont {Szabo}},\ }\href {\doibase 10.1007/JHEP11(2010)077} {\bibfield
  {journal} {\bibinfo  {journal} {JHEP}\ }\textbf {\bibinfo {volume} {11}},\
  \bibinfo {pages} {077} (\bibinfo {year} {2010})},\ \Eprint
  {http://arxiv.org/abs/1007.2580} {arXiv:1007.2580 [hep-lat]} \BibitemShut
  {NoStop}%
\bibitem [{\citenamefont {Borsanyi}\ \emph {et~al.}(2014)\citenamefont
  {Borsanyi}, \citenamefont {Fodor}, \citenamefont {Hoelbling}, \citenamefont
  {Katz}, \citenamefont {Krieg},\ and\ \citenamefont
  {Szabo}}]{Borsanyi:2013bia}%
  \BibitemOpen
  \bibfield  {author} {\bibinfo {author} {\bibfnamefont {S.}~\bibnamefont
  {Borsanyi}}, \bibinfo {author} {\bibfnamefont {Z.}~\bibnamefont {Fodor}},
  \bibinfo {author} {\bibfnamefont {C.}~\bibnamefont {Hoelbling}}, \bibinfo
  {author} {\bibfnamefont {S.~D.}\ \bibnamefont {Katz}}, \bibinfo {author}
  {\bibfnamefont {S.}~\bibnamefont {Krieg}}, \ and\ \bibinfo {author}
  {\bibfnamefont {K.~K.}\ \bibnamefont {Szabo}},\ }\href {\doibase
  10.1016/j.physletb.2014.01.007} {\bibfield  {journal} {\bibinfo  {journal}
  {Phys. Lett. B}\ }\textbf {\bibinfo {volume} {730}},\ \bibinfo {pages} {99}
  (\bibinfo {year} {2014})},\ \Eprint {http://arxiv.org/abs/1309.5258}
  {arXiv:1309.5258 [hep-lat]} \BibitemShut {NoStop}%
\bibitem [{\citenamefont {Bellwied}\ \emph {et~al.}(2015)\citenamefont
  {Bellwied}, \citenamefont {Borsanyi}, \citenamefont {Fodor}, \citenamefont
  {Katz}, \citenamefont {Pasztor}, \citenamefont {Ratti},\ and\ \citenamefont
  {Szabo}}]{Bellwied:2015lba}%
  \BibitemOpen
  \bibfield  {author} {\bibinfo {author} {\bibfnamefont {R.}~\bibnamefont
  {Bellwied}}, \bibinfo {author} {\bibfnamefont {S.}~\bibnamefont {Borsanyi}},
  \bibinfo {author} {\bibfnamefont {Z.}~\bibnamefont {Fodor}}, \bibinfo
  {author} {\bibfnamefont {S.~D.}\ \bibnamefont {Katz}}, \bibinfo {author}
  {\bibfnamefont {A.}~\bibnamefont {Pasztor}}, \bibinfo {author} {\bibfnamefont
  {C.}~\bibnamefont {Ratti}}, \ and\ \bibinfo {author} {\bibfnamefont {K.~K.}\
  \bibnamefont {Szabo}},\ }\href {\doibase 10.1103/PhysRevD.92.114505}
  {\bibfield  {journal} {\bibinfo  {journal} {Phys. Rev. D}\ }\textbf {\bibinfo
  {volume} {92}},\ \bibinfo {pages} {114505} (\bibinfo {year} {2015})},\
  \Eprint {http://arxiv.org/abs/1507.04627} {arXiv:1507.04627 [hep-lat]}
  \BibitemShut {NoStop}%
\bibitem [{\citenamefont {Bors\'anyi}\ \emph {et~al.}(2021)\citenamefont
  {Bors\'anyi}, \citenamefont {Fodor}, \citenamefont {Guenther}, \citenamefont
  {Kara}, \citenamefont {Katz}, \citenamefont {Parotto}, \citenamefont
  {P\'asztor}, \citenamefont {Ratti},\ and\ \citenamefont
  {Szab\'o}}]{Borsanyi:2021sxv}%
  \BibitemOpen
  \bibfield  {author} {\bibinfo {author} {\bibfnamefont {S.}~\bibnamefont
  {Bors\'anyi}}, \bibinfo {author} {\bibfnamefont {Z.}~\bibnamefont {Fodor}},
  \bibinfo {author} {\bibfnamefont {J.~N.}\ \bibnamefont {Guenther}}, \bibinfo
  {author} {\bibfnamefont {R.}~\bibnamefont {Kara}}, \bibinfo {author}
  {\bibfnamefont {S.~D.}\ \bibnamefont {Katz}}, \bibinfo {author}
  {\bibfnamefont {P.}~\bibnamefont {Parotto}}, \bibinfo {author} {\bibfnamefont
  {A.}~\bibnamefont {P\'asztor}}, \bibinfo {author} {\bibfnamefont
  {C.}~\bibnamefont {Ratti}}, \ and\ \bibinfo {author} {\bibfnamefont {K.~K.}\
  \bibnamefont {Szab\'o}},\ }\href {\doibase 10.1103/PhysRevLett.126.232001}
  {\bibfield  {journal} {\bibinfo  {journal} {Phys. Rev. Lett.}\ }\textbf
  {\bibinfo {volume} {126}},\ \bibinfo {pages} {232001} (\bibinfo {year}
  {2021})},\ \Eprint {http://arxiv.org/abs/2102.06660} {arXiv:2102.06660
  [hep-lat]} \BibitemShut {NoStop}%
\bibitem [{\citenamefont {Kovtun}\ \emph {et~al.}(2005)\citenamefont {Kovtun},
  \citenamefont {Son},\ and\ \citenamefont {Starinets}}]{Kovtun:2004de}%
  \BibitemOpen
  \bibfield  {author} {\bibinfo {author} {\bibfnamefont {P.}~\bibnamefont
  {Kovtun}}, \bibinfo {author} {\bibfnamefont {D.~T.}\ \bibnamefont {Son}}, \
  and\ \bibinfo {author} {\bibfnamefont {A.~O.}\ \bibnamefont {Starinets}},\
  }\href {\doibase 10.1103/PhysRevLett.94.111601} {\bibfield  {journal}
  {\bibinfo  {journal} {Phys. Rev. Lett.}\ }\textbf {\bibinfo {volume} {94}},\
  \bibinfo {pages} {111601} (\bibinfo {year} {2005})},\ \Eprint
  {http://arxiv.org/abs/hep-th/0405231} {arXiv:hep-th/0405231} \BibitemShut
  {NoStop}%
\bibitem [{\citenamefont {Grefa}\ \emph {et~al.}(2021)\citenamefont {Grefa},
  \citenamefont {Noronha}, \citenamefont {Noronha-Hostler}, \citenamefont
  {Portillo}, \citenamefont {Ratti},\ and\ \citenamefont
  {Rougemont}}]{Grefa:2021qvt}%
  \BibitemOpen
  \bibfield  {author} {\bibinfo {author} {\bibfnamefont {J.}~\bibnamefont
  {Grefa}}, \bibinfo {author} {\bibfnamefont {J.}~\bibnamefont {Noronha}},
  \bibinfo {author} {\bibfnamefont {J.}~\bibnamefont {Noronha-Hostler}},
  \bibinfo {author} {\bibfnamefont {I.}~\bibnamefont {Portillo}}, \bibinfo
  {author} {\bibfnamefont {C.}~\bibnamefont {Ratti}}, \ and\ \bibinfo {author}
  {\bibfnamefont {R.}~\bibnamefont {Rougemont}},\ }\href {\doibase
  10.1103/PhysRevD.104.034002} {\bibfield  {journal} {\bibinfo  {journal}
  {Phys. Rev. D}\ }\textbf {\bibinfo {volume} {104}},\ \bibinfo {pages}
  {034002} (\bibinfo {year} {2021})},\ \Eprint
  {http://arxiv.org/abs/2102.12042} {arXiv:2102.12042 [nucl-th]} \BibitemShut
  {NoStop}%
\bibitem [{\citenamefont {Gursoy}\ and\ \citenamefont
  {Kiritsis}(2008)}]{Gursoy:2007cb}%
  \BibitemOpen
  \bibfield  {author} {\bibinfo {author} {\bibfnamefont {U.}~\bibnamefont
  {Gursoy}}\ and\ \bibinfo {author} {\bibfnamefont {E.}~\bibnamefont
  {Kiritsis}},\ }\href {\doibase 10.1088/1126-6708/2008/02/032} {\bibfield
  {journal} {\bibinfo  {journal} {JHEP}\ }\textbf {\bibinfo {volume} {02}},\
  \bibinfo {pages} {032} (\bibinfo {year} {2008})},\ \Eprint
  {http://arxiv.org/abs/0707.1324} {arXiv:0707.1324 [hep-th]} \BibitemShut
  {NoStop}%
\bibitem [{\citenamefont {Gursoy}\ \emph {et~al.}(2008)\citenamefont {Gursoy},
  \citenamefont {Kiritsis},\ and\ \citenamefont {Nitti}}]{Gursoy:2007er}%
  \BibitemOpen
  \bibfield  {author} {\bibinfo {author} {\bibfnamefont {U.}~\bibnamefont
  {Gursoy}}, \bibinfo {author} {\bibfnamefont {E.}~\bibnamefont {Kiritsis}}, \
  and\ \bibinfo {author} {\bibfnamefont {F.}~\bibnamefont {Nitti}},\ }\href
  {\doibase 10.1088/1126-6708/2008/02/019} {\bibfield  {journal} {\bibinfo
  {journal} {JHEP}\ }\textbf {\bibinfo {volume} {02}},\ \bibinfo {pages} {019}
  (\bibinfo {year} {2008})},\ \Eprint {http://arxiv.org/abs/0707.1349}
  {arXiv:0707.1349 [hep-th]} \BibitemShut {NoStop}%
\bibitem [{\citenamefont {Gursoy}\ \emph {et~al.}(2009)\citenamefont {Gursoy},
  \citenamefont {Kiritsis}, \citenamefont {Michalogiorgakis},\ and\
  \citenamefont {Nitti}}]{Gursoy:2009kk}%
  \BibitemOpen
  \bibfield  {author} {\bibinfo {author} {\bibfnamefont {U.}~\bibnamefont
  {Gursoy}}, \bibinfo {author} {\bibfnamefont {E.}~\bibnamefont {Kiritsis}},
  \bibinfo {author} {\bibfnamefont {G.}~\bibnamefont {Michalogiorgakis}}, \
  and\ \bibinfo {author} {\bibfnamefont {F.}~\bibnamefont {Nitti}},\ }\href
  {\doibase 10.1088/1126-6708/2009/12/056} {\bibfield  {journal} {\bibinfo
  {journal} {JHEP}\ }\textbf {\bibinfo {volume} {12}},\ \bibinfo {pages} {056}
  (\bibinfo {year} {2009})},\ \Eprint {http://arxiv.org/abs/0906.1890}
  {arXiv:0906.1890 [hep-ph]} \BibitemShut {NoStop}%
\bibitem [{\citenamefont {Gursoy}\ \emph {et~al.}(2010)\citenamefont {Gursoy},
  \citenamefont {Kiritsis}, \citenamefont {Mazzanti},\ and\ \citenamefont
  {Nitti}}]{Gursoy:2010aa}%
  \BibitemOpen
  \bibfield  {author} {\bibinfo {author} {\bibfnamefont {U.}~\bibnamefont
  {Gursoy}}, \bibinfo {author} {\bibfnamefont {E.}~\bibnamefont {Kiritsis}},
  \bibinfo {author} {\bibfnamefont {L.}~\bibnamefont {Mazzanti}}, \ and\
  \bibinfo {author} {\bibfnamefont {F.}~\bibnamefont {Nitti}},\ }\href
  {\doibase 10.1007/JHEP12(2010)088} {\bibfield  {journal} {\bibinfo  {journal}
  {JHEP}\ }\textbf {\bibinfo {volume} {12}},\ \bibinfo {pages} {088} (\bibinfo
  {year} {2010})},\ \Eprint {http://arxiv.org/abs/1006.3261} {arXiv:1006.3261
  [hep-th]} \BibitemShut {NoStop}%
\bibitem [{\citenamefont {Finazzo}\ \emph {et~al.}(2015)\citenamefont
  {Finazzo}, \citenamefont {Rougemont}, \citenamefont {Marrochio},\ and\
  \citenamefont {Noronha}}]{Finazzo:2014cna}%
  \BibitemOpen
  \bibfield  {author} {\bibinfo {author} {\bibfnamefont {S.~I.}\ \bibnamefont
  {Finazzo}}, \bibinfo {author} {\bibfnamefont {R.}~\bibnamefont {Rougemont}},
  \bibinfo {author} {\bibfnamefont {H.}~\bibnamefont {Marrochio}}, \ and\
  \bibinfo {author} {\bibfnamefont {J.}~\bibnamefont {Noronha}},\ }\href
  {\doibase 10.1007/JHEP02(2015)051} {\bibfield  {journal} {\bibinfo  {journal}
  {JHEP}\ }\textbf {\bibinfo {volume} {02}},\ \bibinfo {pages} {051} (\bibinfo
  {year} {2015})},\ \Eprint {http://arxiv.org/abs/1412.2968} {arXiv:1412.2968
  [hep-ph]} \BibitemShut {NoStop}%
\bibitem [{\citenamefont {Rougemont}\ \emph
  {et~al.}(2016{\natexlab{a}})\citenamefont {Rougemont}, \citenamefont
  {Critelli},\ and\ \citenamefont {Noronha}}]{Rougemont:2015oea}%
  \BibitemOpen
  \bibfield  {author} {\bibinfo {author} {\bibfnamefont {R.}~\bibnamefont
  {Rougemont}}, \bibinfo {author} {\bibfnamefont {R.}~\bibnamefont {Critelli}},
  \ and\ \bibinfo {author} {\bibfnamefont {J.}~\bibnamefont {Noronha}},\ }\href
  {\doibase 10.1103/PhysRevD.93.045013} {\bibfield  {journal} {\bibinfo
  {journal} {Phys. Rev. D}\ }\textbf {\bibinfo {volume} {93}},\ \bibinfo
  {pages} {045013} (\bibinfo {year} {2016}{\natexlab{a}})},\ \Eprint
  {http://arxiv.org/abs/1505.07894} {arXiv:1505.07894 [hep-th]} \BibitemShut
  {NoStop}%
\bibitem [{\citenamefont {Rougemont}\ \emph
  {et~al.}(2016{\natexlab{b}})\citenamefont {Rougemont}, \citenamefont
  {Ficnar}, \citenamefont {Finazzo},\ and\ \citenamefont
  {Noronha}}]{Rougemont:2015wca}%
  \BibitemOpen
  \bibfield  {author} {\bibinfo {author} {\bibfnamefont {R.}~\bibnamefont
  {Rougemont}}, \bibinfo {author} {\bibfnamefont {A.}~\bibnamefont {Ficnar}},
  \bibinfo {author} {\bibfnamefont {S.}~\bibnamefont {Finazzo}}, \ and\
  \bibinfo {author} {\bibfnamefont {J.}~\bibnamefont {Noronha}},\ }\href
  {\doibase 10.1007/JHEP04(2016)102} {\bibfield  {journal} {\bibinfo  {journal}
  {JHEP}\ }\textbf {\bibinfo {volume} {04}},\ \bibinfo {pages} {102} (\bibinfo
  {year} {2016}{\natexlab{b}})},\ \Eprint {http://arxiv.org/abs/1507.06556}
  {arXiv:1507.06556 [hep-th]} \BibitemShut {NoStop}%
\bibitem [{\citenamefont {Rougemont}\ \emph {et~al.}(2015)\citenamefont
  {Rougemont}, \citenamefont {Noronha},\ and\ \citenamefont
  {Noronha-Hostler}}]{Rougemont:2015ona}%
  \BibitemOpen
  \bibfield  {author} {\bibinfo {author} {\bibfnamefont {R.}~\bibnamefont
  {Rougemont}}, \bibinfo {author} {\bibfnamefont {J.}~\bibnamefont {Noronha}},
  \ and\ \bibinfo {author} {\bibfnamefont {J.}~\bibnamefont
  {Noronha-Hostler}},\ }\href {\doibase 10.1103/PhysRevLett.115.202301}
  {\bibfield  {journal} {\bibinfo  {journal} {Phys. Rev. Lett.}\ }\textbf
  {\bibinfo {volume} {115}},\ \bibinfo {pages} {202301} (\bibinfo {year}
  {2015})},\ \Eprint {http://arxiv.org/abs/1507.06972} {arXiv:1507.06972
  [hep-ph]} \BibitemShut {NoStop}%
\bibitem [{\citenamefont {Finazzo}\ and\ \citenamefont
  {Rougemont}(2016)}]{Finazzo:2015xwa}%
  \BibitemOpen
  \bibfield  {author} {\bibinfo {author} {\bibfnamefont {S.~I.}\ \bibnamefont
  {Finazzo}}\ and\ \bibinfo {author} {\bibfnamefont {R.}~\bibnamefont
  {Rougemont}},\ }\href {\doibase 10.1103/PhysRevD.93.034017} {\bibfield
  {journal} {\bibinfo  {journal} {Phys. Rev. D}\ }\textbf {\bibinfo {volume}
  {93}},\ \bibinfo {pages} {034017} (\bibinfo {year} {2016})},\ \Eprint
  {http://arxiv.org/abs/1510.03321} {arXiv:1510.03321 [hep-ph]} \BibitemShut
  {NoStop}%
\bibitem [{\citenamefont {Rougemont}\ and\ \citenamefont
  {Finazzo}(2016)}]{Rougemont:2016nyr}%
  \BibitemOpen
  \bibfield  {author} {\bibinfo {author} {\bibfnamefont {R.}~\bibnamefont
  {Rougemont}}\ and\ \bibinfo {author} {\bibfnamefont {S.~I.}\ \bibnamefont
  {Finazzo}},\ }\href {\doibase 10.1103/PhysRevD.93.106005} {\bibfield
  {journal} {\bibinfo  {journal} {Phys. Rev. D}\ }\textbf {\bibinfo {volume}
  {93}},\ \bibinfo {pages} {106005} (\bibinfo {year} {2016})},\ \Eprint
  {http://arxiv.org/abs/1603.00553} {arXiv:1603.00553 [hep-ph]} \BibitemShut
  {NoStop}%
\bibitem [{\citenamefont {Finazzo}\ \emph {et~al.}(2016)\citenamefont
  {Finazzo}, \citenamefont {Critelli}, \citenamefont {Rougemont},\ and\
  \citenamefont {Noronha}}]{Finazzo:2016mhm}%
  \BibitemOpen
  \bibfield  {author} {\bibinfo {author} {\bibfnamefont {S.~I.}\ \bibnamefont
  {Finazzo}}, \bibinfo {author} {\bibfnamefont {R.}~\bibnamefont {Critelli}},
  \bibinfo {author} {\bibfnamefont {R.}~\bibnamefont {Rougemont}}, \ and\
  \bibinfo {author} {\bibfnamefont {J.}~\bibnamefont {Noronha}},\ }\href
  {\doibase 10.1103/PhysRevD.94.054020} {\bibfield  {journal} {\bibinfo
  {journal} {Phys. Rev. D}\ }\textbf {\bibinfo {volume} {94}},\ \bibinfo
  {pages} {054020} (\bibinfo {year} {2016})},\ \bibinfo {note} {[Erratum:
  Phys.Rev.D 96, 019903 (2017)]},\ \Eprint {http://arxiv.org/abs/1605.06061}
  {arXiv:1605.06061 [hep-ph]} \BibitemShut {NoStop}%
\bibitem [{\citenamefont {Attems}\ \emph {et~al.}(2017)\citenamefont {Attems},
  \citenamefont {Casalderrey-Solana}, \citenamefont {Mateos}, \citenamefont
  {Santos-Oliv\'an}, \citenamefont {Sopuerta}, \citenamefont {Triana},\ and\
  \citenamefont {Zilh\~ao}}]{Attems:2016tby}%
  \BibitemOpen
  \bibfield  {author} {\bibinfo {author} {\bibfnamefont {M.}~\bibnamefont
  {Attems}}, \bibinfo {author} {\bibfnamefont {J.}~\bibnamefont
  {Casalderrey-Solana}}, \bibinfo {author} {\bibfnamefont {D.}~\bibnamefont
  {Mateos}}, \bibinfo {author} {\bibfnamefont {D.}~\bibnamefont
  {Santos-Oliv\'an}}, \bibinfo {author} {\bibfnamefont {C.~F.}\ \bibnamefont
  {Sopuerta}}, \bibinfo {author} {\bibfnamefont {M.}~\bibnamefont {Triana}}, \
  and\ \bibinfo {author} {\bibfnamefont {M.}~\bibnamefont {Zilh\~ao}},\ }\href
  {\doibase 10.1007/JHEP01(2017)026} {\bibfield  {journal} {\bibinfo  {journal}
  {JHEP}\ }\textbf {\bibinfo {volume} {01}},\ \bibinfo {pages} {026} (\bibinfo
  {year} {2017})},\ \Eprint {http://arxiv.org/abs/1604.06439} {arXiv:1604.06439
  [hep-th]} \BibitemShut {NoStop}%
\bibitem [{\citenamefont {Attems}\ \emph {et~al.}(2016)\citenamefont {Attems},
  \citenamefont {Casalderrey-Solana}, \citenamefont {Mateos}, \citenamefont
  {Papadimitriou}, \citenamefont {Santos-Oliv\'an}, \citenamefont {Sopuerta},
  \citenamefont {Triana},\ and\ \citenamefont {Zilh\~ao}}]{Attems:2016ugt}%
  \BibitemOpen
  \bibfield  {author} {\bibinfo {author} {\bibfnamefont {M.}~\bibnamefont
  {Attems}}, \bibinfo {author} {\bibfnamefont {J.}~\bibnamefont
  {Casalderrey-Solana}}, \bibinfo {author} {\bibfnamefont {D.}~\bibnamefont
  {Mateos}}, \bibinfo {author} {\bibfnamefont {I.}~\bibnamefont
  {Papadimitriou}}, \bibinfo {author} {\bibfnamefont {D.}~\bibnamefont
  {Santos-Oliv\'an}}, \bibinfo {author} {\bibfnamefont {C.~F.}\ \bibnamefont
  {Sopuerta}}, \bibinfo {author} {\bibfnamefont {M.}~\bibnamefont {Triana}}, \
  and\ \bibinfo {author} {\bibfnamefont {M.}~\bibnamefont {Zilh\~ao}},\ }\href
  {\doibase 10.1007/JHEP10(2016)155} {\bibfield  {journal} {\bibinfo  {journal}
  {JHEP}\ }\textbf {\bibinfo {volume} {10}},\ \bibinfo {pages} {155} (\bibinfo
  {year} {2016})},\ \Eprint {http://arxiv.org/abs/1603.01254} {arXiv:1603.01254
  [hep-th]} \BibitemShut {NoStop}%
\bibitem [{\citenamefont {Critelli}\ \emph {et~al.}(2016)\citenamefont
  {Critelli}, \citenamefont {Rougemont}, \citenamefont {Finazzo},\ and\
  \citenamefont {Noronha}}]{Critelli:2016cvq}%
  \BibitemOpen
  \bibfield  {author} {\bibinfo {author} {\bibfnamefont {R.}~\bibnamefont
  {Critelli}}, \bibinfo {author} {\bibfnamefont {R.}~\bibnamefont {Rougemont}},
  \bibinfo {author} {\bibfnamefont {S.~I.}\ \bibnamefont {Finazzo}}, \ and\
  \bibinfo {author} {\bibfnamefont {J.}~\bibnamefont {Noronha}},\ }\href
  {\doibase 10.1103/PhysRevD.94.125019} {\bibfield  {journal} {\bibinfo
  {journal} {Phys. Rev. D}\ }\textbf {\bibinfo {volume} {94}},\ \bibinfo
  {pages} {125019} (\bibinfo {year} {2016})},\ \Eprint
  {http://arxiv.org/abs/1606.09484} {arXiv:1606.09484 [hep-ph]} \BibitemShut
  {NoStop}%
\bibitem [{\citenamefont {Demircik}\ and\ \citenamefont
  {Gursoy}(2017)}]{Demircik:2016nhr}%
  \BibitemOpen
  \bibfield  {author} {\bibinfo {author} {\bibfnamefont {T.}~\bibnamefont
  {Demircik}}\ and\ \bibinfo {author} {\bibfnamefont {U.}~\bibnamefont
  {Gursoy}},\ }\href {\doibase 10.1016/j.nuclphysb.2017.03.020} {\bibfield
  {journal} {\bibinfo  {journal} {Nucl. Phys. B}\ }\textbf {\bibinfo {volume}
  {919}},\ \bibinfo {pages} {384} (\bibinfo {year} {2017})},\ \Eprint
  {http://arxiv.org/abs/1605.08118} {arXiv:1605.08118 [hep-th]} \BibitemShut
  {NoStop}%
\bibitem [{\citenamefont {Rougemont}\ \emph {et~al.}(2017)\citenamefont
  {Rougemont}, \citenamefont {Critelli}, \citenamefont {Noronha-Hostler},
  \citenamefont {Noronha},\ and\ \citenamefont {Ratti}}]{Rougemont:2017tlu}%
  \BibitemOpen
  \bibfield  {author} {\bibinfo {author} {\bibfnamefont {R.}~\bibnamefont
  {Rougemont}}, \bibinfo {author} {\bibfnamefont {R.}~\bibnamefont {Critelli}},
  \bibinfo {author} {\bibfnamefont {J.}~\bibnamefont {Noronha-Hostler}},
  \bibinfo {author} {\bibfnamefont {J.}~\bibnamefont {Noronha}}, \ and\
  \bibinfo {author} {\bibfnamefont {C.}~\bibnamefont {Ratti}},\ }\href
  {\doibase 10.1103/PhysRevD.96.014032} {\bibfield  {journal} {\bibinfo
  {journal} {Phys. Rev. D}\ }\textbf {\bibinfo {volume} {96}},\ \bibinfo
  {pages} {014032} (\bibinfo {year} {2017})},\ \Eprint
  {http://arxiv.org/abs/1704.05558} {arXiv:1704.05558 [hep-ph]} \BibitemShut
  {NoStop}%
\bibitem [{\citenamefont {Knaute}\ \emph {et~al.}(2018)\citenamefont {Knaute},
  \citenamefont {Yaresko},\ and\ \citenamefont {K\"ampfer}}]{Knaute:2017opk}%
  \BibitemOpen
  \bibfield  {author} {\bibinfo {author} {\bibfnamefont {J.}~\bibnamefont
  {Knaute}}, \bibinfo {author} {\bibfnamefont {R.}~\bibnamefont {Yaresko}}, \
  and\ \bibinfo {author} {\bibfnamefont {B.}~\bibnamefont {K\"ampfer}},\ }\href
  {\doibase 10.1016/j.physletb.2018.01.053} {\bibfield  {journal} {\bibinfo
  {journal} {Phys. Lett. B}\ }\textbf {\bibinfo {volume} {778}},\ \bibinfo
  {pages} {419} (\bibinfo {year} {2018})},\ \Eprint
  {http://arxiv.org/abs/1702.06731} {arXiv:1702.06731 [hep-ph]} \BibitemShut
  {NoStop}%
\bibitem [{\citenamefont {Li}\ \emph {et~al.}(2018)\citenamefont {Li},
  \citenamefont {Chen}, \citenamefont {Li},\ and\ \citenamefont
  {Huang}}]{Li:2017ple}%
  \BibitemOpen
  \bibfield  {author} {\bibinfo {author} {\bibfnamefont {Z.}~\bibnamefont
  {Li}}, \bibinfo {author} {\bibfnamefont {Y.}~\bibnamefont {Chen}}, \bibinfo
  {author} {\bibfnamefont {D.}~\bibnamefont {Li}}, \ and\ \bibinfo {author}
  {\bibfnamefont {M.}~\bibnamefont {Huang}},\ }\href {\doibase
  10.1088/1674-1137/42/1/013103} {\bibfield  {journal} {\bibinfo  {journal}
  {Chin. Phys. C}\ }\textbf {\bibinfo {volume} {42}},\ \bibinfo {pages}
  {013103} (\bibinfo {year} {2018})},\ \Eprint
  {http://arxiv.org/abs/1706.02238} {arXiv:1706.02238 [hep-ph]} \BibitemShut
  {NoStop}%
\bibitem [{\citenamefont {Rougemont}\ \emph {et~al.}(2018)\citenamefont
  {Rougemont}, \citenamefont {Critelli},\ and\ \citenamefont
  {Noronha}}]{Rougemont:2018ivt}%
  \BibitemOpen
  \bibfield  {author} {\bibinfo {author} {\bibfnamefont {R.}~\bibnamefont
  {Rougemont}}, \bibinfo {author} {\bibfnamefont {R.}~\bibnamefont {Critelli}},
  \ and\ \bibinfo {author} {\bibfnamefont {J.}~\bibnamefont {Noronha}},\ }\href
  {\doibase 10.1103/PhysRevD.98.034028} {\bibfield  {journal} {\bibinfo
  {journal} {Phys. Rev. D}\ }\textbf {\bibinfo {volume} {98}},\ \bibinfo
  {pages} {034028} (\bibinfo {year} {2018})},\ \Eprint
  {http://arxiv.org/abs/1804.00189} {arXiv:1804.00189 [hep-ph]} \BibitemShut
  {NoStop}%
\bibitem [{\citenamefont {Aref'eva}\ and\ \citenamefont
  {Rannu}(2018)}]{Arefeva:2018hyo}%
  \BibitemOpen
  \bibfield  {author} {\bibinfo {author} {\bibfnamefont {I.}~\bibnamefont
  {Aref'eva}}\ and\ \bibinfo {author} {\bibfnamefont {K.}~\bibnamefont
  {Rannu}},\ }\href {\doibase 10.1007/JHEP05(2018)206} {\bibfield  {journal}
  {\bibinfo  {journal} {JHEP}\ }\textbf {\bibinfo {volume} {05}},\ \bibinfo
  {pages} {206} (\bibinfo {year} {2018})},\ \Eprint
  {http://arxiv.org/abs/1802.05652} {arXiv:1802.05652 [hep-th]} \BibitemShut
  {NoStop}%
\bibitem [{\citenamefont {Attems}\ \emph {et~al.}(2018)\citenamefont {Attems},
  \citenamefont {Bea}, \citenamefont {Casalderrey-Solana}, \citenamefont
  {Mateos}, \citenamefont {Triana},\ and\ \citenamefont
  {Zilh\~ao}}]{Attems:2018gou}%
  \BibitemOpen
  \bibfield  {author} {\bibinfo {author} {\bibfnamefont {M.}~\bibnamefont
  {Attems}}, \bibinfo {author} {\bibfnamefont {Y.}~\bibnamefont {Bea}},
  \bibinfo {author} {\bibfnamefont {J.}~\bibnamefont {Casalderrey-Solana}},
  \bibinfo {author} {\bibfnamefont {D.}~\bibnamefont {Mateos}}, \bibinfo
  {author} {\bibfnamefont {M.}~\bibnamefont {Triana}}, \ and\ \bibinfo {author}
  {\bibfnamefont {M.}~\bibnamefont {Zilh\~ao}},\ }\href {\doibase
  10.1103/PhysRevLett.121.261601} {\bibfield  {journal} {\bibinfo  {journal}
  {Phys. Rev. Lett.}\ }\textbf {\bibinfo {volume} {121}},\ \bibinfo {pages}
  {261601} (\bibinfo {year} {2018})},\ \Eprint
  {http://arxiv.org/abs/1807.05175} {arXiv:1807.05175 [hep-th]} \BibitemShut
  {NoStop}%
\bibitem [{\citenamefont {Attems}\ \emph {et~al.}(2020)\citenamefont {Attems},
  \citenamefont {Bea}, \citenamefont {Casalderrey-Solana}, \citenamefont
  {Mateos},\ and\ \citenamefont {Zilh\~ao}}]{Attems:2019yqn}%
  \BibitemOpen
  \bibfield  {author} {\bibinfo {author} {\bibfnamefont {M.}~\bibnamefont
  {Attems}}, \bibinfo {author} {\bibfnamefont {Y.}~\bibnamefont {Bea}},
  \bibinfo {author} {\bibfnamefont {J.}~\bibnamefont {Casalderrey-Solana}},
  \bibinfo {author} {\bibfnamefont {D.}~\bibnamefont {Mateos}}, \ and\ \bibinfo
  {author} {\bibfnamefont {M.}~\bibnamefont {Zilh\~ao}},\ }\href {\doibase
  10.1007/JHEP01(2020)106} {\bibfield  {journal} {\bibinfo  {journal} {JHEP}\
  }\textbf {\bibinfo {volume} {01}},\ \bibinfo {pages} {106} (\bibinfo {year}
  {2020})},\ \Eprint {http://arxiv.org/abs/1905.12544} {arXiv:1905.12544
  [hep-th]} \BibitemShut {NoStop}%
\bibitem [{\citenamefont {Aref'eva}\ \emph {et~al.}(2021)\citenamefont
  {Aref'eva}, \citenamefont {Rannu},\ and\ \citenamefont
  {Slepov}}]{Arefeva:2020vae}%
  \BibitemOpen
  \bibfield  {author} {\bibinfo {author} {\bibfnamefont {I.~Y.}\ \bibnamefont
  {Aref'eva}}, \bibinfo {author} {\bibfnamefont {K.}~\bibnamefont {Rannu}}, \
  and\ \bibinfo {author} {\bibfnamefont {P.}~\bibnamefont {Slepov}},\ }\href
  {\doibase 10.1007/JHEP07(2021)161} {\bibfield  {journal} {\bibinfo  {journal}
  {JHEP}\ }\textbf {\bibinfo {volume} {07}},\ \bibinfo {pages} {161} (\bibinfo
  {year} {2021})},\ \Eprint {http://arxiv.org/abs/2011.07023} {arXiv:2011.07023
  [hep-th]} \BibitemShut {NoStop}%
\bibitem [{\citenamefont {Rougemont}(2020)}]{Rougemont:2020had}%
  \BibitemOpen
  \bibfield  {author} {\bibinfo {author} {\bibfnamefont {R.}~\bibnamefont
  {Rougemont}},\ }\href {\doibase 10.1103/PhysRevD.102.034009} {\bibfield
  {journal} {\bibinfo  {journal} {Phys. Rev. D}\ }\textbf {\bibinfo {volume}
  {102}},\ \bibinfo {pages} {034009} (\bibinfo {year} {2020})},\ \Eprint
  {http://arxiv.org/abs/2002.06725} {arXiv:2002.06725 [hep-ph]} \BibitemShut
  {NoStop}%
\bibitem [{\citenamefont {Z\"ollner}\ and\ \citenamefont
  {K\"ampfer}(2021)}]{Zollner:2020nnt}%
  \BibitemOpen
  \bibfield  {author} {\bibinfo {author} {\bibfnamefont {R.}~\bibnamefont
  {Z\"ollner}}\ and\ \bibinfo {author} {\bibfnamefont {B.}~\bibnamefont
  {K\"ampfer}},\ }\href {\doibase 10.3390/particles4020015} {\bibfield
  {journal} {\bibinfo  {journal} {Particles}\ }\textbf {\bibinfo {volume}
  {4}},\ \bibinfo {pages} {159} (\bibinfo {year} {2021})},\ \Eprint
  {http://arxiv.org/abs/2007.14287} {arXiv:2007.14287 [hep-ph]} \BibitemShut
  {NoStop}%
\bibitem [{\citenamefont {Attems}(2021)}]{Attems:2020qkg}%
  \BibitemOpen
  \bibfield  {author} {\bibinfo {author} {\bibfnamefont {M.}~\bibnamefont
  {Attems}},\ }\href {\doibase 10.1007/JHEP08(2021)155} {\bibfield  {journal}
  {\bibinfo  {journal} {JHEP}\ }\textbf {\bibinfo {volume} {08}},\ \bibinfo
  {pages} {155} (\bibinfo {year} {2021})},\ \Eprint
  {http://arxiv.org/abs/2012.15687} {arXiv:2012.15687 [hep-th]} \BibitemShut
  {NoStop}%
\bibitem [{\citenamefont {Ballon-Bayona}\ \emph {et~al.}(2020)\citenamefont
  {Ballon-Bayona}, \citenamefont {Boschi-Filho}, \citenamefont {Capossoli},\
  and\ \citenamefont {Rodrigues}}]{Ballon-Bayona:2020xls}%
  \BibitemOpen
  \bibfield  {author} {\bibinfo {author} {\bibfnamefont {A.}~\bibnamefont
  {Ballon-Bayona}}, \bibinfo {author} {\bibfnamefont {H.}~\bibnamefont
  {Boschi-Filho}}, \bibinfo {author} {\bibfnamefont {E.~F.}\ \bibnamefont
  {Capossoli}}, \ and\ \bibinfo {author} {\bibfnamefont {D.~M.}\ \bibnamefont
  {Rodrigues}},\ }\href {\doibase 10.1103/PhysRevD.102.126003} {\bibfield
  {journal} {\bibinfo  {journal} {Phys. Rev. D}\ }\textbf {\bibinfo {volume}
  {102}},\ \bibinfo {pages} {126003} (\bibinfo {year} {2020})},\ \Eprint
  {http://arxiv.org/abs/2006.08810} {arXiv:2006.08810 [hep-th]} \BibitemShut
  {NoStop}%
\bibitem [{\citenamefont {Ballon-Bayona}\ \emph {et~al.}(2021)\citenamefont
  {Ballon-Bayona}, \citenamefont {Mamani}, \citenamefont {Miranda},\ and\
  \citenamefont {Zanchin}}]{Ballon-Bayona:2021tzw}%
  \BibitemOpen
  \bibfield  {author} {\bibinfo {author} {\bibfnamefont {A.}~\bibnamefont
  {Ballon-Bayona}}, \bibinfo {author} {\bibfnamefont {L.~A.~H.}\ \bibnamefont
  {Mamani}}, \bibinfo {author} {\bibfnamefont {A.~S.}\ \bibnamefont {Miranda}},
  \ and\ \bibinfo {author} {\bibfnamefont {V.~T.}\ \bibnamefont {Zanchin}},\
  }\href {\doibase 10.1103/PhysRevD.104.046013} {\bibfield  {journal} {\bibinfo
   {journal} {Phys. Rev. D}\ }\textbf {\bibinfo {volume} {104}},\ \bibinfo
  {pages} {046013} (\bibinfo {year} {2021})},\ \Eprint
  {http://arxiv.org/abs/2103.14188} {arXiv:2103.14188 [hep-th]} \BibitemShut
  {NoStop}%
\bibitem [{\citenamefont {Cai}\ \emph {et~al.}(2022)\citenamefont {Cai},
  \citenamefont {He}, \citenamefont {Li},\ and\ \citenamefont
  {Wang}}]{Cai:2022omk}%
  \BibitemOpen
  \bibfield  {author} {\bibinfo {author} {\bibfnamefont {R.-G.}\ \bibnamefont
  {Cai}}, \bibinfo {author} {\bibfnamefont {S.}~\bibnamefont {He}}, \bibinfo
  {author} {\bibfnamefont {L.}~\bibnamefont {Li}}, \ and\ \bibinfo {author}
  {\bibfnamefont {Y.-X.}\ \bibnamefont {Wang}},\ }\href@noop {} {\  (\bibinfo
  {year} {2022})},\ \Eprint {http://arxiv.org/abs/2201.02004} {arXiv:2201.02004
  [hep-th]} \BibitemShut {NoStop}%
\bibitem [{\citenamefont {Li}\ \emph {et~al.}(2022)\citenamefont {Li},
  \citenamefont {Liu}, \citenamefont {Liu},\ and\ \citenamefont
  {Fang}}]{Li:2022erd}%
  \BibitemOpen
  \bibfield  {author} {\bibinfo {author} {\bibfnamefont {Y.-Y.}\ \bibnamefont
  {Li}}, \bibinfo {author} {\bibfnamefont {X.-L.}\ \bibnamefont {Liu}},
  \bibinfo {author} {\bibfnamefont {X.-Y.}\ \bibnamefont {Liu}}, \ and\
  \bibinfo {author} {\bibfnamefont {Z.}~\bibnamefont {Fang}},\ }\href {\doibase
  10.1103/PhysRevD.105.034019} {\bibfield  {journal} {\bibinfo  {journal}
  {Phys. Rev. D}\ }\textbf {\bibinfo {volume} {105}},\ \bibinfo {pages}
  {034019} (\bibinfo {year} {2022})},\ \Eprint
  {http://arxiv.org/abs/2201.11427} {arXiv:2201.11427 [hep-ph]} \BibitemShut
  {NoStop}%
\bibitem [{\citenamefont {Dudal}\ \emph {et~al.}(2021)\citenamefont {Dudal},
  \citenamefont {Hajilou},\ and\ \citenamefont {Mahapatra}}]{Dudal:2021jav}%
  \BibitemOpen
  \bibfield  {author} {\bibinfo {author} {\bibfnamefont {D.}~\bibnamefont
  {Dudal}}, \bibinfo {author} {\bibfnamefont {A.}~\bibnamefont {Hajilou}}, \
  and\ \bibinfo {author} {\bibfnamefont {S.}~\bibnamefont {Mahapatra}},\ }\href
  {\doibase 10.1140/epja/s10050-021-00461-4} {\bibfield  {journal} {\bibinfo
  {journal} {Eur. Phys. J. A}\ }\textbf {\bibinfo {volume} {57}},\ \bibinfo
  {pages} {142} (\bibinfo {year} {2021})},\ \Eprint
  {http://arxiv.org/abs/2103.01185} {arXiv:2103.01185 [hep-th]} \BibitemShut
  {NoStop}%
\bibitem [{\citenamefont {Jena}\ \emph {et~al.}(2022)\citenamefont {Jena},
  \citenamefont {Shukla}, \citenamefont {Dudal},\ and\ \citenamefont
  {Mahapatra}}]{Jena:2022nzw}%
  \BibitemOpen
  \bibfield  {author} {\bibinfo {author} {\bibfnamefont {S.~S.}\ \bibnamefont
  {Jena}}, \bibinfo {author} {\bibfnamefont {B.}~\bibnamefont {Shukla}},
  \bibinfo {author} {\bibfnamefont {D.}~\bibnamefont {Dudal}}, \ and\ \bibinfo
  {author} {\bibfnamefont {S.}~\bibnamefont {Mahapatra}},\ }\href@noop {} {\
  (\bibinfo {year} {2022})},\ \Eprint {http://arxiv.org/abs/2202.01486}
  {arXiv:2202.01486 [hep-th]} \BibitemShut {NoStop}%
\bibitem [{\citenamefont {Jarvinen}\ and\ \citenamefont
  {Kiritsis}(2012)}]{Jarvinen:2011qe}%
  \BibitemOpen
  \bibfield  {author} {\bibinfo {author} {\bibfnamefont {M.}~\bibnamefont
  {Jarvinen}}\ and\ \bibinfo {author} {\bibfnamefont {E.}~\bibnamefont
  {Kiritsis}},\ }\href {\doibase 10.1007/JHEP03(2012)002} {\bibfield  {journal}
  {\bibinfo  {journal} {JHEP}\ }\textbf {\bibinfo {volume} {03}},\ \bibinfo
  {pages} {002} (\bibinfo {year} {2012})},\ \Eprint
  {http://arxiv.org/abs/1112.1261} {arXiv:1112.1261 [hep-ph]} \BibitemShut
  {NoStop}%
\bibitem [{\citenamefont {Hoyos}\ \emph {et~al.}(2020)\citenamefont {Hoyos},
  \citenamefont {Jokela}, \citenamefont {Jarvinen}, \citenamefont {Subils},
  \citenamefont {Tarrio},\ and\ \citenamefont {Vuorinen}}]{Hoyos:2020hmq}%
  \BibitemOpen
  \bibfield  {author} {\bibinfo {author} {\bibfnamefont {C.}~\bibnamefont
  {Hoyos}}, \bibinfo {author} {\bibfnamefont {N.}~\bibnamefont {Jokela}},
  \bibinfo {author} {\bibfnamefont {M.}~\bibnamefont {Jarvinen}}, \bibinfo
  {author} {\bibfnamefont {J.~G.}\ \bibnamefont {Subils}}, \bibinfo {author}
  {\bibfnamefont {J.}~\bibnamefont {Tarrio}}, \ and\ \bibinfo {author}
  {\bibfnamefont {A.}~\bibnamefont {Vuorinen}},\ }\href {\doibase
  10.1103/PhysRevLett.125.241601} {\bibfield  {journal} {\bibinfo  {journal}
  {Phys. Rev. Lett.}\ }\textbf {\bibinfo {volume} {125}},\ \bibinfo {pages}
  {241601} (\bibinfo {year} {2020})},\ \Eprint
  {http://arxiv.org/abs/2005.14205} {arXiv:2005.14205 [hep-th]} \BibitemShut
  {NoStop}%
\bibitem [{\citenamefont {Hoyos}\ \emph {et~al.}(2022)\citenamefont {Hoyos},
  \citenamefont {Jokela}, \citenamefont {J\"arvinen}, \citenamefont {Subils},
  \citenamefont {Tarrio},\ and\ \citenamefont {Vuorinen}}]{Hoyos:2021njg}%
  \BibitemOpen
  \bibfield  {author} {\bibinfo {author} {\bibfnamefont {C.}~\bibnamefont
  {Hoyos}}, \bibinfo {author} {\bibfnamefont {N.}~\bibnamefont {Jokela}},
  \bibinfo {author} {\bibfnamefont {M.}~\bibnamefont {J\"arvinen}}, \bibinfo
  {author} {\bibfnamefont {J.~G.}\ \bibnamefont {Subils}}, \bibinfo {author}
  {\bibfnamefont {J.}~\bibnamefont {Tarrio}}, \ and\ \bibinfo {author}
  {\bibfnamefont {A.}~\bibnamefont {Vuorinen}},\ }\href {\doibase
  10.1103/PhysRevD.105.066014} {\bibfield  {journal} {\bibinfo  {journal}
  {Phys. Rev. D}\ }\textbf {\bibinfo {volume} {105}},\ \bibinfo {pages}
  {066014} (\bibinfo {year} {2022})},\ \Eprint
  {http://arxiv.org/abs/2109.12122} {arXiv:2109.12122 [hep-th]} \BibitemShut
  {NoStop}%
\bibitem [{\citenamefont {J\"arvinen}(2022)}]{Jarvinen:2021jbd}%
  \BibitemOpen
  \bibfield  {author} {\bibinfo {author} {\bibfnamefont {M.}~\bibnamefont
  {J\"arvinen}},\ }\href {\doibase 10.1140/epjc/s10052-022-10227-x} {\bibfield
  {journal} {\bibinfo  {journal} {Eur. Phys. J. C}\ }\textbf {\bibinfo {volume}
  {82}},\ \bibinfo {pages} {282} (\bibinfo {year} {2022})},\ \Eprint
  {http://arxiv.org/abs/2110.08281} {arXiv:2110.08281 [hep-ph]} \BibitemShut
  {NoStop}%
\bibitem [{\citenamefont {Demircik}\ \emph {et~al.}(2021)\citenamefont
  {Demircik}, \citenamefont {Ecker},\ and\ \citenamefont
  {J\"arvinen}}]{Demircik:2021zll}%
  \BibitemOpen
  \bibfield  {author} {\bibinfo {author} {\bibfnamefont {T.}~\bibnamefont
  {Demircik}}, \bibinfo {author} {\bibfnamefont {C.}~\bibnamefont {Ecker}}, \
  and\ \bibinfo {author} {\bibfnamefont {M.}~\bibnamefont {J\"arvinen}},\
  }\href@noop {} {\  (\bibinfo {year} {2021})},\ \Eprint
  {http://arxiv.org/abs/2112.12157} {arXiv:2112.12157 [hep-ph]} \BibitemShut
  {NoStop}%
\bibitem [{\citenamefont {Danielewicz}\ and\ \citenamefont
  {Gyulassy}(1985)}]{Danielewicz:1984ww}%
  \BibitemOpen
  \bibfield  {author} {\bibinfo {author} {\bibfnamefont {P.}~\bibnamefont
  {Danielewicz}}\ and\ \bibinfo {author} {\bibfnamefont {M.}~\bibnamefont
  {Gyulassy}},\ }\href {\doibase 10.1103/PhysRevD.31.53} {\bibfield  {journal}
  {\bibinfo  {journal} {Phys. Rev. D}\ }\textbf {\bibinfo {volume} {31}},\
  \bibinfo {pages} {53} (\bibinfo {year} {1985})}\BibitemShut {NoStop}%
\bibitem [{\citenamefont {Gavin}(1985)}]{Gavin:1985ph}%
  \BibitemOpen
  \bibfield  {author} {\bibinfo {author} {\bibfnamefont {S.}~\bibnamefont
  {Gavin}},\ }\href {\doibase 10.1016/0375-9474(85)90190-3} {\bibfield
  {journal} {\bibinfo  {journal} {Nucl. Phys. A}\ }\textbf {\bibinfo {volume}
  {435}},\ \bibinfo {pages} {826} (\bibinfo {year} {1985})}\BibitemShut
  {NoStop}%
\bibitem [{\citenamefont {Aarts}\ and\ \citenamefont
  {Martinez~Resco}(2002)}]{Aarts:2002cc}%
  \BibitemOpen
  \bibfield  {author} {\bibinfo {author} {\bibfnamefont {G.}~\bibnamefont
  {Aarts}}\ and\ \bibinfo {author} {\bibfnamefont {J.~M.}\ \bibnamefont
  {Martinez~Resco}},\ }\href {\doibase 10.1088/1126-6708/2002/04/053}
  {\bibfield  {journal} {\bibinfo  {journal} {JHEP}\ }\textbf {\bibinfo
  {volume} {04}},\ \bibinfo {pages} {053} (\bibinfo {year} {2002})},\ \Eprint
  {http://arxiv.org/abs/hep-ph/0203177} {arXiv:hep-ph/0203177} \BibitemShut
  {NoStop}%
\bibitem [{\citenamefont {Noronha-Hostler}\ \emph {et~al.}(2009)\citenamefont
  {Noronha-Hostler}, \citenamefont {Noronha},\ and\ \citenamefont
  {Greiner}}]{Noronha-Hostler:2008kkf}%
  \BibitemOpen
  \bibfield  {author} {\bibinfo {author} {\bibfnamefont {J.}~\bibnamefont
  {Noronha-Hostler}}, \bibinfo {author} {\bibfnamefont {J.}~\bibnamefont
  {Noronha}}, \ and\ \bibinfo {author} {\bibfnamefont {C.}~\bibnamefont
  {Greiner}},\ }\href {\doibase 10.1103/PhysRevLett.103.172302} {\bibfield
  {journal} {\bibinfo  {journal} {Phys. Rev. Lett.}\ }\textbf {\bibinfo
  {volume} {103}},\ \bibinfo {pages} {172302} (\bibinfo {year} {2009})},\
  \Eprint {http://arxiv.org/abs/0811.1571} {arXiv:0811.1571 [nucl-th]}
  \BibitemShut {NoStop}%
\bibitem [{\citenamefont {Noronha-Hostler}\ \emph {et~al.}(2012)\citenamefont
  {Noronha-Hostler}, \citenamefont {Noronha},\ and\ \citenamefont
  {Greiner}}]{Noronha-Hostler:2012ycm}%
  \BibitemOpen
  \bibfield  {author} {\bibinfo {author} {\bibfnamefont {J.}~\bibnamefont
  {Noronha-Hostler}}, \bibinfo {author} {\bibfnamefont {J.}~\bibnamefont
  {Noronha}}, \ and\ \bibinfo {author} {\bibfnamefont {C.}~\bibnamefont
  {Greiner}},\ }\href {\doibase 10.1103/PhysRevC.86.024913} {\bibfield
  {journal} {\bibinfo  {journal} {Phys. Rev. C}\ }\textbf {\bibinfo {volume}
  {86}},\ \bibinfo {pages} {024913} (\bibinfo {year} {2012})},\ \Eprint
  {http://arxiv.org/abs/1206.5138} {arXiv:1206.5138 [nucl-th]} \BibitemShut
  {NoStop}%
\bibitem [{\citenamefont {Denicol}\ \emph {et~al.}(2013)\citenamefont
  {Denicol}, \citenamefont {Gale}, \citenamefont {Jeon},\ and\ \citenamefont
  {Noronha}}]{Denicol:2013nua}%
  \BibitemOpen
  \bibfield  {author} {\bibinfo {author} {\bibfnamefont {G.~S.}\ \bibnamefont
  {Denicol}}, \bibinfo {author} {\bibfnamefont {C.}~\bibnamefont {Gale}},
  \bibinfo {author} {\bibfnamefont {S.}~\bibnamefont {Jeon}}, \ and\ \bibinfo
  {author} {\bibfnamefont {J.}~\bibnamefont {Noronha}},\ }\href {\doibase
  10.1103/PhysRevC.88.064901} {\bibfield  {journal} {\bibinfo  {journal} {Phys.
  Rev. C}\ }\textbf {\bibinfo {volume} {88}},\ \bibinfo {pages} {064901}
  (\bibinfo {year} {2013})},\ \Eprint {http://arxiv.org/abs/1308.1923}
  {arXiv:1308.1923 [nucl-th]} \BibitemShut {NoStop}%
\bibitem [{\citenamefont {Haas}\ \emph {et~al.}(2014)\citenamefont {Haas},
  \citenamefont {Fister},\ and\ \citenamefont {Pawlowski}}]{Haas:2013hpa}%
  \BibitemOpen
  \bibfield  {author} {\bibinfo {author} {\bibfnamefont {M.}~\bibnamefont
  {Haas}}, \bibinfo {author} {\bibfnamefont {L.}~\bibnamefont {Fister}}, \ and\
  \bibinfo {author} {\bibfnamefont {J.~M.}\ \bibnamefont {Pawlowski}},\ }\href
  {\doibase 10.1103/PhysRevD.90.091501} {\bibfield  {journal} {\bibinfo
  {journal} {Phys. Rev. D}\ }\textbf {\bibinfo {volume} {90}},\ \bibinfo
  {pages} {091501} (\bibinfo {year} {2014})},\ \Eprint
  {http://arxiv.org/abs/1308.4960} {arXiv:1308.4960 [hep-ph]} \BibitemShut
  {NoStop}%
\bibitem [{\citenamefont {Christiansen}\ \emph {et~al.}(2015)\citenamefont
  {Christiansen}, \citenamefont {Haas}, \citenamefont {Pawlowski},\ and\
  \citenamefont {Strodthoff}}]{Christiansen:2014ypa}%
  \BibitemOpen
  \bibfield  {author} {\bibinfo {author} {\bibfnamefont {N.}~\bibnamefont
  {Christiansen}}, \bibinfo {author} {\bibfnamefont {M.}~\bibnamefont {Haas}},
  \bibinfo {author} {\bibfnamefont {J.~M.}\ \bibnamefont {Pawlowski}}, \ and\
  \bibinfo {author} {\bibfnamefont {N.}~\bibnamefont {Strodthoff}},\ }\href
  {\doibase 10.1103/PhysRevLett.115.112002} {\bibfield  {journal} {\bibinfo
  {journal} {Phys. Rev. Lett.}\ }\textbf {\bibinfo {volume} {115}},\ \bibinfo
  {pages} {112002} (\bibinfo {year} {2015})},\ \Eprint
  {http://arxiv.org/abs/1411.7986} {arXiv:1411.7986 [hep-ph]} \BibitemShut
  {NoStop}%
\bibitem [{\citenamefont {Aarts}\ \emph {et~al.}(2015)\citenamefont {Aarts},
  \citenamefont {Allton}, \citenamefont {Amato}, \citenamefont {Giudice},
  \citenamefont {Hands},\ and\ \citenamefont {Skullerud}}]{Aarts:2014nba}%
  \BibitemOpen
  \bibfield  {author} {\bibinfo {author} {\bibfnamefont {G.}~\bibnamefont
  {Aarts}}, \bibinfo {author} {\bibfnamefont {C.}~\bibnamefont {Allton}},
  \bibinfo {author} {\bibfnamefont {A.}~\bibnamefont {Amato}}, \bibinfo
  {author} {\bibfnamefont {P.}~\bibnamefont {Giudice}}, \bibinfo {author}
  {\bibfnamefont {S.}~\bibnamefont {Hands}}, \ and\ \bibinfo {author}
  {\bibfnamefont {J.-I.}\ \bibnamefont {Skullerud}},\ }\href {\doibase
  10.1007/JHEP02(2015)186} {\bibfield  {journal} {\bibinfo  {journal} {JHEP}\
  }\textbf {\bibinfo {volume} {02}},\ \bibinfo {pages} {186} (\bibinfo {year}
  {2015})},\ \Eprint {http://arxiv.org/abs/1412.6411} {arXiv:1412.6411
  [hep-lat]} \BibitemShut {NoStop}%
\bibitem [{\citenamefont {Kadam}\ and\ \citenamefont
  {Mishra}(2014)}]{Kadam:2014cua}%
  \BibitemOpen
  \bibfield  {author} {\bibinfo {author} {\bibfnamefont {G.~P.}\ \bibnamefont
  {Kadam}}\ and\ \bibinfo {author} {\bibfnamefont {H.}~\bibnamefont {Mishra}},\
  }\href {\doibase 10.1016/j.nuclphysa.2014.12.004} {\bibfield  {journal}
  {\bibinfo  {journal} {Nucl. Phys. A}\ }\textbf {\bibinfo {volume} {934}},\
  \bibinfo {pages} {133} (\bibinfo {year} {2014})},\ \Eprint
  {http://arxiv.org/abs/1408.6329} {arXiv:1408.6329 [hep-ph]} \BibitemShut
  {NoStop}%
\bibitem [{\citenamefont {Rose}\ \emph {et~al.}(2018)\citenamefont {Rose},
  \citenamefont {Torres-Rincon}, \citenamefont {Sch\"afer}, \citenamefont
  {Oliinychenko},\ and\ \citenamefont {Petersen}}]{Rose:2017bjz}%
  \BibitemOpen
  \bibfield  {author} {\bibinfo {author} {\bibfnamefont {J.~B.}\ \bibnamefont
  {Rose}}, \bibinfo {author} {\bibfnamefont {J.~M.}\ \bibnamefont
  {Torres-Rincon}}, \bibinfo {author} {\bibfnamefont {A.}~\bibnamefont
  {Sch\"afer}}, \bibinfo {author} {\bibfnamefont {D.~R.}\ \bibnamefont
  {Oliinychenko}}, \ and\ \bibinfo {author} {\bibfnamefont {H.}~\bibnamefont
  {Petersen}},\ }\href {\doibase 10.1103/PhysRevC.97.055204} {\bibfield
  {journal} {\bibinfo  {journal} {Phys. Rev. C}\ }\textbf {\bibinfo {volume}
  {97}},\ \bibinfo {pages} {055204} (\bibinfo {year} {2018})},\ \Eprint
  {http://arxiv.org/abs/1709.03826} {arXiv:1709.03826 [nucl-th]} \BibitemShut
  {NoStop}%
\bibitem [{\citenamefont {Fotakis}\ \emph {et~al.}(2020)\citenamefont
  {Fotakis}, \citenamefont {Greif}, \citenamefont {Greiner}, \citenamefont
  {Denicol},\ and\ \citenamefont {Niemi}}]{Fotakis:2019nbq}%
  \BibitemOpen
  \bibfield  {author} {\bibinfo {author} {\bibfnamefont {J.~A.}\ \bibnamefont
  {Fotakis}}, \bibinfo {author} {\bibfnamefont {M.}~\bibnamefont {Greif}},
  \bibinfo {author} {\bibfnamefont {C.}~\bibnamefont {Greiner}}, \bibinfo
  {author} {\bibfnamefont {G.~S.}\ \bibnamefont {Denicol}}, \ and\ \bibinfo
  {author} {\bibfnamefont {H.}~\bibnamefont {Niemi}},\ }\href {\doibase
  10.1103/PhysRevD.101.076007} {\bibfield  {journal} {\bibinfo  {journal}
  {Phys. Rev. D}\ }\textbf {\bibinfo {volume} {101}},\ \bibinfo {pages}
  {076007} (\bibinfo {year} {2020})},\ \Eprint
  {http://arxiv.org/abs/1912.09103} {arXiv:1912.09103 [hep-ph]} \BibitemShut
  {NoStop}%
\bibitem [{\citenamefont {Fotakis}\ \emph {et~al.}(2021)\citenamefont
  {Fotakis}, \citenamefont {Soloveva}, \citenamefont {Greiner}, \citenamefont
  {Kaczmarek},\ and\ \citenamefont {Bratkovskaya}}]{Fotakis:2021diq}%
  \BibitemOpen
  \bibfield  {author} {\bibinfo {author} {\bibfnamefont {J.~A.}\ \bibnamefont
  {Fotakis}}, \bibinfo {author} {\bibfnamefont {O.}~\bibnamefont {Soloveva}},
  \bibinfo {author} {\bibfnamefont {C.}~\bibnamefont {Greiner}}, \bibinfo
  {author} {\bibfnamefont {O.}~\bibnamefont {Kaczmarek}}, \ and\ \bibinfo
  {author} {\bibfnamefont {E.}~\bibnamefont {Bratkovskaya}},\ }\href {\doibase
  10.1103/PhysRevD.104.034014} {\bibfield  {journal} {\bibinfo  {journal}
  {Phys. Rev. D}\ }\textbf {\bibinfo {volume} {104}},\ \bibinfo {pages}
  {034014} (\bibinfo {year} {2021})},\ \Eprint
  {http://arxiv.org/abs/2102.08140} {arXiv:2102.08140 [hep-ph]} \BibitemShut
  {NoStop}%
\bibitem [{\citenamefont {McLaughlin}\ \emph {et~al.}(2022)\citenamefont
  {McLaughlin}, \citenamefont {Rose}, \citenamefont {Dore}, \citenamefont
  {Parotto}, \citenamefont {Ratti},\ and\ \citenamefont
  {Noronha-Hostler}}]{McLaughlin:2021dph}%
  \BibitemOpen
  \bibfield  {author} {\bibinfo {author} {\bibfnamefont {E.}~\bibnamefont
  {McLaughlin}}, \bibinfo {author} {\bibfnamefont {J.}~\bibnamefont {Rose}},
  \bibinfo {author} {\bibfnamefont {T.}~\bibnamefont {Dore}}, \bibinfo {author}
  {\bibfnamefont {P.}~\bibnamefont {Parotto}}, \bibinfo {author} {\bibfnamefont
  {C.}~\bibnamefont {Ratti}}, \ and\ \bibinfo {author} {\bibfnamefont
  {J.}~\bibnamefont {Noronha-Hostler}},\ }\href {\doibase
  10.1103/PhysRevC.105.024903} {\bibfield  {journal} {\bibinfo  {journal}
  {Phys. Rev. C}\ }\textbf {\bibinfo {volume} {105}},\ \bibinfo {pages}
  {024903} (\bibinfo {year} {2022})},\ \Eprint
  {http://arxiv.org/abs/2103.02090} {arXiv:2103.02090 [nucl-th]} \BibitemShut
  {NoStop}%
\bibitem [{\citenamefont {Soloveva}\ \emph {et~al.}(2021)\citenamefont
  {Soloveva}, \citenamefont {Aichelin},\ and\ \citenamefont
  {Bratkovskaya}}]{Soloveva:2021quj}%
  \BibitemOpen
  \bibfield  {author} {\bibinfo {author} {\bibfnamefont {O.}~\bibnamefont
  {Soloveva}}, \bibinfo {author} {\bibfnamefont {J.}~\bibnamefont {Aichelin}},
  \ and\ \bibinfo {author} {\bibfnamefont {E.}~\bibnamefont {Bratkovskaya}},\
  }\href@noop {} {\  (\bibinfo {year} {2021})},\ \Eprint
  {http://arxiv.org/abs/2108.08561} {arXiv:2108.08561 [hep-ph]} \BibitemShut
  {NoStop}%
\bibitem [{\citenamefont {Kadam}(2022)}]{Kadam:2022xcc}%
  \BibitemOpen
  \bibfield  {author} {\bibinfo {author} {\bibfnamefont {G.}~\bibnamefont
  {Kadam}},\ }\href@noop {} {\  (\bibinfo {year} {2022})},\ \Eprint
  {http://arxiv.org/abs/2201.06816} {arXiv:2201.06816 [nucl-th]} \BibitemShut
  {NoStop}%
\bibitem [{\citenamefont {Kadam}\ \emph {et~al.}(2021)\citenamefont {Kadam},
  \citenamefont {Mishra},\ and\ \citenamefont {Panero}}]{Kadam:2020utt}%
  \BibitemOpen
  \bibfield  {author} {\bibinfo {author} {\bibfnamefont {G.}~\bibnamefont
  {Kadam}}, \bibinfo {author} {\bibfnamefont {H.}~\bibnamefont {Mishra}}, \
  and\ \bibinfo {author} {\bibfnamefont {M.}~\bibnamefont {Panero}},\ }\href
  {\doibase 10.1140/epjc/s10052-021-09596-6} {\bibfield  {journal} {\bibinfo
  {journal} {Eur. Phys. J. C}\ }\textbf {\bibinfo {volume} {81}},\ \bibinfo
  {pages} {795} (\bibinfo {year} {2021})},\ \Eprint
  {http://arxiv.org/abs/2011.02171} {arXiv:2011.02171 [hep-ph]} \BibitemShut
  {NoStop}%
\bibitem [{\citenamefont {Borsanyi}\ \emph {et~al.}(2012)\citenamefont
  {Borsanyi}, \citenamefont {Fodor}, \citenamefont {Katz}, \citenamefont
  {Krieg}, \citenamefont {Ratti},\ and\ \citenamefont
  {Szabo}}]{Borsanyi:2011sw}%
  \BibitemOpen
  \bibfield  {author} {\bibinfo {author} {\bibfnamefont {S.}~\bibnamefont
  {Borsanyi}}, \bibinfo {author} {\bibfnamefont {Z.}~\bibnamefont {Fodor}},
  \bibinfo {author} {\bibfnamefont {S.~D.}\ \bibnamefont {Katz}}, \bibinfo
  {author} {\bibfnamefont {S.}~\bibnamefont {Krieg}}, \bibinfo {author}
  {\bibfnamefont {C.}~\bibnamefont {Ratti}}, \ and\ \bibinfo {author}
  {\bibfnamefont {K.}~\bibnamefont {Szabo}},\ }\href {\doibase
  10.1007/JHEP01(2012)138} {\bibfield  {journal} {\bibinfo  {journal} {JHEP}\
  }\textbf {\bibinfo {volume} {01}},\ \bibinfo {pages} {138} (\bibinfo {year}
  {2012})},\ \Eprint {http://arxiv.org/abs/1112.4416} {arXiv:1112.4416
  [hep-lat]} \BibitemShut {NoStop}%
\bibitem [{\citenamefont {Bazavov}\ \emph {et~al.}(2014)\citenamefont {Bazavov}
  \emph {et~al.}}]{HotQCD:2014kol}%
  \BibitemOpen
  \bibfield  {author} {\bibinfo {author} {\bibfnamefont {A.}~\bibnamefont
  {Bazavov}} \emph {et~al.} (\bibinfo {collaboration} {HotQCD}),\ }\href
  {\doibase 10.1103/PhysRevD.90.094503} {\bibfield  {journal} {\bibinfo
  {journal} {Phys. Rev. D}\ }\textbf {\bibinfo {volume} {90}},\ \bibinfo
  {pages} {094503} (\bibinfo {year} {2014})},\ \Eprint
  {http://arxiv.org/abs/1407.6387} {arXiv:1407.6387 [hep-lat]} \BibitemShut
  {NoStop}%
\bibitem [{\citenamefont {Buchel}\ and\ \citenamefont
  {Liu}(2004)}]{Buchel:2003tz}%
  \BibitemOpen
  \bibfield  {author} {\bibinfo {author} {\bibfnamefont {A.}~\bibnamefont
  {Buchel}}\ and\ \bibinfo {author} {\bibfnamefont {J.~T.}\ \bibnamefont
  {Liu}},\ }\href {\doibase 10.1103/PhysRevLett.93.090602} {\bibfield
  {journal} {\bibinfo  {journal} {Phys. Rev. Lett.}\ }\textbf {\bibinfo
  {volume} {93}},\ \bibinfo {pages} {090602} (\bibinfo {year} {2004})},\
  \Eprint {http://arxiv.org/abs/hep-th/0311175} {arXiv:hep-th/0311175}
  \BibitemShut {NoStop}%
\bibitem [{\citenamefont {Liao}\ and\ \citenamefont
  {Koch}(2010)}]{Liao:2009gb}%
  \BibitemOpen
  \bibfield  {author} {\bibinfo {author} {\bibfnamefont {J.}~\bibnamefont
  {Liao}}\ and\ \bibinfo {author} {\bibfnamefont {V.}~\bibnamefont {Koch}},\
  }\href {\doibase 10.1103/PhysRevC.81.014902} {\bibfield  {journal} {\bibinfo
  {journal} {Phys. Rev. C}\ }\textbf {\bibinfo {volume} {81}},\ \bibinfo
  {pages} {014902} (\bibinfo {year} {2010})},\ \Eprint
  {http://arxiv.org/abs/0909.3105} {arXiv:0909.3105 [hep-ph]} \BibitemShut
  {NoStop}%
\bibitem [{\citenamefont {Gubser}(2006)}]{Gubser:2006bz}%
  \BibitemOpen
  \bibfield  {author} {\bibinfo {author} {\bibfnamefont {S.~S.}\ \bibnamefont
  {Gubser}},\ }\href {\doibase 10.1103/PhysRevD.74.126005} {\bibfield
  {journal} {\bibinfo  {journal} {Phys. Rev. D}\ }\textbf {\bibinfo {volume}
  {74}},\ \bibinfo {pages} {126005} (\bibinfo {year} {2006})},\ \Eprint
  {http://arxiv.org/abs/hep-th/0605182} {arXiv:hep-th/0605182} \BibitemShut
  {NoStop}%
\bibitem [{\citenamefont {Herzog}\ \emph {et~al.}(2006)\citenamefont {Herzog},
  \citenamefont {Karch}, \citenamefont {Kovtun}, \citenamefont {Kozcaz},\ and\
  \citenamefont {Yaffe}}]{Herzog:2006gh}%
  \BibitemOpen
  \bibfield  {author} {\bibinfo {author} {\bibfnamefont {C.~P.}\ \bibnamefont
  {Herzog}}, \bibinfo {author} {\bibfnamefont {A.}~\bibnamefont {Karch}},
  \bibinfo {author} {\bibfnamefont {P.}~\bibnamefont {Kovtun}}, \bibinfo
  {author} {\bibfnamefont {C.}~\bibnamefont {Kozcaz}}, \ and\ \bibinfo {author}
  {\bibfnamefont {L.~G.}\ \bibnamefont {Yaffe}},\ }\href {\doibase
  10.1088/1126-6708/2006/07/013} {\bibfield  {journal} {\bibinfo  {journal}
  {JHEP}\ }\textbf {\bibinfo {volume} {07}},\ \bibinfo {pages} {013} (\bibinfo
  {year} {2006})},\ \Eprint {http://arxiv.org/abs/hep-th/0605158}
  {arXiv:hep-th/0605158} \BibitemShut {NoStop}%
\bibitem [{\citenamefont {Casalderrey-Solana}\ and\ \citenamefont
  {Teaney}(2006)}]{Casalderrey-Solana:2006fio}%
  \BibitemOpen
  \bibfield  {author} {\bibinfo {author} {\bibfnamefont {J.}~\bibnamefont
  {Casalderrey-Solana}}\ and\ \bibinfo {author} {\bibfnamefont
  {D.}~\bibnamefont {Teaney}},\ }\href {\doibase 10.1103/PhysRevD.74.085012}
  {\bibfield  {journal} {\bibinfo  {journal} {Phys. Rev. D}\ }\textbf {\bibinfo
  {volume} {74}},\ \bibinfo {pages} {085012} (\bibinfo {year} {2006})},\
  \Eprint {http://arxiv.org/abs/hep-ph/0605199} {arXiv:hep-ph/0605199}
  \BibitemShut {NoStop}%
\bibitem [{\citenamefont {Gubser}(2007)}]{Gubser:2006qh}%
  \BibitemOpen
  \bibfield  {author} {\bibinfo {author} {\bibfnamefont {S.~S.}\ \bibnamefont
  {Gubser}},\ }\href {\doibase 10.1103/PhysRevD.76.126003} {\bibfield
  {journal} {\bibinfo  {journal} {Phys. Rev. D}\ }\textbf {\bibinfo {volume}
  {76}},\ \bibinfo {pages} {126003} (\bibinfo {year} {2007})},\ \Eprint
  {http://arxiv.org/abs/hep-th/0611272} {arXiv:hep-th/0611272} \BibitemShut
  {NoStop}%
\bibitem [{\citenamefont {Gubser}(2008)}]{Gubser:2006nz}%
  \BibitemOpen
  \bibfield  {author} {\bibinfo {author} {\bibfnamefont {S.~S.}\ \bibnamefont
  {Gubser}},\ }\href {\doibase 10.1016/j.nuclphysb.2007.09.017} {\bibfield
  {journal} {\bibinfo  {journal} {Nucl. Phys. B}\ }\textbf {\bibinfo {volume}
  {790}},\ \bibinfo {pages} {175} (\bibinfo {year} {2008})},\ \Eprint
  {http://arxiv.org/abs/hep-th/0612143} {arXiv:hep-th/0612143} \BibitemShut
  {NoStop}%
\bibitem [{\citenamefont {Liu}\ \emph {et~al.}(2006)\citenamefont {Liu},
  \citenamefont {Rajagopal},\ and\ \citenamefont {Wiedemann}}]{Liu:2006ug}%
  \BibitemOpen
  \bibfield  {author} {\bibinfo {author} {\bibfnamefont {H.}~\bibnamefont
  {Liu}}, \bibinfo {author} {\bibfnamefont {K.}~\bibnamefont {Rajagopal}}, \
  and\ \bibinfo {author} {\bibfnamefont {U.~A.}\ \bibnamefont {Wiedemann}},\
  }\href {\doibase 10.1103/PhysRevLett.97.182301} {\bibfield  {journal}
  {\bibinfo  {journal} {Phys. Rev. Lett.}\ }\textbf {\bibinfo {volume} {97}},\
  \bibinfo {pages} {182301} (\bibinfo {year} {2006})},\ \Eprint
  {http://arxiv.org/abs/hep-ph/0605178} {arXiv:hep-ph/0605178} \BibitemShut
  {NoStop}%
\bibitem [{\citenamefont {Liu}\ \emph {et~al.}(2007)\citenamefont {Liu},
  \citenamefont {Rajagopal},\ and\ \citenamefont {Wiedemann}}]{Liu:2006he}%
  \BibitemOpen
  \bibfield  {author} {\bibinfo {author} {\bibfnamefont {H.}~\bibnamefont
  {Liu}}, \bibinfo {author} {\bibfnamefont {K.}~\bibnamefont {Rajagopal}}, \
  and\ \bibinfo {author} {\bibfnamefont {U.~A.}\ \bibnamefont {Wiedemann}},\
  }\href {\doibase 10.1088/1126-6708/2007/03/066} {\bibfield  {journal}
  {\bibinfo  {journal} {JHEP}\ }\textbf {\bibinfo {volume} {03}},\ \bibinfo
  {pages} {066} (\bibinfo {year} {2007})},\ \Eprint
  {http://arxiv.org/abs/hep-ph/0612168} {arXiv:hep-ph/0612168} \BibitemShut
  {NoStop}%
\bibitem [{\citenamefont {Casalderrey-Solana}\ and\ \citenamefont
  {Teaney}(2007)}]{Casalderrey-Solana:2007ahi}%
  \BibitemOpen
  \bibfield  {author} {\bibinfo {author} {\bibfnamefont {J.}~\bibnamefont
  {Casalderrey-Solana}}\ and\ \bibinfo {author} {\bibfnamefont
  {D.}~\bibnamefont {Teaney}},\ }\href {\doibase 10.1088/1126-6708/2007/04/039}
  {\bibfield  {journal} {\bibinfo  {journal} {JHEP}\ }\textbf {\bibinfo
  {volume} {04}},\ \bibinfo {pages} {039} (\bibinfo {year} {2007})},\ \Eprint
  {http://arxiv.org/abs/hep-th/0701123} {arXiv:hep-th/0701123} \BibitemShut
  {NoStop}%
\bibitem [{\citenamefont {D'Eramo}\ \emph {et~al.}(2011)\citenamefont
  {D'Eramo}, \citenamefont {Liu},\ and\ \citenamefont
  {Rajagopal}}]{DEramo:2010wup}%
  \BibitemOpen
  \bibfield  {author} {\bibinfo {author} {\bibfnamefont {F.}~\bibnamefont
  {D'Eramo}}, \bibinfo {author} {\bibfnamefont {H.}~\bibnamefont {Liu}}, \ and\
  \bibinfo {author} {\bibfnamefont {K.}~\bibnamefont {Rajagopal}},\ }\href
  {\doibase 10.1103/PhysRevD.84.065015} {\bibfield  {journal} {\bibinfo
  {journal} {Phys. Rev. D}\ }\textbf {\bibinfo {volume} {84}},\ \bibinfo
  {pages} {065015} (\bibinfo {year} {2011})},\ \Eprint
  {http://arxiv.org/abs/1006.1367} {arXiv:1006.1367 [hep-ph]} \BibitemShut
  {NoStop}%
\bibitem [{\citenamefont {Kiritsis}\ and\ \citenamefont
  {Pavlopoulos}(2012)}]{Kiritsis:2011ha}%
  \BibitemOpen
  \bibfield  {author} {\bibinfo {author} {\bibfnamefont {E.}~\bibnamefont
  {Kiritsis}}\ and\ \bibinfo {author} {\bibfnamefont {G.}~\bibnamefont
  {Pavlopoulos}},\ }\href {\doibase 10.1007/JHEP04(2012)096} {\bibfield
  {journal} {\bibinfo  {journal} {JHEP}\ }\textbf {\bibinfo {volume} {04}},\
  \bibinfo {pages} {096} (\bibinfo {year} {2012})},\ \Eprint
  {http://arxiv.org/abs/1111.0314} {arXiv:1111.0314 [hep-th]} \BibitemShut
  {NoStop}%
\bibitem [{\citenamefont {Greif}\ \emph {et~al.}(2018)\citenamefont {Greif},
  \citenamefont {Fotakis}, \citenamefont {Denicol},\ and\ \citenamefont
  {Greiner}}]{Greif:2017byw}%
  \BibitemOpen
  \bibfield  {author} {\bibinfo {author} {\bibfnamefont {M.}~\bibnamefont
  {Greif}}, \bibinfo {author} {\bibfnamefont {J.~A.}\ \bibnamefont {Fotakis}},
  \bibinfo {author} {\bibfnamefont {G.~S.}\ \bibnamefont {Denicol}}, \ and\
  \bibinfo {author} {\bibfnamefont {C.}~\bibnamefont {Greiner}},\ }\href
  {\doibase 10.1103/PhysRevLett.120.242301} {\bibfield  {journal} {\bibinfo
  {journal} {Phys. Rev. Lett.}\ }\textbf {\bibinfo {volume} {120}},\ \bibinfo
  {pages} {242301} (\bibinfo {year} {2018})},\ \Eprint
  {http://arxiv.org/abs/1711.08680} {arXiv:1711.08680 [hep-ph]} \BibitemShut
  {NoStop}%
\bibitem [{\citenamefont {Hohenberg}\ and\ \citenamefont
  {Halperin}(1977)}]{Hohenberg:1977ym}%
  \BibitemOpen
  \bibfield  {author} {\bibinfo {author} {\bibfnamefont {P.~C.}\ \bibnamefont
  {Hohenberg}}\ and\ \bibinfo {author} {\bibfnamefont {B.~I.}\ \bibnamefont
  {Halperin}},\ }\href {\doibase 10.1103/RevModPhys.49.435} {\bibfield
  {journal} {\bibinfo  {journal} {Rev. Mod. Phys.}\ }\textbf {\bibinfo {volume}
  {49}},\ \bibinfo {pages} {435} (\bibinfo {year} {1977})}\BibitemShut
  {NoStop}%
\bibitem [{\citenamefont {Iqbal}\ and\ \citenamefont
  {Liu}(2009)}]{Iqbal:2008by}%
  \BibitemOpen
  \bibfield  {author} {\bibinfo {author} {\bibfnamefont {N.}~\bibnamefont
  {Iqbal}}\ and\ \bibinfo {author} {\bibfnamefont {H.}~\bibnamefont {Liu}},\
  }\href {\doibase 10.1103/PhysRevD.79.025023} {\bibfield  {journal} {\bibinfo
  {journal} {Phys. Rev. D}\ }\textbf {\bibinfo {volume} {79}},\ \bibinfo
  {pages} {025023} (\bibinfo {year} {2009})},\ \Eprint
  {http://arxiv.org/abs/0809.3808} {arXiv:0809.3808 [hep-th]} \BibitemShut
  {NoStop}%
\bibitem [{\citenamefont {Kapusta}\ and\ \citenamefont
  {Torres-Rincon}(2012)}]{Kapusta:2012zb}%
  \BibitemOpen
  \bibfield  {author} {\bibinfo {author} {\bibfnamefont {J.~I.}\ \bibnamefont
  {Kapusta}}\ and\ \bibinfo {author} {\bibfnamefont {J.~M.}\ \bibnamefont
  {Torres-Rincon}},\ }\href {\doibase 10.1103/PhysRevC.86.054911} {\bibfield
  {journal} {\bibinfo  {journal} {Phys. Rev. C}\ }\textbf {\bibinfo {volume}
  {86}},\ \bibinfo {pages} {054911} (\bibinfo {year} {2012})},\ \Eprint
  {http://arxiv.org/abs/1209.0675} {arXiv:1209.0675 [nucl-th]} \BibitemShut
  {NoStop}%
\bibitem [{\citenamefont {Jain}(2010)}]{Jain:2009pw}%
  \BibitemOpen
  \bibfield  {author} {\bibinfo {author} {\bibfnamefont {S.}~\bibnamefont
  {Jain}},\ }\href {\doibase 10.1007/JHEP03(2010)101} {\bibfield  {journal}
  {\bibinfo  {journal} {JHEP}\ }\textbf {\bibinfo {volume} {03}},\ \bibinfo
  {pages} {101} (\bibinfo {year} {2010})},\ \Eprint
  {http://arxiv.org/abs/0912.2228} {arXiv:0912.2228 [hep-th]} \BibitemShut
  {NoStop}%
\bibitem [{\citenamefont {Adamczyk}\ \emph {et~al.}(2017)\citenamefont
  {Adamczyk} \emph {et~al.}}]{STAR:2017sal}%
  \BibitemOpen
  \bibfield  {author} {\bibinfo {author} {\bibfnamefont {L.}~\bibnamefont
  {Adamczyk}} \emph {et~al.} (\bibinfo {collaboration} {STAR}),\ }\href
  {\doibase 10.1103/PhysRevC.96.044904} {\bibfield  {journal} {\bibinfo
  {journal} {Phys. Rev. C}\ }\textbf {\bibinfo {volume} {96}},\ \bibinfo
  {pages} {044904} (\bibinfo {year} {2017})},\ \Eprint
  {http://arxiv.org/abs/1701.07065} {arXiv:1701.07065 [nucl-ex]} \BibitemShut
  {NoStop}%
\bibitem [{\citenamefont {Adamczyk}\ \emph {et~al.}(2018)\citenamefont
  {Adamczyk} \emph {et~al.}}]{STAR:2017ieb}%
  \BibitemOpen
  \bibfield  {author} {\bibinfo {author} {\bibfnamefont {L.}~\bibnamefont
  {Adamczyk}} \emph {et~al.} (\bibinfo {collaboration} {STAR}),\ }\href
  {\doibase 10.1103/PhysRevLett.121.032301} {\bibfield  {journal} {\bibinfo
  {journal} {Phys. Rev. Lett.}\ }\textbf {\bibinfo {volume} {121}},\ \bibinfo
  {pages} {032301} (\bibinfo {year} {2018})},\ \Eprint
  {http://arxiv.org/abs/1707.01988} {arXiv:1707.01988 [nucl-ex]} \BibitemShut
  {NoStop}%
\bibitem [{\citenamefont {McLaughlin}\ \emph {et~al.}(2021)\citenamefont
  {McLaughlin}, \citenamefont {Rose}, \citenamefont {Parotto}, \citenamefont
  {Ratti},\ and\ \citenamefont {Noronha-Hostler}}]{McLaughlin:2021dlq}%
  \BibitemOpen
  \bibfield  {author} {\bibinfo {author} {\bibfnamefont {E.}~\bibnamefont
  {McLaughlin}}, \bibinfo {author} {\bibfnamefont {J.}~\bibnamefont {Rose}},
  \bibinfo {author} {\bibfnamefont {P.}~\bibnamefont {Parotto}}, \bibinfo
  {author} {\bibfnamefont {C.}~\bibnamefont {Ratti}}, \ and\ \bibinfo {author}
  {\bibfnamefont {J.}~\bibnamefont {Noronha-Hostler}},\ }\href@noop {} {\
  (\bibinfo {year} {2021})},\ \Eprint {http://arxiv.org/abs/2103.03329}
  {arXiv:2103.03329 [nucl-th]} \BibitemShut {NoStop}%
\bibitem [{\citenamefont {Xu}\ \emph {et~al.}(2018)\citenamefont {Xu},
  \citenamefont {Bernhard}, \citenamefont {Bass}, \citenamefont {Nahrgang},\
  and\ \citenamefont {Cao}}]{Xu:2017obm}%
  \BibitemOpen
  \bibfield  {author} {\bibinfo {author} {\bibfnamefont {Y.}~\bibnamefont
  {Xu}}, \bibinfo {author} {\bibfnamefont {J.~E.}\ \bibnamefont {Bernhard}},
  \bibinfo {author} {\bibfnamefont {S.~A.}\ \bibnamefont {Bass}}, \bibinfo
  {author} {\bibfnamefont {M.}~\bibnamefont {Nahrgang}}, \ and\ \bibinfo
  {author} {\bibfnamefont {S.}~\bibnamefont {Cao}},\ }\href {\doibase
  10.1103/PhysRevC.97.014907} {\bibfield  {journal} {\bibinfo  {journal} {Phys.
  Rev. C}\ }\textbf {\bibinfo {volume} {97}},\ \bibinfo {pages} {014907}
  (\bibinfo {year} {2018})},\ \Eprint {http://arxiv.org/abs/1710.00807}
  {arXiv:1710.00807 [nucl-th]} \BibitemShut {NoStop}%
\bibitem [{\citenamefont {Dorau}\ \emph {et~al.}(2020)\citenamefont {Dorau},
  \citenamefont {Rose}, \citenamefont {Pablos},\ and\ \citenamefont
  {Elfner}}]{Dorau:2019ozd}%
  \BibitemOpen
  \bibfield  {author} {\bibinfo {author} {\bibfnamefont {P.}~\bibnamefont
  {Dorau}}, \bibinfo {author} {\bibfnamefont {J.-B.}\ \bibnamefont {Rose}},
  \bibinfo {author} {\bibfnamefont {D.}~\bibnamefont {Pablos}}, \ and\ \bibinfo
  {author} {\bibfnamefont {H.}~\bibnamefont {Elfner}},\ }\href {\doibase
  10.1103/PhysRevC.101.035208} {\bibfield  {journal} {\bibinfo  {journal}
  {Phys. Rev. C}\ }\textbf {\bibinfo {volume} {101}},\ \bibinfo {pages}
  {035208} (\bibinfo {year} {2020})},\ \Eprint
  {http://arxiv.org/abs/1910.07027} {arXiv:1910.07027 [nucl-th]} \BibitemShut
  {NoStop}%
\bibitem [{\citenamefont {Burke}\ \emph {et~al.}(2014)\citenamefont {Burke}
  \emph {et~al.}}]{JET:2013cls}%
  \BibitemOpen
  \bibfield  {author} {\bibinfo {author} {\bibfnamefont {K.~M.}\ \bibnamefont
  {Burke}} \emph {et~al.} (\bibinfo {collaboration} {JET}),\ }\href {\doibase
  10.1103/PhysRevC.90.014909} {\bibfield  {journal} {\bibinfo  {journal} {Phys.
  Rev. C}\ }\textbf {\bibinfo {volume} {90}},\ \bibinfo {pages} {014909}
  (\bibinfo {year} {2014})},\ \Eprint {http://arxiv.org/abs/1312.5003}
  {arXiv:1312.5003 [nucl-th]} \BibitemShut {NoStop}%
\bibitem [{\citenamefont {Shi}\ \emph {et~al.}(2019)\citenamefont {Shi},
  \citenamefont {Liao},\ and\ \citenamefont {Gyulassy}}]{Shi:2018izg}%
  \BibitemOpen
  \bibfield  {author} {\bibinfo {author} {\bibfnamefont {S.}~\bibnamefont
  {Shi}}, \bibinfo {author} {\bibfnamefont {J.}~\bibnamefont {Liao}}, \ and\
  \bibinfo {author} {\bibfnamefont {M.}~\bibnamefont {Gyulassy}},\ }\href
  {\doibase 10.1088/1674-1137/43/4/044101} {\bibfield  {journal} {\bibinfo
  {journal} {Chin. Phys. C}\ }\textbf {\bibinfo {volume} {43}},\ \bibinfo
  {pages} {044101} (\bibinfo {year} {2019})},\ \Eprint
  {http://arxiv.org/abs/1808.05461} {arXiv:1808.05461 [hep-ph]} \BibitemShut
  {NoStop}%
\bibitem [{\citenamefont {Baier}(2003)}]{Baier:2002tc}%
  \BibitemOpen
  \bibfield  {author} {\bibinfo {author} {\bibfnamefont {R.}~\bibnamefont
  {Baier}},\ }\href {\doibase 10.1016/S0375-9474(02)01429-X} {\bibfield
  {journal} {\bibinfo  {journal} {Nucl. Phys. A}\ }\textbf {\bibinfo {volume}
  {715}},\ \bibinfo {pages} {209} (\bibinfo {year} {2003})},\ \Eprint
  {http://arxiv.org/abs/hep-ph/0209038} {arXiv:hep-ph/0209038} \BibitemShut
  {NoStop}%
\bibitem [{\citenamefont {Andr\'es}\ \emph {et~al.}(2016)\citenamefont
  {Andr\'es}, \citenamefont {Armesto}, \citenamefont {Luzum}, \citenamefont
  {Salgado},\ and\ \citenamefont {Zurita}}]{Andres:2016iys}%
  \BibitemOpen
  \bibfield  {author} {\bibinfo {author} {\bibfnamefont {C.}~\bibnamefont
  {Andr\'es}}, \bibinfo {author} {\bibfnamefont {N.}~\bibnamefont {Armesto}},
  \bibinfo {author} {\bibfnamefont {M.}~\bibnamefont {Luzum}}, \bibinfo
  {author} {\bibfnamefont {C.~A.}\ \bibnamefont {Salgado}}, \ and\ \bibinfo
  {author} {\bibfnamefont {P.}~\bibnamefont {Zurita}},\ }\href {\doibase
  10.1140/epjc/s10052-016-4320-5} {\bibfield  {journal} {\bibinfo  {journal}
  {Eur. Phys. J. C}\ }\textbf {\bibinfo {volume} {76}},\ \bibinfo {pages} {475}
  (\bibinfo {year} {2016})},\ \Eprint {http://arxiv.org/abs/1606.04837}
  {arXiv:1606.04837 [hep-ph]} \BibitemShut {NoStop}%
\bibitem [{\citenamefont {Moore}(2020)}]{Moore:2020pfu}%
  \BibitemOpen
  \bibfield  {author} {\bibinfo {author} {\bibfnamefont {G.~D.}\ \bibnamefont
  {Moore}},\ }in\ \href@noop {} {\emph {\bibinfo {booktitle} {{Criticality in
  QCD and the Hadron Resonance Gas}}}}\ (\bibinfo {year} {2020})\ \Eprint
  {http://arxiv.org/abs/2010.15704} {arXiv:2010.15704 [hep-ph]} \BibitemShut
  {NoStop}%
\bibitem [{\citenamefont {Bernhard}\ \emph {et~al.}(2016)\citenamefont
  {Bernhard}, \citenamefont {Moreland}, \citenamefont {Bass}, \citenamefont
  {Liu},\ and\ \citenamefont {Heinz}}]{Bernhard:2016tnd}%
  \BibitemOpen
  \bibfield  {author} {\bibinfo {author} {\bibfnamefont {J.~E.}\ \bibnamefont
  {Bernhard}}, \bibinfo {author} {\bibfnamefont {J.~S.}\ \bibnamefont
  {Moreland}}, \bibinfo {author} {\bibfnamefont {S.~A.}\ \bibnamefont {Bass}},
  \bibinfo {author} {\bibfnamefont {J.}~\bibnamefont {Liu}}, \ and\ \bibinfo
  {author} {\bibfnamefont {U.}~\bibnamefont {Heinz}},\ }\href {\doibase
  10.1103/PhysRevC.94.024907} {\bibfield  {journal} {\bibinfo  {journal} {Phys.
  Rev. C}\ }\textbf {\bibinfo {volume} {94}},\ \bibinfo {pages} {024907}
  (\bibinfo {year} {2016})},\ \Eprint {http://arxiv.org/abs/1605.03954}
  {arXiv:1605.03954 [nucl-th]} \BibitemShut {NoStop}%
\bibitem [{\citenamefont {Everett}\ \emph
  {et~al.}(2021{\natexlab{a}})\citenamefont {Everett} \emph
  {et~al.}}]{JETSCAPE:2020shq}%
  \BibitemOpen
  \bibfield  {author} {\bibinfo {author} {\bibfnamefont {D.}~\bibnamefont
  {Everett}} \emph {et~al.} (\bibinfo {collaboration} {JETSCAPE}),\ }\href
  {\doibase 10.1103/PhysRevLett.126.242301} {\bibfield  {journal} {\bibinfo
  {journal} {Phys. Rev. Lett.}\ }\textbf {\bibinfo {volume} {126}},\ \bibinfo
  {pages} {242301} (\bibinfo {year} {2021}{\natexlab{a}})},\ \Eprint
  {http://arxiv.org/abs/2010.03928} {arXiv:2010.03928 [hep-ph]} \BibitemShut
  {NoStop}%
\bibitem [{\citenamefont {Nijs}\ and\ \citenamefont {van~der
  Schee}(2021)}]{Nijs:2021clz}%
  \BibitemOpen
  \bibfield  {author} {\bibinfo {author} {\bibfnamefont {G.}~\bibnamefont
  {Nijs}}\ and\ \bibinfo {author} {\bibfnamefont {W.}~\bibnamefont {van~der
  Schee}},\ }\href@noop {} {\  (\bibinfo {year} {2021})},\ \Eprint
  {http://arxiv.org/abs/2110.13153} {arXiv:2110.13153 [nucl-th]} \BibitemShut
  {NoStop}%
\bibitem [{\citenamefont {Son}\ and\ \citenamefont
  {Stephanov}(2004)}]{Son:2004iv}%
  \BibitemOpen
  \bibfield  {author} {\bibinfo {author} {\bibfnamefont {D.~T.}\ \bibnamefont
  {Son}}\ and\ \bibinfo {author} {\bibfnamefont {M.~A.}\ \bibnamefont
  {Stephanov}},\ }\href {\doibase 10.1103/PhysRevD.70.056001} {\bibfield
  {journal} {\bibinfo  {journal} {Phys. Rev. D}\ }\textbf {\bibinfo {volume}
  {70}},\ \bibinfo {pages} {056001} (\bibinfo {year} {2004})},\ \Eprint
  {http://arxiv.org/abs/hep-ph/0401052} {arXiv:hep-ph/0401052} \BibitemShut
  {NoStop}%
\bibitem [{\citenamefont {Feng}\ \emph {et~al.}(2018)\citenamefont {Feng},
  \citenamefont {Greiner}, \citenamefont {Shi},\ and\ \citenamefont
  {Xu}}]{Feng:2018anl}%
  \BibitemOpen
  \bibfield  {author} {\bibinfo {author} {\bibfnamefont {B.}~\bibnamefont
  {Feng}}, \bibinfo {author} {\bibfnamefont {C.}~\bibnamefont {Greiner}},
  \bibinfo {author} {\bibfnamefont {S.}~\bibnamefont {Shi}}, \ and\ \bibinfo
  {author} {\bibfnamefont {Z.}~\bibnamefont {Xu}},\ }\href {\doibase
  10.1016/j.physletb.2018.05.030} {\bibfield  {journal} {\bibinfo  {journal}
  {Phys. Lett. B}\ }\textbf {\bibinfo {volume} {782}},\ \bibinfo {pages} {262}
  (\bibinfo {year} {2018})},\ \Eprint {http://arxiv.org/abs/1802.02494}
  {arXiv:1802.02494 [hep-ph]} \BibitemShut {NoStop}%
\bibitem [{\citenamefont {Noronha-Hostler}\ \emph {et~al.}(2014)\citenamefont
  {Noronha-Hostler}, \citenamefont {Noronha},\ and\ \citenamefont
  {Grassi}}]{Noronha-Hostler:2014dqa}%
  \BibitemOpen
  \bibfield  {author} {\bibinfo {author} {\bibfnamefont {J.}~\bibnamefont
  {Noronha-Hostler}}, \bibinfo {author} {\bibfnamefont {J.}~\bibnamefont
  {Noronha}}, \ and\ \bibinfo {author} {\bibfnamefont {F.}~\bibnamefont
  {Grassi}},\ }\href {\doibase 10.1103/PhysRevC.90.034907} {\bibfield
  {journal} {\bibinfo  {journal} {Phys. Rev. C}\ }\textbf {\bibinfo {volume}
  {90}},\ \bibinfo {pages} {034907} (\bibinfo {year} {2014})},\ \Eprint
  {http://arxiv.org/abs/1406.3333} {arXiv:1406.3333 [nucl-th]} \BibitemShut
  {NoStop}%
\bibitem [{\citenamefont {Ryu}\ \emph {et~al.}(2015)\citenamefont {Ryu},
  \citenamefont {Paquet}, \citenamefont {Shen}, \citenamefont {Denicol},
  \citenamefont {Schenke}, \citenamefont {Jeon},\ and\ \citenamefont
  {Gale}}]{Ryu:2015vwa}%
  \BibitemOpen
  \bibfield  {author} {\bibinfo {author} {\bibfnamefont {S.}~\bibnamefont
  {Ryu}}, \bibinfo {author} {\bibfnamefont {J.~F.}\ \bibnamefont {Paquet}},
  \bibinfo {author} {\bibfnamefont {C.}~\bibnamefont {Shen}}, \bibinfo {author}
  {\bibfnamefont {G.~S.}\ \bibnamefont {Denicol}}, \bibinfo {author}
  {\bibfnamefont {B.}~\bibnamefont {Schenke}}, \bibinfo {author} {\bibfnamefont
  {S.}~\bibnamefont {Jeon}}, \ and\ \bibinfo {author} {\bibfnamefont
  {C.}~\bibnamefont {Gale}},\ }\href {\doibase 10.1103/PhysRevLett.115.132301}
  {\bibfield  {journal} {\bibinfo  {journal} {Phys. Rev. Lett.}\ }\textbf
  {\bibinfo {volume} {115}},\ \bibinfo {pages} {132301} (\bibinfo {year}
  {2015})},\ \Eprint {http://arxiv.org/abs/1502.01675} {arXiv:1502.01675
  [nucl-th]} \BibitemShut {NoStop}%
\bibitem [{\citenamefont {Monnai}\ and\ \citenamefont
  {Hirano}(2009)}]{Monnai:2009ad}%
  \BibitemOpen
  \bibfield  {author} {\bibinfo {author} {\bibfnamefont {A.}~\bibnamefont
  {Monnai}}\ and\ \bibinfo {author} {\bibfnamefont {T.}~\bibnamefont
  {Hirano}},\ }\href {\doibase 10.1103/PhysRevC.80.054906} {\bibfield
  {journal} {\bibinfo  {journal} {Phys. Rev. C}\ }\textbf {\bibinfo {volume}
  {80}},\ \bibinfo {pages} {054906} (\bibinfo {year} {2009})},\ \Eprint
  {http://arxiv.org/abs/0903.4436} {arXiv:0903.4436 [nucl-th]} \BibitemShut
  {NoStop}%
\bibitem [{\citenamefont {Dusling}\ and\ \citenamefont
  {Sch\"afer}(2012)}]{Dusling:2011fd}%
  \BibitemOpen
  \bibfield  {author} {\bibinfo {author} {\bibfnamefont {K.}~\bibnamefont
  {Dusling}}\ and\ \bibinfo {author} {\bibfnamefont {T.}~\bibnamefont
  {Sch\"afer}},\ }\href {\doibase 10.1103/PhysRevC.85.044909} {\bibfield
  {journal} {\bibinfo  {journal} {Phys. Rev. C}\ }\textbf {\bibinfo {volume}
  {85}},\ \bibinfo {pages} {044909} (\bibinfo {year} {2012})},\ \Eprint
  {http://arxiv.org/abs/1109.5181} {arXiv:1109.5181 [hep-ph]} \BibitemShut
  {NoStop}%
\bibitem [{\citenamefont {Noronha-Hostler}\ \emph {et~al.}(2013)\citenamefont
  {Noronha-Hostler}, \citenamefont {Denicol}, \citenamefont {Noronha},
  \citenamefont {Andrade},\ and\ \citenamefont
  {Grassi}}]{Noronha-Hostler:2013gga}%
  \BibitemOpen
  \bibfield  {author} {\bibinfo {author} {\bibfnamefont {J.}~\bibnamefont
  {Noronha-Hostler}}, \bibinfo {author} {\bibfnamefont {G.~S.}\ \bibnamefont
  {Denicol}}, \bibinfo {author} {\bibfnamefont {J.}~\bibnamefont {Noronha}},
  \bibinfo {author} {\bibfnamefont {R.~P.~G.}\ \bibnamefont {Andrade}}, \ and\
  \bibinfo {author} {\bibfnamefont {F.}~\bibnamefont {Grassi}},\ }\href
  {\doibase 10.1103/PhysRevC.88.044916} {\bibfield  {journal} {\bibinfo
  {journal} {Phys. Rev. C}\ }\textbf {\bibinfo {volume} {88}},\ \bibinfo
  {pages} {044916} (\bibinfo {year} {2013})},\ \Eprint
  {http://arxiv.org/abs/1305.1981} {arXiv:1305.1981 [nucl-th]} \BibitemShut
  {NoStop}%
\bibitem [{\citenamefont {Everett}\ \emph
  {et~al.}(2021{\natexlab{b}})\citenamefont {Everett} \emph
  {et~al.}}]{JETSCAPE:2020mzn}%
  \BibitemOpen
  \bibfield  {author} {\bibinfo {author} {\bibfnamefont {D.}~\bibnamefont
  {Everett}} \emph {et~al.} (\bibinfo {collaboration} {JETSCAPE}),\ }\href
  {\doibase 10.1103/PhysRevC.103.054904} {\bibfield  {journal} {\bibinfo
  {journal} {Phys. Rev. C}\ }\textbf {\bibinfo {volume} {103}},\ \bibinfo
  {pages} {054904} (\bibinfo {year} {2021}{\natexlab{b}})},\ \Eprint
  {http://arxiv.org/abs/2011.01430} {arXiv:2011.01430 [hep-ph]} \BibitemShut
  {NoStop}%
\bibitem [{\citenamefont {Bernhard}\ \emph {et~al.}(2019)\citenamefont
  {Bernhard}, \citenamefont {Moreland},\ and\ \citenamefont
  {Bass}}]{Bernhard:2019bmu}%
  \BibitemOpen
  \bibfield  {author} {\bibinfo {author} {\bibfnamefont {J.~E.}\ \bibnamefont
  {Bernhard}}, \bibinfo {author} {\bibfnamefont {J.~S.}\ \bibnamefont
  {Moreland}}, \ and\ \bibinfo {author} {\bibfnamefont {S.~A.}\ \bibnamefont
  {Bass}},\ }\href {\doibase 10.1038/s41567-019-0611-8} {\bibfield  {journal}
  {\bibinfo  {journal} {Nature Phys.}\ }\textbf {\bibinfo {volume} {15}},\
  \bibinfo {pages} {1113} (\bibinfo {year} {2019})}\BibitemShut {NoStop}%
\bibitem [{\citenamefont {Monnai}\ \emph {et~al.}(2017)\citenamefont {Monnai},
  \citenamefont {Mukherjee},\ and\ \citenamefont {Yin}}]{Monnai:2016kud}%
  \BibitemOpen
  \bibfield  {author} {\bibinfo {author} {\bibfnamefont {A.}~\bibnamefont
  {Monnai}}, \bibinfo {author} {\bibfnamefont {S.}~\bibnamefont {Mukherjee}}, \
  and\ \bibinfo {author} {\bibfnamefont {Y.}~\bibnamefont {Yin}},\ }\href
  {\doibase 10.1103/PhysRevC.95.034902} {\bibfield  {journal} {\bibinfo
  {journal} {Phys. Rev. C}\ }\textbf {\bibinfo {volume} {95}},\ \bibinfo
  {pages} {034902} (\bibinfo {year} {2017})},\ \Eprint
  {http://arxiv.org/abs/1606.00771} {arXiv:1606.00771 [nucl-th]} \BibitemShut
  {NoStop}%
\bibitem [{\citenamefont {Dore}\ \emph {et~al.}(2022)\citenamefont {Dore},
  \citenamefont {Karthein}, \citenamefont {Mroczek}, \citenamefont {Parotto},
  \citenamefont {Noronha-Hostler},\ and\ \citenamefont {Ratti}}]{Dore:2021sbl}%
  \BibitemOpen
  \bibfield  {author} {\bibinfo {author} {\bibfnamefont {T.}~\bibnamefont
  {Dore}}, \bibinfo {author} {\bibfnamefont {J.}~\bibnamefont {Karthein}},
  \bibinfo {author} {\bibfnamefont {D.}~\bibnamefont {Mroczek}}, \bibinfo
  {author} {\bibfnamefont {P.}~\bibnamefont {Parotto}}, \bibinfo {author}
  {\bibfnamefont {J.}~\bibnamefont {Noronha-Hostler}}, \ and\ \bibinfo {author}
  {\bibfnamefont {C.}~\bibnamefont {Ratti}},\ }\href {\doibase
  10.1051/epjconf/202225910001} {\bibfield  {journal} {\bibinfo  {journal} {EPJ
  Web Conf.}\ }\textbf {\bibinfo {volume} {259}},\ \bibinfo {pages} {10001}
  (\bibinfo {year} {2022})},\ \Eprint {http://arxiv.org/abs/2109.05098}
  {arXiv:2109.05098 [nucl-th]} \BibitemShut {NoStop}%
\bibitem [{\citenamefont {Onuki}(1997)}]{PhysRevE.55.403}%
  \BibitemOpen
  \bibfield  {author} {\bibinfo {author} {\bibfnamefont {A.}~\bibnamefont
  {Onuki}},\ }\href {\doibase 10.1103/PhysRevE.55.403} {\bibfield  {journal}
  {\bibinfo  {journal} {Phys. Rev. E}\ }\textbf {\bibinfo {volume} {55}},\
  \bibinfo {pages} {403} (\bibinfo {year} {1997})}\BibitemShut {NoStop}%
\bibitem [{\citenamefont {Buchel}\ and\ \citenamefont
  {Pagnutti}(2010)}]{Buchel:2009mf}%
  \BibitemOpen
  \bibfield  {author} {\bibinfo {author} {\bibfnamefont {A.}~\bibnamefont
  {Buchel}}\ and\ \bibinfo {author} {\bibfnamefont {C.}~\bibnamefont
  {Pagnutti}},\ }\href {\doibase 10.1016/j.nuclphysb.2010.03.016} {\bibfield
  {journal} {\bibinfo  {journal} {Nucl. Phys. B}\ }\textbf {\bibinfo {volume}
  {834}},\ \bibinfo {pages} {222} (\bibinfo {year} {2010})},\ \Eprint
  {http://arxiv.org/abs/0912.3212} {arXiv:0912.3212 [hep-th]} \BibitemShut
  {NoStop}%
\bibitem [{\citenamefont {Natsuume}\ and\ \citenamefont
  {Okamura}(2011)}]{Natsuume:2010bs}%
  \BibitemOpen
  \bibfield  {author} {\bibinfo {author} {\bibfnamefont {M.}~\bibnamefont
  {Natsuume}}\ and\ \bibinfo {author} {\bibfnamefont {T.}~\bibnamefont
  {Okamura}},\ }\href {\doibase 10.1103/PhysRevD.83.046008} {\bibfield
  {journal} {\bibinfo  {journal} {Phys. Rev. D}\ }\textbf {\bibinfo {volume}
  {83}},\ \bibinfo {pages} {046008} (\bibinfo {year} {2011})},\ \Eprint
  {http://arxiv.org/abs/1012.0575} {arXiv:1012.0575 [hep-th]} \BibitemShut
  {NoStop}%
\end{thebibliography}%
\end{document}